\documentstyle[11pt]{article}
  \advance\oddsidemargin  by -1in
  \advance\evensidemargin by -1in
  \advance\topmargin      by -0.5in
\begin{document}
\newcommand {\bbox} {\vrule height7pt width4pt depth1pt}
\newcommand {\ra} {\rangle}
\newcommand {\la} {\langle}
\title{QUANTUM COMPUTATION\footnote{To appear in {\it Annual Reviews of Computational Physics} VI, Edited by Dietrich Stauffer, World Scientific, 1998}}
\author{Dorit Aharonov\\ {\small Departments of Physics and Computer Science,}
\\{\small  The Hebrew University, Jerusalem, Israel}
}

\maketitle

\begin{abstract}
\it{

In the last few years, theoretical study of  
  quantum systems serving as computational devices 
has achieved tremendous progress.
We now have strong theoretical evidence that 
 quantum computers, if built, might be used as a dramatically powerful
computational tool, capable of performing tasks which seem intractable
for classical computers.
This review is about to tell the story of theoretical quantum computation.
I left out the developing topic of experimental 
realizations of the model,  and  neglected  
other closely related topics which are quantum information
and quantum communication.
 As a result of narrowing the scope of this paper, I hope 
it has gained the benefit of being 
an almost self contained
 introduction to the exciting field of quantum computation.

The review begins with background on theoretical computer science, 
Turing machines and Boolean circuits. In light of these models, 
I define  quantum computers, and discuss 
the issue of universal quantum gates.
Quantum algorithms, including  Shor's factorization algorithm and Grover's 
algorithm for searching databases, are explained.
I will devote much attention to understanding what the origins of the quantum 
computational power are, and what the  limits of this power are. 
Finally, I describe the recent theoretical results which show 
that quantum computers 
 maintain their complexity power even in the 
presence of noise, inaccuracies and finite precision.
 This question cannot be separated from that of quantum complexity, 
because any realistic model will inevitably
 be subject to such inaccuracies.
I tried to put all results in their context, asking what the implications to 
other issues in computer science and physics are. 
In the end of this review I make these connections explicit, 
 discussing the possible implications of
 quantum computation on fundamental physical questions,
such as the transition from quantum to classical physics.}

\end{abstract}

\newtheorem{theo}{Theorem}
\newtheorem{lemm}{Lemma}
\newtheorem{conj}{conjecture}
\newtheorem{deff}{Definition}
\newtheorem{coro}{Corollary}

\section{Overview}

Since ancient times, humanity has been   seeking tools to 
help us perform tasks which 
 involve calculations.
Such are computing the 
area of a land,  computing 
the stresses on rods in bridges, 
or finding the shortest route 
from one place to another.
A common feature of all these tasks is 
their structure:
\begin{quote}
\bf ~~~~~~~~~~~~~~~            Input ------$>$ Computation ------$>$ Output
\end{quote}

The computation part of the process is inevitably performed by 
a dynamical physical system, evolving in time.
In this sense, the question of what can be computed,
is intermingled with the physical question of which systems can be physically 
realized.
If one wants to
 perform a certain computation task, one should seek 
the appropriate physical system,
 such that the evolution in time of the system corresponds to 
the desired computation process.
If such a system is initialized according to the input, its 
final state will correspond to the desired output.

A very nice such example was invented by Gaud\'{i}, a great Spanish 
architect, who lived around the turn of the 
century. 
His design of the holy family church, ({\it la sagrada familia})
in Barcelona is a masterpiece of art, and is still in the process 
of building, after almost a hundred years. The church resembles 
a sand palace, with a tremendous complexity of delicate thin but tall
 towers and 
arcs.  
Since the plan of the  church was so complicated,  
 towers and arcs emerging from unexpected places,
 leaning on other arcs and towers, 
it is practically impossible to solve the set of equations 
which corresponds to the requirement of equilibrium in this complex.
Instead of solving this impossible task, Gaud\'{i} thought of the following 
ingenious idea: 
For each arc he desired in his complex, 
he took a rope, of length  proportional to the length of the arc.
He tied the edges of one rope to 
the middle of some other rope, or  where the arcs were supposed 
to lean on each other. 
Then he just tied the edges of the ropes corresponding to the lowest 
arcs, to the ceiling. 
All the computation was instantaneously done by gravity!
The set of arcs arranged itself such that the whole
complex is in equilibrium, but upside down. Everything was there, the 
angles between the different arcs, the radii of the arcs.
Putting a mirror under the whole thing, he could simply see the design of
 the whole church! \cite{gaudi}.

Many examples of analog computers exist, which were invented 
to solve one complicated task.
Such are the differential analyzer invented by Lord Kelvin in 
1870\cite{kelvin}, which uses friction, wheels, 
and pressure to draw the solution of an input differential equations.
The spaghetti sort is another example, and there 
 are many more\cite{vergis}. 
Are these  systems ``computers''?
We do not want 
to construct and build a completely different machine for each task that 
we have to compute. We would rather have a general
 purpose machine, which is ``universal''. 
A mathematical model for a ``universal'' computer  was 
defined long before the invention of computers and is
called the Turing machine\cite{turing}.
Let me describe this model briefly. A Turing machine consists of 
an infinite tape, a head that reads and writes on the tape,
 a machine with finitely many possible states,
 and a transition function $\delta$.
Given what the head reads at time $t$, and the machine's state at time $t$, 
 $\delta$  
determines  what the head will write, to which direction 
 it will move and what will be the new 
machine's state at time $t+1$.
 The Turing machine model seems to capture the entire concept of 
computability, 
according to the following thesis\cite{church}:

\begin{quote}
{\bf Church Turing Thesis:} A
 Turing machine can compute any function computable by a reasonable 
physical device
\end{quote}

What does ``reasonable physical device'' mean? 
This thesis is a physical statement, and as such it cannot be proven.
But one knows a physically unreasonable device when one sees it. 
Up till now there are no candidates for counterexamples to this thesis
(but see Ref. \cite{geroch}).
All physical systems, (including quantum systems), seem to 
have a simulation by a Turing Machine.

It is an astonishing fact that  there are families of functions which 
cannot be computed. In fact, most of the functions cannot be computed.
There are trivial reasons for this:
 There are more functions
than there are ways to compute them.
The reason for this is that the set of  Turing machines is  
countable, where as the set of {\it families} of functions is not.
In spite of the simplicity of this argument (which can be formalized using 
the {\it diagonal argument}) this observation came  
 as a complete surprise in the 1930's when it was first discovered.
The subject of computability of functions is a cornerstone in 
computational complexity. However, 
in the theory of computation, we are interested not only  
in the question of  which functions can be computed,
 but  mainly in the {\it cost}
of computing these functions.
The cost, or {\it computational complexity}, 
is measured naturally  by the physical resources invested in order to
 solve the problem, such as 
time, space, energy, etc.
A fundamental question in computation complexity 
 is how   the cost function behaves as a function of the input size, $n$, 
and in particular  whether it is 
exponential or polynomial in $n$.
In computer science problems
 which can only be solved in exponential cost 
 are regarded intractable, and 
any of the readers who has ever 
 tried to perform an exponentially slow simulation will appreciate 
this characterization. The class of tractable problems constitutes of those 
problems which have polynomial solutions.

It is worthwhile to reconsider what it means to {\it solve} a problem.
One of the most important conceptual breakthroughs in modern 
mathematics was the 
understanding\cite{rabin79} that sometimes it is advantageous to
 relax the requirements that a solution be always correct, 
 and allow some (negligible ) probability for an error.
This gave rise to much more rapid solutions to 
different problems, which make use of random coin flips, 
such as the Miller-Rabin randomized algorithm to test whether 
an integer is prime or not\cite{fft}.
Here is a simple example of the advantage of probabilistic algorithms:

\begin{quote}
 we have access to a database of
 $N$ bits, and we are told that they are either all equal,
(``constant'') 
or half are $0$ and half are $1$ (``balanced'').
We are asked to distinguish between the two cases.
\end{quote}

A deterministic algorithm will have to observe  $N/2+1$ bits in order to
 always give a correct 
answer. To solve this problem 
probabilistically, toss a random $i$ between $1$ to $N$,  observe the $i'$th
 bit, and repeat this experiment $k$ times. 
If  two different bits are found, the answer is ``balanced'', and 
if all bits are equal, the answer is  ``constant''. 
Of course, there is a chance that we are wrong when declaring ``constant'',
but this chance can be made arbitrarily small.  The probability for an error
 equals 
the chance of tossing a fair coin $k$ times  and getting 
always $0$, and it decreases exponentially with $k$.
For example, in order for the error probability to be less than  $10^{-10}$, 
  $k=100$ suffices. In general, for any desired confidence, a constant 
$k$ will do. This  
 is a very helpful shortcut if $N$ is very large.
Hence,  if we allow negligible probability of error, 
 we can do much better!

The class of tractable problems is now considered 
as those problems solvable with a negligible 
probability for error 
in polynomial time. 
These solutions will be computed by a probabilistic Turing machine,
which is defined exactly
as a deterministic Turing machine, except 
that the transition function 
can change  the configuration in one of several possible ways, 
randomly. The modern Church thesis refines the Church thesis
and asserts that 
the probabilistic Turing machine
 captures the entire concept of computational
complexity:

\begin{quote}{\bf  The modern Church thesis}:
A probabilistic Turing machine can simulate  
 any reasonable 
physical device in polynomial cost.
\end{quote}

It is worthwhile considering a few models which 
 might seem to contradict this thesis at first sight. 
One  such model is the  DNA computer which enables a solution
of $NP$-complete problems (these are hard problems to be defined later) 
 in polynomial time\cite{adleman2, lipton}. However, the cost of the 
solution is exponential 
 because the number of molecules in the system grows exponentially 
with the size of the computation.
Vergis et al\cite{vergis} suggested a machine which 
 seems to be able to solve instantaneously an $NP$-complete
problem 
using a construction of  rods and 
balls, which is designed  such that the
 structure moves according to the solution 
to the problem.
A careful consideration\cite{simon2}
reveals that though we tend to think of rigid rods as transferring 
the motion instantaneously, there will be a
  time delay in the rods, which will 
accumulate and cause an exponential 
overall delay. Shamir\cite{shamir} showed how to 
 factorize
 an integer in polynomial time {\em and} space, 
but using another physical resource exponentially, namely
 precision. In fact, J. Simon showed that 
 extremely hard problems 
(The class of problems called Polynomial space, which 
are harder than NP problems) can be solved   with polynomial 
cost in time and space\cite{jsimon}, but with exponential precision.
Hence all these suggestions for  computational models
do not provide counterexamples for the modern Church thesis, since they 
require exponential physical resources.
However, note that all the suggestions mentioned above rely on classical 
physics.

In the early 80's Benioff\cite{benioff1,benioff2} and Feynman\cite{feynman2}
 started to 
discuss the question of whether 
computation can be done in the scale of quantum physics.
In classical computers, the elementary information unit is 
a {\it bit}, i.e. a value which is either $0$ or $1$. 
The quantum analog of a bit would be  a two state 
particle, called a quantum bit or a {\bf qubit}.
 A two state quantum system is 
described by a unit vector in the Hilbert space $C^2$, where 
$C$ are the complex numbers.  
One of the two states will be denoted by $|0\ra$,
and corresponds to the vector $(1,0)$.
The other state, which is orthogonal to the first one, 
will be denoted by $|1\ra=(0,1)$. These two states constitute an 
orthogonal basis to the Hilbert space.
To build a computer, we need to compose a large number 
of these two state particles.
When $n$ such qubits are composed to one system,
their Hilbert space is the tensor product of $n$ spaces: 
$C^2\otimes C^2\otimes \cdots \otimes C^2$.
To understand this space better, it is best to think of it as 
the space spanned by its basis. 
As the natural basis for this space, 
we take the basis consisting of $2^n$ vectors, 
which is sometimes called the computational basis:

  \begin{eqnarray}
   |0\ra\otimes|0\ra\otimes\cdots\otimes|0\ra~\\
  |0\ra\otimes|0\ra\otimes\cdots\otimes|1\ra\nonumber~\\
   \vdots~~~~~~~~~~~~~~~~~~~~~\nonumber\\
   |1\ra\otimes|1\ra\otimes\cdots\otimes|1\ra\nonumber.
   \end{eqnarray}

Naturally  classical strings of bits will correspond  
to quantum states:
\begin{equation}
i_1i_2...i_n \longleftrightarrow |i_1\ra\otimes|i_2\ra\otimes\cdots\otimes|i_n\ra\equiv|i_1....i_n\ra
\end{equation}

How can one perform computation using qubits?
Suppose, e.g., that we want to compute the function  
$f:i_1i_2...i_n \longmapsto f(i_1,....i_n)$,
from $n$ bits to $n$ bits. 
We would like the system to evolve according to 
 the time evolution operator $U$: 
\begin{equation}\label{f}
|i_1i_2...i_n\ra \longmapsto U|i_1i_2...i_n\ra=|f(i_1,....i_n)\ra.
\end{equation}
We therefore have  to find the Hamiltonian ${\cal H}$
 which generates this evolution according to 
    Schr$\ddot{{\rm o}}$dinger's equation:
         \(i\hbar\frac{d}{dt}|\Psi(t)\ra={\cal H }|\Psi(t)\ra\).
This means that we have to solve for ${\cal H}$ given the desired $U$:

\begin{equation}\label{evu}
|\Psi_f\ra=
\exp\left(-\frac{i}{\hbar}\int{\cal{H}}dt\right)|\Psi_0\ra=
 U|\Psi_0\ra
\end{equation}

A solution for 	 ${\cal H}$ always exists,
 as long as the linear operator $U$ is unitary.
It is important to pay attention to the unitarity restriction.
Note that the quantum analog of a classical operation will be unitary 
only if $f$ is one-to-one, or reversible.
Hence, reversible classical function can be implemented by 
a physical Hamiltonian.  
Researchers investigated 
 the question of reversible classical 
functions   in connection with completely different problems, e.g.
 the  problem of whether 
computation can be done without generating heat
(which is inevitable  in irreversible operations)
and as a solution to the ``maxwell demon'' paradox\cite{landauer2,bennett2,bennett4,keyes2}.
It turns out  that any classical function can be represented
as a reversible function\cite{lecerf,bennett1}
on a few more bits, and the computation of $f$ 
can be made reversible without losing much in efficiency. Moreover,
 if  $f$ can be computed classically by 
polynomially many elementary reversible steps,
 the corresponding  $U$ is also decomposable into a sequence of polynomially  
many elementary unitary operations.
We see that quantum systems can imitate 
 all computations 
which can be done by classical systems, and  do not lose much in
 efficiency.

 Quantum computation is interesting not because it can imitate 
classical computation, but because it can probably do much more.
In a seminal paper\cite{feynman1}, Feynman
 pointed out  the fact that quantum systems of $n$ particles seem 
exponentially hard to simulate by 
classical devices.
In other words, quantum systems do not seem to
obey the modern Church thesis, i.e.  they do not seem to be 
        polynomially equivalent 
to classical systems!
If quantum systems are hard to simulate, then quantum systems, 
 harnessed as 
computational devices, might be dramatically more powerful 
than other computational devices.

Where can the ``quantumness'' of the particles be used?
When I described how quantum systems  imitate  
 classical computation, the quantum particles were either 
in the state $|0\ra$ or $|1\ra$. However, quantum theory asserts that 
a quantum system, like   Schr$\ddot{\rm{o}}$dinger's cat, 
need not be in one of the basis states $|0\ra$ and $|1\ra$,
 but can also  be in a 
{\it linear superposition} of those.
Such a superposition can be written as:
\begin{equation}
c_0 |0\ra + c_1 |1\ra
\end{equation}
where $c_0,c_1$ are complex numbers and $|c_0|^2+|c_1|^2=1$.
The wave function, or superposition, of $n$ such 
quantum bits,  can be in  a superposition 
 of all of the $2^n$ possible basis states!
Consider for example the following state of $3$ particles, known as 
the GHZ state\cite{ghz}: 
\begin{equation}
\frac{1}{\sqrt{2}}(|000\ra +|111\ra)
\end{equation}
What is the superposition describing the first qubit?
The answer is that there is no such superposition.
Each one of the $3$ qubits does not have a state of its own;
 the state of the system is not a tensor product of the states of 
each particle, but is some superposition which describes 
quantum correlations between these particles. Such particles are said 
to be quantumly  {\it entangled}.
The Einstein Podolski Rosen paradox\cite{epr}, 
and Bell inequalities\cite{bell,bell1, clauser,ghz},
 correspond to this 
puzzling quantum feature by which a quantum particle does not have a state of 
its own.
Because of the entanglement or quantum correlations 
 between the $n$ quantum particles, the state of the system 
cannot be specified by simply describing the state of each of the $n$ 
particles.
Instead,  the state of $n$ quantum bits
 is a complicated superposition of all 
$2^n$ basis states, so
 $2^n$ complex coefficients are needed
in order to describe it. 
This exponentiality of the Hilbert space is a crucial ingredient 
in quantum computation. 
To gain more understanding of 
the advantages of the  exponentiality of the space, 
 consider the following
 superposition of $n$ quantum bits.
\begin{equation}
\frac{1}{\sqrt{2^n}}\sum_{i_1,i_2,...,i_n=0}^{1} |i_1,i_2,...,i_n\ra
\end{equation}
This is a uniform superposition of all possible basis states of $n$ qubits.
If we  now apply the unitary operation which computes
 $f$, from equation \ref{f},  to this state, 
 we will get, simply  from linearity of quantum mechanics:
\begin{equation}
\frac{1}{\sqrt{2^n}}\sum_{i_1,i_2,...,i_n=0}^{1}
 |i_1,i_2,...,i_n\ra\longmapsto
\frac{1}{\sqrt{2^n}}\sum_{i_1,i_2,...,i_n=0}^{1} |f(i_1,
i_2,...,i_n)\ra.
\end{equation} 
Applying $U$ once  computes $f$ simultaneously 
 on all the $2^n$ possible inputs $i$, 
which is an enormous power of parallelism!

It is tempting to think that exponential parallelism 
immediately implies exponential computational power, 
but this is not the case. In fact, 
classical computations can be viewed as having exponential 
parallelism as well-- we will devote much attention to this later on.
The problem lies in the question of how to 
 extract the 
exponential information out of the system. 
In quantum computation, in order
 to extract quantum information one has to {\it observe} the system. 
 The measurement process  causes the famous {\it collapse of the wave
 function}. In a nutshell, this means that after the measurement
the state is projected to 
 only one of the exponentially many possible states, 
so that the exponential 
amount of information which has been computed is completely lost!
In order to gain advantage of
 exponential parallelism, one needs to combine it with another
 quantum feature,  known as interference. 
Interference allows the exponentially many computations 
done in parallel to cancel each other, just like destructive 
interference of waves or light.  The goal is to 
 arrange the  cancelation such that only 
the computations which we are interested in remain, and all the rest 
cancel out.  The combination of exponential parallelism and interference 
is  what makes 
quantum computation powerful, and  plays an important role in quantum algorithms.

A quantum algorithm is a sequence of elementary unitary steps, 
which manipulate the initial quantum state $|i\ra$ (for an input $i$)
such that a measurement of the final state of the system
yields the correct output.
The first quantum algorithm which combines interference and exponentiality
to solve a problem faster than classical computers, was  
discovered by   Deutsch and Jozsa\cite{deutsch3}.
This algorithm addresses the problem we have encountered before
in connection with probabilistic algorithms: 
Distinguish between   ``constant'' and ``balanced'' databases.
The quantum algorithm solves this 
problem  {\it exactly}, in polynomial  cost.
 As we have seen, classical  computers cannot do this,
and must release the restriction of exactness.
Deutsch and Jozsa made use of the most powerful tool in quantum algorithms, 
the {\it Fourier transform}, which indeed manifests
 interference and  exponentiality. 
 Simon's algorithm\cite{simon} uses similar techniques, and was the
seed for the most important quantum algorithm
known today: Shor's algorithm.

Shor's algorithm (1994) is a polynomial quantum algorithm for 
factoring integers, and for finding the  logarithm over 
a finite field\cite{shor1}.
For both problems, the best  known classical 
           algorithms are exponential.
 However, 
there is no proof that  classical efficient algorithms 
do not exist.
Shor's result is regarded as extremely important both theoretically and 
practically,  mainly due to  the fact that 
the assumption that factorization is hard
 lies in the heart of the $RSA$ cryptographic system 
\cite{rsa,fft}. A cryptosystem is supposed to be
a secure way to transform information such that an eavesdropper 
will not be able to learn in reasonable time
significant information about the message sent.
The RSA cryptosystem is used very heavily: The CIA uses it, 
 the security embedded into Netscape and the Explorer Web
browsers is based on RSA, banks use RSA for internal security
as well as securing external connections.
However, RSA can be cracked  
by any one who has an efficient  algorithm  for factoring. It is therefore understandable why the publication of the factorization algorithm caused a rush of excitement
all over the world.

It is important that the quantum computation power 
does not rely on unreasonable precision but a polynomial amount of precision 
in the computational elements is enough\cite{bv}.
This means that the new model requires 
physically reasonable resources, in terms of time, space, and
precision,   but yet  it is (possibly) exponentially stronger than 
the ordinary model of probabilistic Turing machine.
As such, it is the only model which really threatens the 
modern Church thesis.

There are a few major developing directions of research in the area of
 quantum computation.
In $1995$ Grover\cite{grover1} discovered an algorithm which searches 
an unsorted database of $N$ items and finds a specific
 item in $\sqrt{N}$ time steps. 
This result is surprising, because intuitively, 
one cannot search the database without going through all the items.
Grover's solution is
 quadratically better than any possible
classical algorithms, and  
 was followed by numerous extensions
and  applications\cite{boyer1,grover2,grover3,durr,brassard3,brassard4}, 
all achieving polynomial advantage over classical algorithms.
 A promising new branch in quantum complexity theory is the study 
            of a class of problems which
 is the quantum analog of the complexity 
            class NP\cite{kitaevNP}. 
 Another interesting direction in quantum computation
is concerned with quantum computers simulating 
efficiently other physical systems such as 
many body Fermi systems\cite{zalka1,abrams,
weisner2,bogosian}.
This direction pursues the original suggestion
by Feynman\cite{feynman1}, who noticed that quantum systems are 
hard to simulate by classical devices. 
An important direction
 of investigation is the search for a different, perhaps
stronger, quantum computation model.
For example, consider the introduction of 
slight non-linearities into quantum mechanics. This is 
 completely hypothetical, as all experiments verify the linearity 
of quantum mechanics. However, such slight non linearities 
  would imply 
extremely strong quantum algorithms\cite{ abrams2}.
A very interesting quantum computation model 
which  is based on anyons, and uses  non-local features of quantum
 mechanics,
 was suggested by
Kitaev\cite{kitaev3}. 
A possibly 
much stronger model, based on quantum field theory, 
was sketched recently by Freedman, but it has not been rigorously defined yet\cite{freedman}.
One other direction is oracle results in 
 quantum complexity.
This direction   compares 
 quantum complexity  power and   classical complexity power
 when the two models are allowed to have access to an oracle, 
i.e. a black box which can compute a certain (possibly 
difficult) function in one step 
\cite{bv, bbbv,
bert2,bert3}. In fact, the result 
of Bernstein and Vazirani\cite{bv}  
from $1993$  demonstrating a superpolynomial 
gap between quantum and classical computational comlexity 
with an access to a certain oracle
initialized the sequence of results leading to the Shor's 
algorithm. An important recent result\cite{beals2}
in quantum complexity 
 shows  that quantum computers 
have no more than polynomial advantage 
 in terms of number of accesses to the inputs.  
As of now, we are very far from  understanding  the computational power
of quantum systems. In particular, it is not
 known whether quantum systems can efficiently solve 
 $NP$ complete problems or not.

Quantum information theory, a subject which 
is intermingled with quantum computation, 
provides a bunch of quantum magic tricks,
 which might be used to construct more 
powerful quantum algorithms.
Probably the first ``quantum pearl'' that one encounters in 
quantum mechanics 
 is the Einstein Podolsky Rosen
paradox,
 which, as is best explained by Bell's inequalities, 
establishes the existence of correlations between quantum particles, 
which are stronger than any classical model can explain. 
Another ``quantum pearl'' 
 which builds on quantum  entanglement,
  is teleportation\cite{bennett13}. 
This is an amazing quantum recipe which  
 enables two parties (Alice and Bob) which are far apart,
 to transfer an unknown quantum state of a  particle in Alice's hands   
onto a particle in Bob's hand, without sending the actual particle. 
This can be done if Alice and Bob share a pair of particles
which interacted in the past and therefore are quantumly
entangled. 
Such quantum effects already serve as ingredients
in different computation and communication tasks.
Entanglement can be used, for example, in order 
to gain advantage in communication. If two parties, Alice and Bob, 
want to communicate, they can save bits of communication 
if they  share entangled pairs of qubits\cite{cleve2,cleve3, cleve4,barenco2}. 
Teleportation can be viewed as a quantum computation\cite{brassard5},
  and beautiful connections 
            were drawn\cite{bennett14} between teleportation and 
quantum algorithms which are used to correct quantum noise. 
 All these are uses 
of quantum effects in quantum computation. However, 
 I  believe that the
 full potential of 
 quantum mechanics in the context of complexity and 
algorithmic problems is yet 
to be revealed.

Despite the impressive progress in quantum computation,
 a menacing question still remained.
Quantum information is extremely fragile, due to inevitable interactions 
between the system and its environment.   
These interactions cause the system to lose part of its quantum nature, 
a process called {\it decoherence}\cite{stern1,zurek1}.
In addition, quantum elementary operations (called {\it gates})
 will inevitably suffer from inaccuracies.
Will physical realizations of the model 
of quantum  computation  still be as powerful as the ideal model?
In classical computation, it was already shown by von-Neumann\cite{neumann}
how to compute when the elements of the computation are faulty, 
using redundant information. 
Indeed, nowadays error corrections are seldom used in 
computers because of extremely high reliability of the elements, 
but quantum elements are much more fragile, 
and it is almost certain that quantum error corrections will be 
necessary in future quantum computers.
It was shown that if the errors are not corrected
during quantum computation, 
 they soon accumulate and ruin the entire 
computation\cite{decoherence,decoherence2,barenco6, miquel1}.
Hence, a method to correct the effect of quantum noise is necessary.
Physicists
 were pessimistic about the question of whether 
such a correction method exists\cite{landauer1,unroh1}.
The  reason is that  quantum information in general cannot be 
cloned\cite{dieks,wootters,barnum2}, and so 
the information cannot be simply protected by redundancy, as is 
done classically. Another problem is that  in contrast to the
 discreteness of digital computers,
a quantum system can be in a superposition of 
eigenstates with continuous coefficients.
Since the range of allowed coefficients is continuous, it seems
impossible to     
distinguish between bona fide information and information 
which has been contaminated.

As opposed to the  physical intuition, it turns out that
clever techniques enable  
quantum information to be protected.
The conceptual breakthrough in quantum error corrections  
 was the understanding that quantum errors, 
which are continuous, can be viewed as a discrete process in which one out 
of  four quantum operations occurs.
Moreover, these errors can be viewed as classical errors, called bit 
flips, and quantum errors, called phase flips.
 Bit flip errors can be corrected using classical error correction techniques. Fortunately, 
phase flips transform to bit flips, using the familiar 
Fourier transform.
 This understanding allowed using classical 
error correction codes techniques in the quantum setting.
Shor was the first to  present  
 a scheme that reduces the affect of noise and inaccuracies, 
building on the discretization of errors\cite{shor2}.
As in classical error correcting codes, quantum states 
of $k$ qubits are {\it encoded} on states of more qubits.  
 Spreading the state of a few qubits on more 
qubits, allows correction of the information, if part of it 
has been contaminated.
These ideas were extended \cite{calshor,steane1}
to  show that a quantum state of $k$ qubits
 can be encoded on $n$ qubits, such that if the $n$ qubits 
are sent through a noisy channel, the original state of the 
$k$ qubits can be recovered. $k/n$ tends asymptotically to
a constant {\it transmission rate} which is non zero. 
 This is analogous to Shannon's result from noisy 
classical communication\cite{shannon}. 
Many different examples of quantum 
 error correcting codes 
followed\cite{steane2,laflamme2,chuang1,knill5,
rains2, leung}, 
and a group theoretical framework for most quantum codes was
 established\cite{gf4,calderbank3,gottesman2}.

Resilient quantum computation is more complicated than simply 
protecting quantum information which is sent through 
a noisy quantum channel. Naturally, to protect the information 
we would compute on encoded states.
There are two problems with noisy computation on encoded states.
 The first is that the error correction 
is done with faulty gates, which cause errors  themselves\cite{barenco7}.
We should be careful that the error correction does not cause 
more harm than it helps.
The second problem is that 
  when computing on encoded states, qubits interact with each
other through the  gates, and this way
 errors can {\it propagate} through the gates, 
from one qubit to another. The error can spread in this way 
to the entire set of qubits very quickly. 
In order to deal with these problems, 
the idea is to perform computation and error correction 
in a {\it distributed manner}, such that each qubit 
can effect only a small number 
of other qubits.
 Kitaev\cite{kitaev2}
 showed how to perform the computation
of error correction with faulty gates. Shor discovered\cite{shor3} how 
 to perform a general computation
 in the presence of noise, under the unphysical assumption that 
 the noise decreases (slowly) with
 the size of the computation. 
 A more physically reasonable assumption would be that the devices 
           used in the laboratory have a constant amount of noise,
 independent of 
the size of the computation. 
 To achieve fault tolerance against such noise, we  apply a 
concatenation of Shor's scheme. We encode the state once,
and then encode the encoded state, and so on for 
for several levels.
 This technique enabled
  the proof of the {\it threshold theorem}\cite{knill1,
knill2,gottesman5,aharonov1,
kitaev3,preskill2}, which asserts that
it is possible to perform resilient quantum computation 
for as long as we wish,
 if the noise is smaller than 
a certain {\it threshold}. 
Decoherence and imprecision are therefore  no longer considered
insurmountable obstacles to realizing a quantum computation.

In accord with these theoretical optimistic
 results,   attempts at implementations of quantum circuits  are now
 being carried out all 
over the world.  Unfortunately, the progress in this direction is much slower 
than the impressive pace in which  theoretical quantum computation
has progressed. 
The reason is that
handling quantum systems experimentally is extremely difficult.
Entanglement
is a necessary ingredient in quantum computers,
but experimentally, it is a fragile property which 
is difficult to create and preserve\cite{cirac3}.
So far, entangled pairs 
of photons were created successfully\cite{kwiat,tittel}, and 
entanglement features such as violation 
of Bell inequalities were demonstrated
\cite{aspect1,aspect2}. Even entangled pairs of atoms were 
created\cite{hagley}. 
However  quantum computation is advantageous only when   
macroscopically many particles are entangled\cite{jozsa2,aharonov2},
 a task which seems impossible as of now.
Promising experimental developments come from the closely related
subject of quantum cryptography\cite{brassard6,bennett13,brassard2}. 
Quantum communication was successfully tested\cite{hughes,mattle}. 
Bouwmeester {\it et. al.} have recently reported
on experimental realization  of quantum teleportation\cite{bouwmeester} . 
Suggestions  for  implementations
of quantum computation \cite{
cirac1,cory4,gershenfeld,lloyd3,div-rev,jones2,berman1,
loss, jones2,pellizzari,privman,steane6} include  
 quantum dots,
cold trapped ions and nuclear magnetic resonance, and some of these 
suggestions were already implemented
\cite{monroe2,turchette,mattle,gershenfeld,cory2}.
Unfortunately, these implementations were so far limited to three qubits.
With three qubits it is possible to perform partial error correction, 
and successful implementation of error correction of phases 
using NMR was reported\cite{cory3,chuang2}. 
Using nuclear magnetic resonance 
techniques, a quantum algorithm was implemented which achieves 
proven advantage over classical algorithms\cite{chuang3}.
It should be noted, however, that all these suggestions for implementation
 suffer from severe 
problems. In nuclear magnetic resonance   the signal-to-noise ratio
 decays exponentially
with the number of qubits\cite{warren1}, though 
            a theoretical solution to this problem was given recently\cite{umesh}.
Other implementations do not allow 
parallel operations, which 
are necessary for error resilience\cite{aharonov2}. 
In all the above 
systems controlling    thousands of qubits seems hopeless at present.
Never the less, the experimental successes encourage 
our hope that the ambitious task 
of realizing quantum computation might be possible.

The exciting developments in quantum computation
give rise to deep new  open questions in both the fields
of computer science and physics. 
In particular,  computational complexity questions  
 shed new light on old questions in fundamental 
quantum physics such as the 
transition from quantum to classical physics, and the measurement 
process. I shall discuss these interesting topics at the end of the paper. 

We will start with 
 a survey of the important concepts connected to computation, in section $2$.
            The model of quantum computation is  defined in section $3$. 
            Section $4$ discusses elementary quantum operations. 
            Section $5$ describes basic quantum algorithms by Deutsch and Jozsa's and by Simon.
            Shor's factorization algorithm is presented in section $6$, 
            while Fourier transforms are discussed separately in section $7$, 
            together with an alternative factorization algorithm by Kitaev.  
            Grover's database search and variants are explained in section $8$. 
            Section $9$ discusses the origins for the power of  quantum computation, 
            while section $10$ discusses weaknesses of quantum computers.
            Sections $11,12$ and $13$ are devoted to noise, error correction and 
            fault tolerant computation. In Section $14$ I conclude with a few
            remarks of a philosophical flavor.

\section{What is a Computer?}

             Let us discuss now the basic notions of computational complexity theory: 
             Turing machines, Boolean circuits, computability and computational complexity. 
             The important complexity classes $P$ and $NP$ are also defined in this section.
             For more background, consult \cite{fft,papa}. 
             We begin with the definition of a Turing machine:

\begin{deff}
A Turing machine (TM) is  a triplet $M=(\Sigma,K,\delta)$.  
\begin{enumerate}
\item $\Sigma=\{\sqcup,0,1,...\}$ is a finite set of symbols
which we call the alphabet.  
$\sqcup$  is a special symbol called the blank symbol.
\item  $K$ is a finite set of ``machine states'', 
with two special states: $s\in K$  
the initial state and $h\in K$ the final state.
\item A transition function $
\delta: 
K\times \Sigma \longmapsto  K\times
   \Sigma\times \{-1,0,1\} $
\end{enumerate}
\end{deff}

The machine works as follows:
the tape has a head which  can read and write on the tape
during the computation.
The tape is thus  used as working space, or memory.
The computation  starts with an input of $n$ symbols written
in positions $[1,...n]$ 
 on  the tape,  
 all symbols except these $n$ symbols  
are blank ($\sqcup$), the head is initially at position $1$, 
and the state is initially $s$.
Each time step, the machine evolves according to the transition function 
$\delta$ in the following way. 
If the current state of the machine is $q$ and the 
symbol in the current place of the tape is $\sigma$, 
and  $\delta(q, \sigma)=(q',\sigma',\epsilon)$, 
then 
 the machine state is changed to  $q'$, the symbol under the head
is replaced by $\sigma'$
 and the tape head moves one step in direction 
$\epsilon$. (if $\epsilon=0$ the head doesn't move).
Here is a schematic description of a Turing machine:

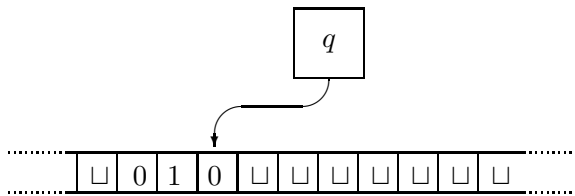
\setlength{\unitlength}{0.030in}

\begin{picture}(40,35)(-60,0)

\put(20,0){\line(1,0){80}}
\qbezier[10](10,0)(15,0)(20,0)
\qbezier[10](100,0)(105,0)(110,0)
\put(22,0){\line(0,1){7}}\put(26,3){\makebox(0,0){$\sqcup$}}
\put(29,0){\line(0,1){7}}\put(33,3){\makebox(0,0){0}}
\put(36,0){\line(0,1){7}}\put(39,3){\makebox(0,0){1}}
\put(43,0){\line(0,1){7}}\put(46,3){\makebox(0,0){0}}
\put(50,0){\line(0,1){7}}\put(54,3){\makebox(0,0){$\sqcup$}}
\put(57,0){\line(0,1){7}}\put(61,3){\makebox(0,0){$\sqcup$}}
\put(64,0){\line(0,1){7}}\put(68,3){\makebox(0,0){$\sqcup$}}
\put(71,0){\line(0,1){7}}\put(75,3){\makebox(0,0){$\sqcup$}}
\put(78,0){\line(0,1){7}}\put(82,3){\makebox(0,0){$\sqcup$}}
\put(85,0){\line(0,1){7}}\put(89,3){\makebox(0,0){$\sqcup$}}
\put(92,0){\line(0,1){7}}\put(96,3){\makebox(0,0){$\sqcup$}}
\put(20,7){\line(1,0){80}}
\qbezier[10](10,7)(15,7)(20,7)
\qbezier[10](100,7)(105,7)(110,7)
\put(60,20){\framebox(12,12){$q$}}
\put(51,20){\oval(30,10)[br]}
\put(51,10){\oval(10,10)[tl]}
\put(46,10){\vector(0,-1){2}}
\end{picture}

{~}

Note that the operation of the Turing machine is local: 
It depends only on the current state of
 the machine and the symbol written in the current 
position of the tape.
Thus the operation of the machine is a sequence of {\it elementary
steps} which require a constant amount of effort. 
If the machine gets to ``final state'', $h$, we say that 
the machine has ``halted''.
 What is written at that time on the tape should contain 
the output. (Typically, the output will be given 
in the form ``yes'' or ``no''.)  One can easily construct examples in which 
the machine never halts  on a given input, 
for example by entering into an infinite loop.

According to the definition above, there are many possible Turing machines, 
each designed to compute a specific task, according to the transition 
function. 
However, there exists one Turing machine, $U$ which 
when presented with an input, it interprets this input 
as a description of another  Turing 
machine, $M$, concatenated with the description of the input to $M$, 
call it $x$. $U$ will simulate  efficiently
the behavior of $M$ when presented with the input $x$, and 
we write $U(M,x)=M(x)$. This $U$ is a called a 
 {\it universal} Turing machine. 
More precisely, the description of $M$ should be given with 
some fixed notation. Without loss of generality, all the symbols 
and states of $M$ can be given numbers from $1$ to $|K|+|\Sigma|$. 
The description of $M$ should contain $|K|$, $|\Sigma|$
and the transition function, which will be described 
by a set of rules (which is finite) of the form $((q, \sigma)
(q',\sigma',\epsilon))$.
For this, $U$'s set of symbols will contain the symbols
$''(``$ and $'')''$ apart from $\sqcup,0,1$.
$U$ will contain a few machine states, such as:
``$q_1$: now reading input'', ``$q_2$: looking for an appropriate
rule to apply'' and so on. I will not go through the details, 
but it is convincing that  with such a finite  set of states, 
$U$ can simulate the operation of any $M$ on any input $x$,  because the entire set of rules of 
the transition function is written on the tape.

The existence of a universal Turing machine leads naturally to the 
deep and beautiful subject of {\it non-computability.}
  A function is non-computable if it cannot be computed by a Turing machine, 
             i.e. there is no  Turing machine which for any given input, halts and 
             outputs the correct answer.
The most famous example is the HALTING problem. 
The problem is this:
Given a description of a Turing machine $M$ and its input $x$, 
will $M$ halt on $x$?

\begin{theo}
There is no Turing machine that solves the HALTING problem
on all inputs $(M,x)$.
\end{theo}

{\bf Proof:}
The proof of this theorem is conceptually puzzling. It uses the so called 
diagonal argument.
Assume that $H$ is a Turing machine, such that 
$H(M,x)$ is $''yes'' $ if $M(x)$ halts and $''no''$ otherwise.
Modify $H$  to obtain $\tilde{H}$, such that

\begin{eqnarray*}
H(M,M)=''\rm{yes}''& 
\longmapsto &\tilde{H}(M)~ \rm{enters ~ an ~infinite~ loop}.\\
H(M,M)=''\rm{no} ''& \longmapsto &\tilde{H}(M)=''\rm{yes}''.
\end{eqnarray*}

 The modification is done easily by replacing a few rules in the transition 
             function of $H$. 
             A  rule which writes "yes" on the tape and causes $H$ to halt  
             is replaced by a rule that takes the machine into an infinite loop. 
             A rule which  writes "no" on the tape and causes $H$ to halt 
             is replaced by a rule that writes "yes" on the tape and than halts $H$.  
             This way, $\tilde{H}$ is a "twisted" version of $H$. Now, does 
              $\tilde{H}(\tilde{H})$ halt or not? We obtain a contradiction in both ways.
Suppose it does halt. This means that $H(\tilde{H},\tilde{H})=''no''$
so  $\tilde{H}(\tilde{H})$ does not halt!
If  $\tilde{H}(\tilde{H})$ does not halt, this means 
  $H(\tilde{H},\tilde{H})=''yes''$ so  $\tilde{H}(\tilde{H})$ does halt! $\bbox$

This beautiful proof shows that there are functions which cannot 
be computed. The Turing machine is actually used to {\it define}
which functions are computable and which are not.

It is sometimes 
more convenient to use another universal model, 
 which is polynomially equivalent to Turing machines, called the Boolean
circuit model. We will use the quantum analog of this model 
throughout this review.
 A Boolean circuit is a directed acyclic graph, 
with nodes which are associated with Boolean functions. 
These nodes are sometimes called {\it logical gates}.
A node with $n$ input wires and $m$ output wires is associated 
with a function $f:\{0,1\}^n \longmapsto \{0,1\}^m$. 
Here is a simple example:

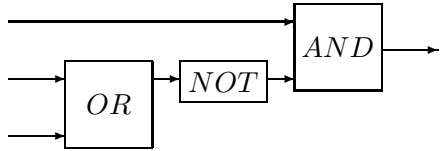
\setlength{\unitlength}{0.030in}
\begin{picture}(40,40)(-80,5)
\put(0,15){\vector(1,0){10}}
\put(0,25){\vector(1,0){10}}
\put(0,35){\vector(1,0){50}}
\put(10,13){\framebox(15,15){$OR$}}
\put(25,25){\vector(1,0){5}}
\put(30,21){\framebox(15,7){$NOT$}}
\put(45,25){\vector(1,0){5}}
\put(50,23){\framebox(15,15){$AND$}}
\put(65,30){\vector(1,0){10}}
\end{picture}
{~}

Given some  string of bits as input, 
the wires carry the values of the bits, until a node is reached.
The node computes a logical function of the bits 
(this function can be NOT, OR, AND, etc.)
 The output wires of the node,   
 carry the output bits to the next node, until the 
computation ends at the output wires.
The input wires can carry {\it constants} which do not vary with 
the different inputs to the circuit, but
 are part of the hardware of the circuit.
In a Turing machine the transition function is local, 
so the operation is a sequence of elementary steps.
In the circuit model the same requirement translates to the 
fact that the  gates are local, i.e. that the number of 
wires which each node operates on is bounded above  by a constant.

To measure the cost of the 
 computation we can use different parameters:
$S$, the number of gates in the circuit,
or $T$, the time, or {\it depth} of the circuit.
In this review, we will mainly be considered with $S$, 
the number of gates. 
We will be interested 
in the behavior of the cost, $S$, as a function of the size of 
the input, i.e. the number of wires input 
to the circuit,
which we will usually denote by $n$. 
To find the cost function $S(n)$, we will 
look at a function $f$ as a family of functions $\{f_n\}_{n=1}^{\infty}$,
 computed
by  a family of circuits $\{C_n\}_{n=1}^{\infty}$,
 each operating on $n$ input bits; $S(n)$ will be the size of the circuit $C_n$.

I would like to remark here on an important distinction between the model 
of Turing machines  and that of circuits.
  A lot of information can get into the circuit 
through the hardware. 
 If we do not specify how long it takes to  design the hardware,
such circuits can compute even non-computable functions.
This can be easily seen by an example. Define the circuit $C_n$ 
to be a very simple circuit, which outputs a constant bit regardless 
of the $n$ input bits. This constant bit will be $0$ or $1$ according to 
whether the $n'$th Turing machine, $M_n$ (ordered according to the 
numerical description of Turing machines)  halts on the input $M_n$ or not.
The family of circuits  $\{C_n\}_{n=1}^{\infty}$ computes the non-computable HALTING problem
with all the circuits having only one gate! 
This unreasonable computational power of circuits is due
 to the fact that we haven't specified who constructs
the hardware of the circuit.
  We want  to avoid such absurdity and concentrate on interesting 
            and realistic cases. We will therefore require that the hardware of the circuits
  which compute  $\{f_n\}_{n=1}^{\infty}$ can be designed with  polynomial cost by a 
Turing machine. The Turing machine is given as an input the integer $n$, and outputs 
the specification of the circuit $C_n$. 
This model is called the ``uniform circuit model'', as opposed to the 
``non uniform'' one, which is too strong.
The models of uniform Boolean circuits and Turing machines are polynomially 
equivalent. This means that given a Turing machine which computes
in polynomial time 
$f(x)$, there is a family of polynomial
circuits $\{C_n\}_{n=0}^{\infty}$, specified by a polynomial Turing machine, 
such that $C_n$ computes $f_n$.
This correspondence is true also in reverse order, i.e. given the family 
of circuits there is a Turing machine that simulates them.  
Therefore the complexity of a computation does not 
depend (except for polynomial factors) on the model used.
From now on, we will work only in the uniform circuit model.

One of the main questions in this review  is whether the cost of the computation 
            grows like a 
polynomial in $n$ or an exponential in $n$.
This distinction might seem arbitrary, but is better understood 
in the context of the complexity classes $P$ and $NP$.
 The complexity class $P$ is essentially the class of "easy" problems, 
              which can be solved  with polynomial cost:

\begin{deff} {\bf: Complexity class P}

\noindent $ f=\{f_n\}_{n=1}^{\infty}\in P$ if there exists a uniform family of circuits
$\{C_n\}_{n=1}^{\infty}$ of poly($n$) size, where $C_n$ computes the function $f_n(x)$ for 
all $x\in \{0,1\}^n$.
\end{deff}
  
          The class of {\it Non-deterministic Polynomial time} (in short, $NP$)
           is a class of much harder problems.
          For a problem to be in $NP$, we do not require that there exists a
          polynomial algorithm that solves it. We merely require that there exists an
          algorithm  which can verify that a solution is correct in polynomial time. 
         Another way to view this is that the algorithm is provided with
          the input for the problem and a {\it hint}, but the hint may be misleading. 
          The algorithm should solve the problem in polynomial time when the hint is good,
        but it should not be mislead by bad hints.  
           In the formal definition which follows, $y$ plays the role of the hint.

 \begin{deff} {\bf: Complexity class NP}

\noindent $f=\{f_n\}_{n=1}^{\infty}\in NP$ if 
 there exists a  uniform family of  circuits,
$\{C_n\}_{n=1}^{\infty}$, of poly($n$) size, such that
 
$~~~~~~~$ If $x$ satisfies $f_n(x)=''yes''$ $\longmapsto$  
 there exists a string $y$ of $\rm{poly}(n)$ size such that $C_n(x,y)=1$, 

$~~~~~$ If $x$ satisfies $f_n(x)=''no''$ there is no such $y$, i.e. for all $y's$, 
  $C_n(x,y)=''no''$. 
\end{deff}

 To understand this formal definition better,  
        let us consider the following example for  an $NP$ problem which is  called 
{\it satisfiability}:

\begin{quote}
{\bf Input}: A formula of $n$ Boolean variables,
$X_1,...X_n$, 
of the form  
\[ g(X_1,...X_n)=( X_{i}\cup \neg  X_{j} \cup X_k )\bigcap 
 ( X_{m}\cup  \neg X_{i})....\]
which is the logical AND of poly$(n)$ clauses, each
clause is the logical OR of poly$(n)$ variables or their negation.

{\bf Output}: $f(g)=1$ if there exists a satisfying assignment of the 
variables $X_1,...X_n$ so that $g(X_1,...X_n)$ is true.
Else, $f(g)=0$.
\end{quote}

To see that satisfiability is in $NP$, define 
 the circuit $C_n$ to
 get as input the specification of the 
 formula $g$ and a possible assignment $X_1,...X_n$.
The circuit will output  $C_n(g,X_1,...X_n)=g(X_1,...X_n)$.
It is easy to see that these circuits satisfy the requirements of 
the definition of $NP$ problems.
However, nobody knows how to build  a polynomial 
circuit which gets  $g$ as an input, 
and finds whether a satisfying assignment exists. 
It seems impossible to find a satisfying assignment  
without literally checking  all $2^n$ possibilities.
Hence satisfiability is not known to be in $P$.

 Satisfiability
           belongs to a very important subclass of $NP$, namely the $NP$ {\it complete}
           problems. These are the hardest problems in $NP$, 
        in the sense that if we know how to solve an NP-complete problem efficiently,
we can solve any problem in $NP$ with only polynomial slowdown.
In other words, a problem $f$ is $NP$-complete if any
 NP problem can be {\it reduced}
to $f$ in polynomial time. 
    Garey and Johnson\cite{grey} give hundreds of examples of
          $NP$-complete problems, all of which  
          are {\it reducible} one to another with polynomial slowdown, 
        and therefore they are all equivalently hard.
As of now, the best known algorithm for any
 $NP$-complete  problem is exponential, and 
the widely believed 
 conjecture is that there is no polynomial algorithm, 
 i.e. $P\not= NP$. 
 Perhaps the most important open question in complexity theory 
 today,  is proving this conjecture.

Another interesting class consists of those problems solvable   with negligible 
probability for error in polynomial time by a probabilistic Turing machine.
 This machine is defined exactly as deterministic TM,
except that  
 the transition function  can change the configuration in one of several possible ways, randomly.
Equivalently, we can define {\it randomized circuits},
 which are Boolean circuits
with the advantage that apart from the input of $n$ bits, they also get 
as input random bits which they can use as random coin flips.
The class of problems solvable by uniform polynomial randomized circuits 
with bounded error probability is called 
 $BPP$ ({\it bounded probability polynomial}):
\begin{deff}
$f=\{f_n\}_{n=1}^{\infty}\in BPP$ if 
 there exists a family of uniform randomized circuits,
$\{C_n\}_{n=1}^{\infty}$, of poly($n$) size
such that $\forall x\in \{0,1\}^n,$  probability$(C_n(x,y)=f_n(x))\ge2/3$, where 
            the probability is measured with respect to a uniformly random $y$.
\end{deff}
Until the appearance of quantum computers, 
the modern Church thesis which asserts that a
 probabilistic Turing machine, or equivalently randomized uniform 
circuits, can simulate  
 any reasonable 
physical device in polynomial time,
held with no counterexamples.
The quantum model, which I will define in the next chapter, is the only model 
which seems to be qualitatively different from all the others.
We can define the quantum complexity classes:
\begin{deff}  
The complexity classes $QP$ and $BQP$ are defined like 
 $P$ and $BPP$, respectively, only with quantum circuits.
\end{deff}
It is known that $P\subseteq QP$ and $BPP \subseteq BQP$,
as we will see very soon.




\section{The Model of Quantum Computation}
Deutsch was the first to define a rigorous model of quantum computation, 
first of quantum Turing machines\cite{deutsch1} and then of quantum 
circuits\cite{deutsch2}.
I will  describe first the model of quantum circuits, which is much simpler.
At the end of the chapter, I present the model of quantum Turing machines, 
for completeness.
 For background on basic 
quantum mechanics such as Hilbert spaces, Schr$\ddot{\rm{o}}$dinger
equation and measurements I recommend to consult the books by Sakurai\cite{sakurai}, and by Cohen-Tanoudji\cite{cohentan}. As for more advanced material, the book by Peres\cite{peres}
would be a good reference. 
However, I will give here all the necessary definitions.

A quantum circuit is a system built of two state 
quantum particles, called  qubits.
We will work with $n$ qubits, 
the state of which is a unit vector in the complex Hilbert space
 ${\cal C}^2\otimes {\cal C}^2\otimes \cdots\otimes {\cal C}^2$.
As the natural basis for this space, 
we take the basis consisting of $2^n$ vectors:
\begin{eqnarray}
            |0\ra\otimes|0\ra\otimes\cdots\otimes|0\ra~\\
          |0\ra\otimes|0\ra\otimes\cdots\otimes|1\ra\nonumber~\\
            \vdots~~~~~~~~~~~~~~~~~~~~~\nonumber\\
         |1\ra\otimes|1\ra\otimes\cdots\otimes|1\ra\nonumber.
            \end{eqnarray}

For brevity, we will sometimes omit the tensor product, and denote 
 \begin{eqnarray}
       |i_1\ra\otimes|i_2\ra\otimes\cdots\otimes|i_n\ra=|i_1,i_2,...,i_n\ra\equiv|i\ra
       \end{eqnarray} 

where $i_1,i_2,...,i_n$ is the binary representation of the integer 
$i$, a number between $0$ and $2^n-1$.
This is an important step, as this representation allows us to use 
our quantum system to encode integers. This is where the quantum system 
starts being a computer.
The general state  which describes this system is a complex unit
vector in the Hilbert space, sometimes called the {\it superposition:}
\begin{equation}
       \sum_{i=0}^{2^n-1} c_i |i\ra
        \end{equation} 
       where $\sum_i |c_i|^2=1$.
The initial state  will correspond to the ``input''
 for the computation. Let us agree that for an input string $i$, 
the initial state of the system will be $|i\ra$: 
\begin{equation}
i \longmapsto  |i\ra 
\end{equation}
We will then perform ``elementary operations'' on the system.
These operations will correspond to the computational steps 
in the computation, just like logical gates are the elementary steps
in classical computers.
In the meantime we will assume 
that all the operations are performed on an isolated system,
so the evolution can always be described by a unitary matrix
operating on the state of the system.
Recall that a unitary matrix satisfies $UU^{\dagger}=I$, where  $U^{\dagger}$ is the 
          transposed complex conjugate of $U$.

\begin{deff}  A {\it quantum  gate} on $k$ qubits is 
  a unitary matrix $U$ of dimensions $2^k\times 2^k$.
\end{deff}

Here is an example of a simple quantum gate, operating on one qubit.
\begin{equation}
NOT=\left(\begin{array}{ll}
0&1\\
1&0
\end{array}
\right)
\end{equation}
  Recalling that in our notation 
         $|0\ra=(1,0)$ and $|1\ra=(0,1)$, we have that  $NOT|0\ra=|1\ra$ and 
         $NOT|1\ra=|0\ra$.
Hence, this gate flips the bit, and thus it is justified to call 
this gate the $NOT$ gate.
The $NOT$ gate can operate on superpositions as well. From linearity
of the operation, 
\[NOT(c_0|0\ra+c_1|1\ra)= c_0|1\ra+c_1|0\ra.\]

This linearity is responsible for the quantum parallelism (see Margolus\cite{margolus2})
which we
 will encounter  in all powerful quantum algorithms. 
When the NOT gate operates on the first qubit in a system of $n$ qubits, 
in the state  
$\sum_i c_i |i_1i_2...i_n\ra $
this state transforms to 
$\sum_i c_i (NOT|i_1\ra)|i_2...i_n\ra=\sum_i c_i |\neg i_1i_2...i_n\ra $.
      Formally, the time evolution of the system is described by a unitary matrix, 
         which is a tensor product of the gate operating on the first qubit and the identity
         matrix $I$ operating on the rest of the qubits.

Another important quantum gate is
the {\it controlled} $NOT$ gate acting on two qubits, 
which computes the classical function:
 $(a,b)\longmapsto (a, a \oplus b)$  where $a \oplus b = (a+b)$ mod $2$ and $a,b \in {0,1}$.
This function can be represented by the matrix operating on 
all $4$ configurations of $2$ bits:
 \begin{equation}
CNOT=\left(\begin{array}{llll}
1&0&0&0\\
0&1&0&0\\
0&0&0&1\\
0&0&1&0
\end{array}
\right)
\end{equation}
The above matrix, as all matrices in this review, is written in the computational
basis in lexicographic order.
This gate  is also called the 
 {\it exclusive or} or XOR gate (On its importance see \cite{divincenzo2}.) 
The XOR gate applies a $NOT$ on the second bit, called the {\it target} bit, 
conditioned that the first {\it control} bit is $1$.
If a black circle denotes the bit we condition upon, 
we can denote the XOR gate by:

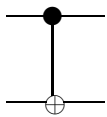
\setlength{\unitlength}{0.030in}

\begin{picture}(40,25)(-90,10)

\put(5,15){\line(1,0){7}}
\put(15,15){\line(1,0){7}}
\put(5,30){\line(1,0){17}}

\put(13,30){\circle*{3}}

\put(13,16){\line(0,1){14}}

\put(12,13){\makebox(3,3){$\oplus$}}
\end{picture}

In the same way, all classical Boolean functions can be transformed to quantum gates.
The matrix representing a classical
 gate which computes a reversible function, (in particular 
the number of inputs to the gate equals the number of outputs)  is a permutation on all the possible 
classical strings. Such a permutation is easily seen to be unitary.
Of course, not all functions are reversible, but 
they can easily be converted to reversible functions,
 by writing down the input bits instead of erasing them.
For a function $f$,from $n$ bits to $m$ bits, 
we get the reversible function from $m+n$ bits to $m+n$ bits:
 \begin{equation}\label{rev}\begin{array}{c}
          f: i\longmapsto f(i)\\  
        \Downarrow\\
         f_r: (i,j)\longmapsto (i,f(i)\oplus j). 
           \end{array}\end{equation}
       
Applying this method, for example, to
the logical AND gate, 
\( (a,b) \longmapsto ab \) 
 it will become 
the known Toffoli gate\cite{toffoli} $ (a,b,c)\longmapsto (a,b,c\oplus ab),$ 
which is described by the unitary matrix on three 
qubits:
\begin{equation}\label{tof}
T=    \left( \begin{array}{llllllll}
    1&&&&&&&\\
 & 1 &&&&&&\\
 & &1&&&&&\\
& &&1&&&&\\
& &&&1&&&\\
& &&&&1&&\\
& &&&&& 0& 1\\
& &&&&& 1& 0
\end{array}\right)
\end{equation}
The Toffoli  gate applies NOT on the last bit, conditioned that 
the other bits are $1$, so we can describe it by the following diagram:

\setlength{\unitlength}{0.030in}

\begin{picture}(40,45)(-80,5)

\put(0,15){\line(1,0){8}}
\put(23,15){\line(1,0){7}}
\put(0,30){\line(1,0){30}}
\put(0,45){\line(1,0){30}}

\put(15,30){\circle*{3}}
\put(15,45){\circle*{3}}

\put(15,22){\line(0,1){25}}

\put(8,10){\framebox(15,12){$NOT$}}

\end{picture}

Quantum gates can perform more complicated tasks  than simply computing 
classical functions.
An example of such a quantum gate, which is 
not a classical gate in disguise, is 
  a gate which applies a general rotation on one qubit:

\begin{equation}
G_{\theta,\phi}=\left(\begin{array}{ll}
\cos(\theta) & \sin(\theta)e^{i\phi}\\
-\sin(\theta)e^{-i\phi} & \cos(\theta)
\end{array}
\right)
\end{equation}


 To perform a quantum computation, we apply a sequence of elementary quantum gates 
        on the qubits in our system. Suppose now, that we have applied all the quantum gates 
         in our algorithm, and  the computation has come to an end. 
The state which was initially a basis state has been 
{\it rotated} to the state $|\alpha\ra\in C^{2^n}$.
We now want  to extract the output from this state.
This is done by the process of {\it measurement}.
    The notion of measurement in quantum mechanics is puzzling. For example, consider 
           a measurement of a qubit in the state $|\alpha\rangle=c_0|0\rangle +c_1|1\rangle$.
           This qubit is neither in the state $|0\rangle$ nor in $|1\rangle$. 
Yet, the {\it measurement postulate}
 asserts that when the state of this qubit is observed, 
it must decide on one  of the two possibilities.
 This decision is made non-deterministically. 
           The classical outcome of the measurement
            would be $0$ with probability $|c_0|^2$ and $1$ with probability 
            $|c_1|^2$. After the measurement, the state of the qubit is 
            either  $|0\rangle$ or $|1\rangle$, in consistency with the classical outcome
            of the measurement.  
Geometrically, this process can be interpreted as 
a projection of the state on one of the two  orthogonal subspaces, 
$S_0$ and $S_1$, 
where $S_0=\rm{span}\{|0\ra\}$ and $S_1=\rm{span}\{|1\ra\}$,
and a measurement of the state of the qubit  $|\alpha\rangle$
is actually  an observation in which of the subspaces the 
state is, in spite of the fact that the state might be in neither. 
  The probability that the decision is $S_0$ 
is the norm squared of the projection of $|\alpha\ra$
 on $S_0$, and likewise for 
$1$. Due to the fact that the norm of $|\alpha\ra$ is one, 
these probabilities add up to one.
After the measurement $|\alpha\ra$ 
 is projected to the space $S_0$ if the answer is $0$, and to the space $S_1$
if the answer is $1$.
 This projection is the famous {\it collapse} of the wave function.
             Now what if we measure a qubit in a system of $n$ qubits? 
 Again, we project the state onto one of two subspaces, $S_0$ and $S_1$, 
             where $S_a$ is the subspace spanned by all 
            basis states in which the measured qubit is $a$. 
The rule is that
  if the measured superposition is 
$\sum_i c_i |i_1,...i_n\ra$,
a measurement of the first qubit will give the outcome $0$ with probability
$\rm{Prob}(0)=\sum_{i_2,...i_n} |c_{0,i_2,...i_n}|^2$, and the 
superposition will collapse to 
\[\frac{1}{\rm{Prob}(0)} \sum_{i_2,...i_n} c_{0,i_2,...i_n}|0,i_2,...i_n\ra,\]
and likewise with $1$.
Here is a simple example:
Given the state of two qubits:
 \[\frac{1}{\sqrt{3}} \big(|00\ra+|01\ra-|11\ra\big),\]
the probability to measure $0$ in the left qubit is $2/3$,
and  the probability to measure $1$  is $1/3$.
After measuring the left qubit,  the state has collapsed to 
  $\frac{1}{\sqrt{2}} \big(|00\ra+|01\ra\big)$ with probability  Pr$(0)=2/3$ 
 and to $ -|11\ra$ with probability Pr$(1)=1/3$. 
Thus,  the resulting state 
depends on the outcome of the measurement. 
After the collapse, the projected state is renormalized back to $1$.

  We can now summarize  the definition of the model of quantum circuits.  
           A quantum circuit is a directed acyclic 
          graph, where each node in the graph is associated a quantum gate. 
          This is exactly the definition from section $2$ of classical Boolean circuits, 
           except that the gates are quantum. The input for the circuit  is a basis 
           state, which  evolves in time according to the operation  of the quantum gate.
At the end of the computation we apply measurements on 
the output qubits (The order does not  matter). The string of classical outcome
 bits is  the classical
 output of the quantum computation.
This output is in general probabilistic.
This concludes the definition of the model. 

Let us now
 build a repertoire of quantum computations step by step.
We have seen that classical gates can be implemented 
quantumly, by making the computation reversible.
More explicitly, 
\begin{lemm}
Let $f$  be a function from $n$ bits to $m$ bits, 
computed by a Boolean circuit $C$ of size $S$.
There exists a quantum circuit $Q$ which computes the 
unitary transformation on $n+m$ qubits:
$|0^b,i,j\ra \longmapsto |0^b,i,f(i)\oplus j\ra$.
$b$ and the size of $Q$ are linear in $S$.
\end{lemm}
{\bf Proof:} Replace
each gate in $C$ by its reversible extension, 
according to equation \ref{rev}. We will add $b$ extra  bits for this
 purpose. The input for this circuit is thus $(0^b,i)$.
The modified  $C$, denoted by $ \tilde{C},$ 
 can be viewed as a quantum circuit since all its nodes 
correspond to unitary matrices. 
The function that it computes is still not the required function, 
because the input $i$ is not necessarily part of the output as  it should be. 
    To solve this problem, we add to $\tilde{C}$ $m$ extra wires, or qubits. 
            The input to these wires is $0$. At the end of 
              the sequence of gates of  $\tilde{C}$,  we copy 
the $m$ ``result'' qubits in  $\tilde{C}$ on
 these $m$ blank qubits by applying $m$ CNOT gates. 
We  now apply 
in reverse order
the reversed gates of all the gates  applied so far, except the $CNOT$ gates.
This will reverse all operations, and retain the input  $(0^b,i)$,
while the $m$ last qubits contain the desired $f(i)$. $\bbox$

The state of the system is always a basis state
 during the computation which is described in the proof.
Hence measurements of the final state will yield exactly the expected 
result. This shows that  any computation which can be done 
classically can also be done quantumly with the same efficiency, 
i.e. the same order of number of gates. We have shown:
\begin{theo}
$P\subseteq QP$
\end{theo}

 In the process of 
conversion to reversible operations,
 each gate is replaced by a gate operating on more qubits.
This means that making circuits reversible costs 
 in adding a linear number  of extra  qubits.
In \cite{bennett9}, Bennett 
used a nice pebbling argument, to show that
the space cost can be decreased to a logarithmic factor with 
only a minor cost in time: $T\longmapsto T^{1+\epsilon}$.
 Thus the above conversion to quantum 
circuit can be made very efficient.

To implement classical computation we must also show how 
to implement probabilistic algorithms. For this we need a quantum subroutine 
that generates a random bit. This is done easily by measurements.
We define the Hadamard gate which acts on one qubit. 
It is an extremely useful gate in quantum  algorithms.

\begin{equation}\label{hadamard}
H=\left(\begin{array}{ll}
\frac{1}{\sqrt{2}}&\frac{1}{\sqrt{2}}\\
\frac{1}{\sqrt{2}}&-\frac{1}{\sqrt{2}}
\end{array}
\right)
\end{equation}
Applying this gate on a qubit in the state $|0\ra$ or $|1\ra$, 
we get a superposition:
\(\frac{1}{\sqrt{2}}(|0\ra\pm|1\ra)\).
A measurement of this qubit yields a random bit.
Any classical circuit that uses random bits can be converted to  
a quantum circuit by replacing the gates with 
reversible gates and adding the ``quantum random bit'' subroutine 
when needed.
 Note that here we allow measuring in the middle of the computation.
This shows that:
\begin{theo}
$BPP\subseteq BQP$
\end{theo}

The repertoire of classical algorithms can therefore be simulated efficiently by
  quantum computers.
But quantum systems feature  characteristics which are far more interesting.
We will encounter these possibilities when we discuss 
quantum algorithms.


Let  me define here also the 
 model of  quantum Turing Machine\cite{deutsch1,bv,solovay} ($QTM$) which 
 is the quantum analog of the classical TM.
The difference is that all the degrees of freedom become quantum:
Each cell in the tape, the state of the machine, and the reading head 
of the tape can all be in linear  superpositions
of their different possible classical states.

\begin{deff}
A quantum Turing machine is specified by the following items:
\begin{enumerate}
\item A finite alphabet $\Sigma=\{\sqcup,0,1...\}$ 
where $\sqcup$  is the blank symbol.
\item A finite set $K=\{q_0,...q_s\}$ of ``machine states'', 
with $h,s\in K$ two special states.
\item A transition function $
\delta: 
Q\times \Sigma\times Q\times
   \Sigma\times \{-1,0,1\} \longmapsto {\cal C}$
\end{enumerate}
\end{deff}

As in classical TM, the tape
is associated  a head that reads and writes on that tape.
A classical configuration, $c$, of the Turing machine is specified by 
the head's position, the contents of the tape and the machine's state.
The Hilbert space of the QTM is defined as the vector space, spanned
by all possible classical configurations
$\{|c\ra\}$.  The dimension of this space is infinite.
The computation  starts with the QTM in a basis state $|c\ra$, 
which corresponds to the following classical configuration:
An input of $n$ symbols is written
in positions $1,..,n$ 
 on the tape,
 all symbols except these $n$ symbols  
are blank ($\sqcup$,)  and the head is at position $1$.
Each time step, the machine evolves according 
to an infinite unitary matrix  which is defined in the following way.
            $U_{c,c'}$, the probability  amplitude  to transform from configuration 
        $c$ to $c'$   is determined by the transition function $\delta$.
If in $c$, the state of the machine is $q$ and the 
symbol in the current place of the tape head is $\sigma$
then 
$\delta(q,\sigma,q',\sigma',\epsilon)$
is the probability amplitude to go from $c$ to $c'$, 
where $c'$ is equal to $c$ everywhere except 
locally.  The machine state in $c'$, $q$,   is changed to  $q'$, the symbol under the head
is changed to 
$\sigma'$ and the tape head  moves one step in direction 
$\epsilon$.
Note that the operation of the Turing machine is local, i.e. it
 depends only on the current state of
 the machine and the symbol now read by the tape.
Unitarity of infinite matrices is not easy to check, and conditions  for unitarity
 were  given by Bernstein and Vazirani\cite{bv}.

In my opinion, the QTM model is less appealing than the model of quantum circuits, 
           for a few reasons. First, QTMs involve infinite unitary matrices. 
Second, it seems very unlikely that a physical quantum computer, 
 will resemble this model, because the head, 
or the apparatus executing the quantum operations, is most 
likely to be classical in its position and state.
Another point is that the QTM model is a sequential 
model, which means that it is able to apply only one operation at each time step.
Aharonov and Ben-Or showed\cite{aharonov2}  that a sequential model is fundamentally
incapable of operating fault tolerantly in the presence of noise.
Above all, constructing algorithms is much simpler 
in the circuit model. For these reasons
 I will restrict  this review to quantum circuits.
The model of quantum circuits, just like that of classical circuits, 
has  a ``uniform'' and ``non-uniform'' versions. 
Again, we will restrict ourselves to the uniform model, i.e. 
quantum circuits which can be designed in polynomial time on a classical 
Turing Machine.
 Yao\cite{yao} showed that uniform 
 quantum circuits are polynomially equivalent to 
quantum Turing machines, 
by a proof which is surprisingly complicated.
This proof  enables us the freedom of choosing whichever model 
is more convenient for us.

Another model worth mentioning in this context is 
 the quantum cellular automaton\cite{margolus2,watrous2,durr1,vandam2}. 
This model resembles quantum circuits, but is 
different in the fact that the operations are homogeneous, or periodic,
 in space and in time. The definition of this model is subtle
and, unlike  the case of 
 quantum circuits, it is not trivial to decide whether 
a given quantum cellular automaton obeys the rules of quantum mechanics or
 not\cite{durr1}.
Another interesting quantum model is that of a finite state 
quantum automaton, which is similar to a quantum Turing machine
except it can only read and not write, so it has no memory. 
It is therefore a very limited 
model. In this model Watrous\cite{watrous} showed an interesting algorithm 
which uses interference, and  is able to compute a function which cannot be 
 computed in the analogous classical model.

%


\section{Universal Quantum Gates}

What kind of elementary gates can be used in a quantum computation program?
We would like to write our program using elementary steps:
i.e., the algorithm should be a sequence of steps, 
each potentially implementable in the laboratory.
 It seems that achieving controlled interactions between a large 
number of qubits in one elementary step is extremely difficult.
Therefore it is reasonable to require an ``elementary 
gate'' to operate on a small number 
of qubits, (independent of $n$ which can be very large.)
We want our computer to be able to compute any 
function. The set of elementary  gates used should thus be
 {\it universal}. 
For classical reversible computation, there exists a single 
universal gate\cite{fredkin, toffoli}, called the Toffoli gate, which we have already encountered.
This gate computes the function
\[a,b,c \longmapsto a,b,ab\oplus c.\] 
The claim is that any reversible function can be represented as 
a concatenation of the Toffoli gate on different inputs. 
For example, to construct the logical AND gate on $a,b$, we simply 
input $c=0$, and the last bit will contain $ab\oplus 0=AND(a,b)$. 
To implement the NOT gate on the third bit
we set the first two bits to be equal to $1$. 
We now have what is well known to be a universal set 
of gates, The NOT and AND gates.
In the quantum case, the notion of universality is slightly 
more complicated, because operations are continuous. We need not require that 
all operations are achieved exactly, but a very good approximation suffices.
The notion of approximation is very important in quantum computation. 
Frequently operations are approximated instead of achieved exactly, 
without significantly damaging  the correctness of the computation.
\begin{deff} {\bf Approximation:}

\noindent A unitary matrix $U$ is said to be approximated to within $\epsilon$ 
by a unitary matrix  $U'$ if $|U-U'|\le \epsilon$.
\end{deff}
The 
norm we use is the one induced by the Euclidean norm
on vectors in the Hilbert space.

Note that unitary transformations can be thought of as rigid rotations 
of the Hilbert space. This means that angles between vectors 
are preserved 
during the computation.The result of using  $U'$ 
 instead of $U$, where $|U-U'|\le \epsilon$,
  is that the state is tilted by an angle  of order $\epsilon$
       from the correct state.
 However this angle does not grow during the computation, 
because the rotation is rigid.
The state always remains within $\epsilon$ angle from the correct state.
Therefore the overall error in the entire computation is additive:
it is at most the sum of the errors in all the gates.
This shows that 
the accuracy to which the gates should be approximated
is not very large. 
If $S$ gates are used in the circuit, it suffices to approximate
each gate  to within $O(\frac{1}{S})$,
in order that the computation is correct with constant probability\cite{bv}.

We can now define the notion of universal gates, which approximate 
any possible quantum operation:

\begin{deff}
{\bf   Universal Set of Gates:}

A set of quantum gates, $\cal{G}$, is called {\it universal}
 if for any $\epsilon$ and any unitary matrix $U$ on any number of bits,
$U$ can be  approximated to within $\epsilon >0$ by 
a sequence of gates from $\cal{G}$. In other words, the subgroup 
generated by $\cal{G}$ is dense in the group of unitary operators, 
                     $U(n)$, for all $n$. 
\end{deff}

Deutsch was the first to show a universal elementary gate, 
 which operates 
 on three 
qubits\cite{deutsch2}.
  Bernstein and Vazirani\cite{bv} gave another proof of universality in  terms of $QTM$.
It was then shown by DiVincenzo that two-qubit gates
are universal\cite{twobit}. 
This is an important result, since it seems impossible to 
 control interactions
 between three particles, whereas two particle interactions are likely to be 
 much easier to implement.
It was a surprising achievement, since in reversible classical computation, 
which is a special case of quantum computation, 
there is no set of two bit gates which is universal.
Note that one qubit gate is certainly not enough to construct 
all operations.
Barenco\cite{barenco1} and Deutsch {\it et.al}\cite{deutsch4} 
 showed that almost any two-bit gate is universal (See also Lloyd
 \cite{lloyd2,lloyd4}).
An improvement of DiVincenzo's result was achieved later by Barenco 
{\it et.al}\cite{barenco4}, 
where it was shown that the classical controlled not gate,
together with all one-qubit gates  construct a universal set as well!
In fact, one 1-qubit gate and the controlled not gate will do.  
This is perhaps the simplest and  most economic set constructed so far.
Implementation of one qubit gates are feasible, 
and experimentalists have already implemented a controlled not gate
\cite{turchette}.
However, there are other possible sets of gates.
Adleman, {\it et. al.}\cite{adleman} and Solovay\cite{solovay}
 suggested a set of gates, 
where all entries of the matrices
 are $\pm\frac{3}{5}$ and $\pm\frac{4}{5}$ and 
$\pm 1$.
Other universal sets of gates were suggested in connection 
with fault tolerant quantum computation\cite{shor3,aharonov1,knill2}.

 Why do we need so many possible universal sets to choose from? 
Universal sets of gates are our computer languages.
At the lowest level,
 we need {\it quantum  assembly}, the machine language by which 
everything will be implemented. For this  purpose, we will use 
the set which consists of the easiest gates to implement in the laboratory.
Probably,  the set of one and two qubit gates will be most appropriate.
Another incentive is  analyzing the complexity power of quantum computers.
For this the set suggested by Solovay and 
by Adleman {\it et. al.} seems more appropriate.
(Fortnow recently reported on bounds  
using this set\cite{fortnow}).
We will see that for error correction purposes, we will need a completely
different universal set of gates.
An important question should arise here.
If our computer is built using one set, how can we design algorithms 
using another set, and analyze the computational power using a third set?
The answer is that since they are all universal sets, 
there is a way to {\it translate} between all these 
languages. A gate from one set can be approximated  by a sequence of gates 
from another set.
It turns out that in all  the universal sets described here, the 
approximation to within $\epsilon$ of an operation on $k$ qubits takes
 $\rm{poly}(\rm{log}(\frac{1}{\epsilon}),2^k)$ gates from the set.
As long as the gates are local (i.e  $k$ is constant) the 
translation between different universal sets is efficient.




Now that the concept of a universal set of gates is understood, 
I would like to  present an example of a simple universal set of gates.
It relies on the proof of Deutsch's universal gate.
The idea underlying Deutsch's universal gate is that 
Reversible computation is a special case of quantum computation. 
It is therefore natural that 
 universal quantum computation can be achieved by generalizing 
universal reversible computation.
Deutsch showed how to generalize  
Toffoli's gate so that it becomes  a universal gate for quantum computation:

{~}

{~}

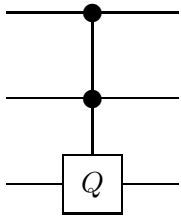
\setlength{\unitlength}{0.030in}
\begin{picture}(40,40)(-80,5)
\put(0,15){\line(1,0){10}}
\put(20,15){\line(1,0){10}}
\put(0,30){\line(1,0){30}}
\put(0,45){\line(1,0){30}}
\put(15,30){\circle*{3}}
\put(15,45){\circle*{3}}
\put(15,20){\line(0,1){25}}
\put(10,10){\framebox(10,10){$Q$}}
\end{picture}

The $NOT$ matrix in the original Toffoli gate (see equation \ref{tof})
 is replaced 
by another 
 unitary matrix on one qubit, $Q$, such that
 $Q^n$ can approximate any $2\otimes 2$
matrix.
I will present here a modification of Deutsch's proof, using two gates 
of the above form.
Define:
\begin{equation}
    U= \left( \begin{array}{ll}
  \rm{cos}(2\pi\alpha)&\rm{sin}(2\pi\alpha)\\
 -\rm{sin}(2\pi\alpha)&\rm{cos}(2\pi\alpha)
\end{array}\right),
W= \left( \begin{array}{ll}
    1&0\\
0& e^{i2\pi\alpha}
\end{array}\right).
\end{equation} 
We have freedom in choosing $\alpha$, except we require that
           the sequence  
\newline\(\alpha ~\rm{mod}~1, 2\alpha ~\rm{mod}~1, 3\alpha ~\rm{mod}~1,...\) 
            hits the $\epsilon$-neighborhood of  any number in $[0,1]$, within 
           poly($\frac{1}{\epsilon}$) steps. Clearly, $\alpha$ should  be irrational, 
   but not all irrational numbers satisfy this property.
           It is not very difficult to see that   an irrational root of a polynomial of degree $2$
           satisfies the required property. 
Let $U_3$ $(W_3)$ be the generalized Toffoli gate with $U$ ($W$)
playing the role of the conditioned matrix, $Q$, respectively.
Then
\begin{theo}
$\{U_3, W_3\}$ is a universal set of quantum gates. 
\end{theo}

\noindent{\bf Proof:}
First, note that according to the choice of $\alpha$, $U$  approximates any rotation in the real plane, 
and $W$ approximates any rotation in the complex plane.
 Given an $8\times 8$ unitary matrix $U$,
let us denote its $8$ eigenvectors as $|\psi_j\ra$ with corresponding
 eigenvalues
$e^{i\theta_j}$.
$U$ is determined by \(U|\psi_j\ra= e^{i\theta_j}|\psi_j\ra.\)
Define:
\begin{equation}
 U_k |\psi_j\ra = \left\{\begin{array}{ll}
|\psi_j\ra &\mbox{if $k\ne j$} \\
 e^{i\theta_k}|\psi_k\ra & \mbox{if $k= j$}\end{array}\right.
\end{equation}
Then $U=U_7U_6....U_0$.
$U_k$ can be achieved by first taking $|\psi_k\ra$ to $|111\ra$, 
  by a transformation which we will denote by $R$. 
          After $R$ we apply $W$ the correct number of times to approximate
 $|111\ra \longmapsto e^{i\theta_k}|111\ra$  and then
 we take $|111\ra$ to $|\psi_k\ra$ by applying the reverse transformation of $R$, $R^{-1}$.

It is left to show how to apply $R$, i.e.  how to take a general state 
$|\psi\ra=\sum_{i=0}^{7} c_i |i\ra$ to $|111\ra$.
For this, note that $U_3^n$ can approximate 
the Toffoli gate, and therefore can approximate  
 all permutations on basis states.
To apply  
 $|\psi\ra\longmapsto |111\ra$,  first 
turn the coefficient on the coordinate $|110\ra$ to $0$. 
This is done by
applying $W$ an appropriate number of times so that the phase 
in the coefficient of $|110\ra$ will equal that of $|111\ra$. 
The coefficients now become $c_6=r_6e^{i\phi}, c_7=r_7e^{i\phi}$.
Let $\theta$ be such that $r_6=r\rm{sin}(\theta),r_7=r\rm{cos}{\theta}$.
Now apply $U$ an  appropriate number of times
to approximate a rotation by $-\theta$. This will transform all the 
weight of  $|110\ra$ to $|111\ra$. 
In the same way we transform  the weight
from all coordinates to  $|111\ra$, using permutations between 
coordinates.  This achieves  $|\psi\ra\longmapsto |111\ra$, i.e. the transformation $R$. 
   $R^{-1}$ is constructed in the same way.

We have shown that all three qubit operations can be approximated.
For operations on more qubits, note that the group generated
by $\{U_m,W_m\}$ is dense in all operations on $m$ bits, by the same 
reasoning. To create $U_m$ ( $W_m$)
 from  $U_3$ ( $W_3$) use recursion: compute the logical AND 
of the first two bits by a Toffoli gate writing it on an extra bit, 
and then apply $U_{m-1}$ ( $W_{m-1}$).
The reader can verify that the approximation is polynomially fast,
i.e. for fixed $m$, 
 any unitary matrix on $m$ qubits
 can be approximated to within $\epsilon$ by $poly(\frac{1}{\epsilon})$
applications of   the gates $U_3$ and $W_3$. 
$\Box$

{~}

The generalized Toffoli gates operate on three qubits.
 Barenco {\it et. al.}\cite{barenco4} show an explicit sequence of two bit gates 
 which  constructs any matrix on three qubits, 
 of the form of a generalized Toffoli gate:

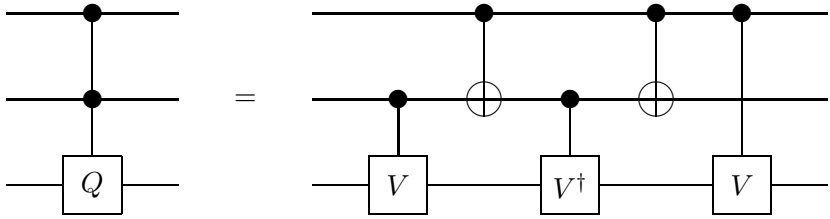
\setlength{\unitlength}{0.030in}

\begin{picture}(40,60)(-10,0)

\put(0,15){\line(1,0){10}}
\put(20,15){\line(1,0){10}}
\put(0,30){\line(1,0){30}}
\put(0,45){\line(1,0){30}}

\put(15,30){\circle*{3}}
\put(15,45){\circle*{3}}

\put(15,20){\line(0,1){25}}

\put(10,10){\framebox(10,10){$Q$}}

\end{picture}
\begin{picture}(20,60)(0,0)

\put(0,0){\makebox(20,60){$=$}}

\end{picture}
\begin{picture}(90,60)(0,0)

\put(0,15){\line(1,0){10}}
\put(20,15){\line(1,0){20}}
\put(50,15){\line(1,0){20}}
\put(80,15){\line(1,0){10}}
\put(0,30){\line(1,0){90}}
\put(0,45){\line(1,0){90}}

\put(30,30){\circle{6}}
\put(60,30){\circle{6}}

\put(15,30){\circle*{3}}
\put(45,30){\circle*{3}}
\put(30,45){\circle*{3}}
\put(60,45){\circle*{3}}
\put(75,45){\circle*{3}}

\put(75,20){\line(0,1){25}}
\put(15,20){\line(0,1){10}}
\put(30,27){\line(0,1){18}}
\put(45,20){\line(0,1){10}}
\put(60,27){\line(0,1){18}}

\put(10,10){\framebox(10,10){$V$}}
\put(70,10){\framebox(10,10){$V$}}
\put(40,10){\framebox(10,10){$V^{\scriptsize \dag}$}}

\end{picture}

 where $V=\sqrt{Q}$.
Thus, two bit gates are universal.
$\Box$

{~}

  It was further shown\cite{barenco4} that one-qubit matrix conditioned 
on one other qubit  can be expressed as a sequence of one-qubit 
matrices and $CNOT$s. So the generalized Toffoli gate of 
Deutsch can be written as a finite sequence of one-qubit gates and 
$CNOT$s. This shows that $\{One-qubit~ gates, CNOT\}$ is universal.

  The description above shows how to  approximate unitary matrices
            using poly($\frac{1}{\epsilon}$) gates from the universal set. 
             In fact, an exponentially faster approximation is possible 
              due to a theorem by Kitaev \cite{kitaev0}, 
              which was also proved by Solovay\cite{solovay}:
              \begin{theo} 
              Let the matrices $U_1,...U_r\in SU(n)$ generate a dense subset in  $SU(n)$. 
               Then any matrix $U\in SU(n)$ can be approximated to within $\epsilon$ by a product
                of poly$(\rm{log}(\frac{1}{\epsilon}))$ matrices from 
 $U_1,...U_r,U_1^{\dagger},...U_r^{\dagger}$.
               \end{theo}
                  $SU(n)$ is the set of $n\times n$ unitary matrices
              with determinant $1$.
               Given a universal quantum set, we can easily convert it to a set in  $SU(n)$
               by multiplying each matrix with an overall complex scalar
              of absulute value $1$, namely a phase.  
               This overall phase does not effect the result of any
                measurement, 
                so any gate can be multiplied by a phase  without affecting the computation.
                We thus have: 
                       
               \begin{coro}
               The approximation rate of 
              any universal set of quantum gates is exponential. 
               \end{coro}

              The idea of the proof of the theorem is to
               construct finer and finer nets of points in $SU(n)$. 
  The $2k$'th net is constructed 
                 by taking commutators of points from the $k$'th net. 
               Each point in the $k'$th net is a product of  a linear  (in $k$) 
                  number of gates from the set of gates. It turns out that   
                the distance between two adjacent points in the net
                  decreases exponentially with $k$. $\bbox$

Having chosen the set of gates to write algorithms with,
actually writing the algorithm  
in this  assembler-like language
 seems like a very tedious task!
Just like higher languages in ordinary computer programming, 
it is desirable that quantum operations which are commonly used can 
be treated as black boxes,
 without rewriting them from the beginning with elementary 
gates.
Steps in this direction were made by \cite{barenco4,barenco3,beckman1,
nielsen1,vedral}.

\section{Quantum Algorithms}
The first and simplest quantum algorithm which 
achieves advantage over classical algorithms 
was presented by Deutsch and Jozsa\cite{deutsch2}.
Deutsch and Jozsa's algorithm
 addresses a problem which we have encountered before, in the context 
of probabilistic algorithms.
\begin{quote}
  $f$ is a Boolean function from $\{1,N\}$ to $\{0,1\}$. Assume $N=2^n$ for 
         some integer $n$. 
We are promised that $f(i)$  are either all equal to $0$,
(``constant'') 
or half are $0$ and half are $1$ (``balanced'').
We are asked to distinguish between the two cases.
\end{quote}

The question is presented in the {\it oracle} setting. This means that the circuit does not get $f(1),....f(N)$ as input. 
         Instead, the circuit has access to an {\it oracle} for $f$. 
         A {\it query} to the oracle is a gate with $n$ 
         input wires carrying an  integer $i\in \{1,n\}$ in 
         bit representation. The output from the oracle gate is $f(i)$. 
         A quantum query to the oracle means applying  the unitary transformation
         \(|i\ra|j\ra\longmapsto |i\ra|j \oplus f(i)\ra\).
The cost is measured by the number of queries to the oracle.
A classical algorithm 
that solves this question exactly will need $O(N)$ queries.
The quantum algorithm of Deutsch and Jozsa solves the problem 
exactly,  with merely one quantum query!
The algorithm makes use of a transformation known as the
discrete Fourier transform over the group $Z_2^n$.
\begin{equation}
|i\ra \stackrel{Foriuer Transform}{\longrightarrow} \frac{1}{\sqrt{N}}
\sum_j (-1)^{i\cdot j}|j\ra
\end{equation}
where  $i,j$ are strings of length $n$, 
and $i\cdot j=\sum_{k=1}^n i_kj_k ~mod ~2$, 
 the inner product of $i$ and $j$ modulo $2$.
Meanwhile, we need only one easily verified fact about the Fourier transform over  $Z_2^n$:
To apply this transformation on $n$ qubits, 
we simply apply the Hadamard transform 
$H$ from equation \ref{hadamard} 
on each of the $n$ qubits.
Note also that the reversed Fourier transform, $FT^{-1}$ is 
equal to the $FT$. 
We now turn to solve Deutsch and Jozsa's problem.
We will work with two registers, one will hold a number between 
$1$ to $N$ and therefore will consist of $n$ qubits,
and the other register will consist of one qubit that will carry 
the value of the function.



{~}

\frame{\begin{minipage}
[70mm]{100mm}
~\\\raggedright~\\
\raggedright{\bf $~~~~~$ Deutsch and Jozsa's Algorithm}\\~\\
\raggedright$~~~~~~|0^n\ra\otimes |1\ra$\\~\\
\raggedright$~~~~~$ Apply Fourier transform on first register.\\
\raggedright$~~~~~$ Apply $H$  on last qubit\\
\center{ $\Downarrow$}\\~\\
\raggedright
\(~~~~~\frac{1}{\sqrt{N}}\sum_{i=1}^{N} |i>\otimes(\frac{1}{\sqrt{2}}
 |0\ra - \frac{1}{\sqrt{2}}|1\ra)\)
\\~\\\raggedright
$~~~~~$ Call oracle, \(|i\ra|j\ra\longmapsto |i\ra|j \oplus f(i)\ra\).\\
 \center{ $\Downarrow$}\\~\\
\raggedright\(~~~\frac{1}{\sqrt{N}}\sum_{i=1}^{N} (-1)^{f(i)}|i>\otimes (\frac{1}{\sqrt{2}} 
|0\ra -  \frac{1}{\sqrt{2}}|1\ra)\)\\~\\\raggedright
$~~~~~$ Apply reversed Fourier transform on first register\\
\center{$\Downarrow$}\\~\\\raggedright
\(~~~~~|\psi\ra\otimes 
(\frac{1}{\sqrt{2}} |0\ra -  \frac{1}{\sqrt{2}}|1\ra)\)\\~\\
 $~~~~~$ Measure first register\\
\center{$\Downarrow$}\\~\\
\raggedright $~~~~~$ If outcome equals $0^n$, output ``constant''\\
\raggedright
$~~~~~$ Else, output ``balanced''
~\\~
\end{minipage}}

$~$

To see why this algorithm indeed works, let us denote by $|\psi_c\ra$
the vector $|\psi\ra$ in the case ``constant'', 
and $|\psi_b\ra$
the vector $|\psi\ra$ in the case ``balanced''.
Note that if $f(i)$ is constant, the second Fourier 
transform merely undoes the first Fourier transform, so $|\psi_c\ra=|0^n\ra$.
On the other hand,  if $f(i)$ is balanced,
 the vector \[\frac{1}{\sqrt{N}}\sum_{i=1}^{N} (-1)^{f(i)}|i>\]
is orthogonal to 
\[\frac{1}{\sqrt{N}}\sum_{i=1}^{N}|i>.\]
Since unitary operations preserve angles between vectors,
$|\psi_b\ra$ is orthogonal to $|\psi_c\ra$. Hence the probability to 
measure $0^n$ in the ``balanced'' case is zero. 
Hence, the algorithm gives the correct answer with probability $1$.
This algorithm shows the advantage of exact quantum complexity over 
exact classical complexity. 
However, when the restriction to exact solution is released, this advantage is 
gone. A classical probabilistic machine can solve the problem
 using a constant number of queries - though not by one query!
(This was shown in the overview).

Let me remark that discussing exact solutions is problematic
 in the context of quantum algorithms,  
 because of the continuous characteristics of quantum operators. 
Almost all quantum computations cannot be achieved exactly, when using a finite universal set of gates; the set of 
unitary operations is continuous, while the set of achievable operations using
a finite universal set of gates  is countable. 
Moreover, the notion of exact quantum algorithms is not robust, 
because the set of 
 problems that have exact solution depend very strongly on the universal set 
of gates. The function AND, for example,cannot be computed exactly by Deutsch's 
universal machine!

In the next algorithm, due to Simon, the exponential advantage is achieved 
even without requiring exact solutions. 
The problem can be specified as follows:

\begin{quote}
{\bf Simon's Problem:}

$f$ is a function from $\{1,N\}$ to $\{1,N\}$, where $N=2^n$.
 We are promised that one of two cases occurs:

Either all $f(i)$
are different, i.e. $f$ is ``one to one'',

 or

$f$  satisfies that $\exists s, f(i)=f(j)$ if and only if $i=j$ or $i=j\oplus s$, 
i.e $f$ is ``two to one''.

We are asked to distinguish between the two cases.
\end{quote}

Here a classical computer will need order of $O(N)$ queries, even when an error is allowed.  
Simon's quantum algorithm can solve this question with the expected number 
of queries being $O(\rm{log}(N))$. (In fact, 
Brassard {\it et.al.} improved this result from expected  $O(\rm{log}(N))$ queries
to worst case  $O(\rm{log}(N))$ queries\cite{brassard4}.)

We will work with two registers of $n$ qubits; both will hold an integer  between 
$1$ to $N$. The first register will carry numbers 
in  the range of the function.
The second register will carry 
the value of the function.

\frame{\begin{minipage}
[70mm]{160mm}
~\\\raggedright~\\\raggedright
$~~~~~~~~~~~~~~~~~~~~~~~~~~~$
{\bf   Simon's Algorithm}\\
~\\\raggedright$~~~~~~|0^n\ra\otimes |0^n\ra$\\~\\
\raggedright$~~~~~$ Apply Fourier transform on first register.\\
\center{ $\Downarrow$}\\~\\
\raggedright
\(~~~~~\frac{1}{\sqrt{N}}\sum_{i=1}^{N} |i>\otimes |0^n\ra\)
\\~\\\raggedright
$~~~~~$ Call oracle\\
 \center{ $\Downarrow$}\\~\\
\raggedright\(~~~~\frac{1}{\sqrt{N}}\sum_{i=1}^{N} |i>\otimes|f(i)\ra\)
\\~\\\raggedright
$~~~~~$ Apply Fourier transform on first register.\\
\center{$\Downarrow$}\\~\\
\raggedright\(~~~~\frac{1}{N}\sum_{k=1}^{N} |k>\otimes \sum_{i=1}^{N}(-1)^{i\cdot k}|f(i)\ra\)\\~\\
\raggedright
 $~~~~~$ Measure first register. Let $k_1$ be the outcome.\\
$~~~~~$ Repeat the previous steps $cn$ times to get $k_1$, $k_2$,..., $k_{cn}$.\\
 \center{ $\Downarrow$}\\~\\
\raggedright $~~~~~$Apply Gauss elimination to find a 
non-trivial solution for $s$ in the set of equations: 
\begin{eqnarray}
k_1\cdot s=0~ mod~ 2\nonumber\\
k_2\cdot s=0~mod ~2\nonumber\\
\vdots    \nonumber\\
k_{cn}\cdot s=0~mod~2\nonumber
\end{eqnarray}
\\ \center{ $\Downarrow$}\\~\\
\raggedright
$~~~~~$ If found, output ``two to one''. If not, declare ``one to one''.
~\\~
\end{minipage}}

{\bf Proof of correctness:}
To see why this algorithm works, let us analyze the probability to measure $k_1=k$, in the two cases.
In the case of ``one to one'', the probability to measure  $k_1=k$     is independent of $k$: 
 \begin{equation}
             \rm{Prob}(k_1=k)=\sum_i \left| \frac{(-1)^{i\cdot k}}{N}\right|^2 =\frac{1}{N}.
            \end{equation} 

  The above formula is derived by computing the squared norm of the projection
            of the measured vector 
             on $|k\rangle\otimes |f(i)\rangle$  and summing over all possible $f(i)$. 
            If we do the same thing in the "two to one" case, the projection 
             on $|k\rangle\otimes |f(i)\rangle$ will consist of two terms: 
             one comes from $i$ and the other from $i\oplus s$, since  $f(i)=f(i\oplus s)$. 
             Hence, in the following sum we divide by $2$ to correct for  
             the fact that every term is counted twice. 
In the case ``two to one'', we derive:

             \begin{equation}
           \rm{Prob}(k_1=k)=\frac{1}{2}\sum_i
             \frac{1}{N^2}|(-1)^{i\cdot k}+(-1)^{(i\oplus s) \cdot k}|^2 
              =\left\{\begin{array}{ll}\frac{2}{N} & \mbox{if $k\cdot s=0~ mod~ 2$}\\
		0 & \mbox{otherwise}\end{array}
		\right.
		\end{equation}

So we will only measure $k$ which is orthogonal to $s$.
In order to distinguish between the cases, we repeat the experiment 
many times, and observe whether the space spanned by the random vectors 
is the whole space or a subspace.
If we perform a large enough number of trials, we can be almost sure 
that in the ``one to one'' case, the vectors will 
span the whole space. Hence finding a non trivial solution will mean that 
we are in the ``two to one'' case.
A more precise argument follows.
Let $V$ be a vector space of dimension $n$ over $Z_2$.
 Let $S\subset V$ be the subspace spanned by the vectors,  $k_1,....k_t$,                                     which were measured at the first $t$ trials. 
If $S$ is not equal to $V$, a random vector $k_{t+1}$  from $V$ will be in $S$ with probability
at most $\frac{1}{2}$. 
Hence, with probability greater than half, the dimension of span$\{S,k_{t+1}\}$ is larger than that of $S$.
By Chernoff's law\cite{chernoff},
 the probability  the vectors will 
not span the whole space after $cn$ trials is exponentially small in $n$. $\Box$

{~}





This algorithm is exponentially more efficient than any  
randomized classical algorithm!
This seems like an extremely strong result, but
it is very important to notice  here that the problem is stated in the oracle setting and that  the algorithm does not apply for 
        any oracle, but only on oracles from  a restricted set:
either ``balanced'' or ``constant'' functions.
This restriction is called in complexity theory a ``promise'' to 
the algorithm: the algorithm is ``promised'' that the oracle 
is from some restricted subset.
          We will see later, in section $10$, that without such a "promise", quantum computation 
          and  classical computation are polynomially equivalent in terms of number of queries 
        to the oracle. This shows that in the absence of a promise, i.e. full range input,    
          the quantum advantage is exhibited not in the number of accesses to the input, 
     but in  the way the  information is processed. We will see an example for this in the next section, 
          in Shor's factorization algorithm.

\section{Shor's Algorithm for Factoring Integers}
 Shor's algorithm is  the most  important algorithmic result 
 in quantum computation.  The algorithm
 builds on ideas that already appear in 
Deutsch and Jozsa's algorithm and in Simon's algorithm, 
and like these algorithms, the basic ingredient of the algorithm is 
the Fourier transform.
The problem can be stated as follows:
\begin{quote}{\bf Input:} An integer N \\
{\bf Output:} A non-trivial factor of N, if exists.\nonumber
\end{quote}

There is no proof that there is no polynomial classical  factorization algorithm
The problem is even  not known to be $NP$-complete.
However, factorization  is regarded as hard, because many people have tried 
to solve it efficiently and failed.
In $1994$, Shor  published a polynomial (in $\rm{log}(N)$ )
  quantum algorithm for solving 
this problem \cite{shor1}.
This result is regarded as extremely important both theoretically and 
practically, although there is no proof that a classical algorithm does not exist.
The reason for the importance of this algorithm is mainly the fact that 
the security of the RSA cryptosystem, which is so widely used, 
is based on the assumed hardness of factoring integers.
Before explaining the algorithm, 
I would like to explain here in short how this cryptosystem works. 

A cryptosystem is a secure way to transform information such that an  eavesdropper
will not have any information about the message sent.
In the RSA method,
 the receiver, Bob, who will get the message, sends first a public 
key to Alice. 
Alice uses this key to encode her message, and sends it to Bob.
Bob is the only one who can encode the message, assuming factoring is hard.

{~}

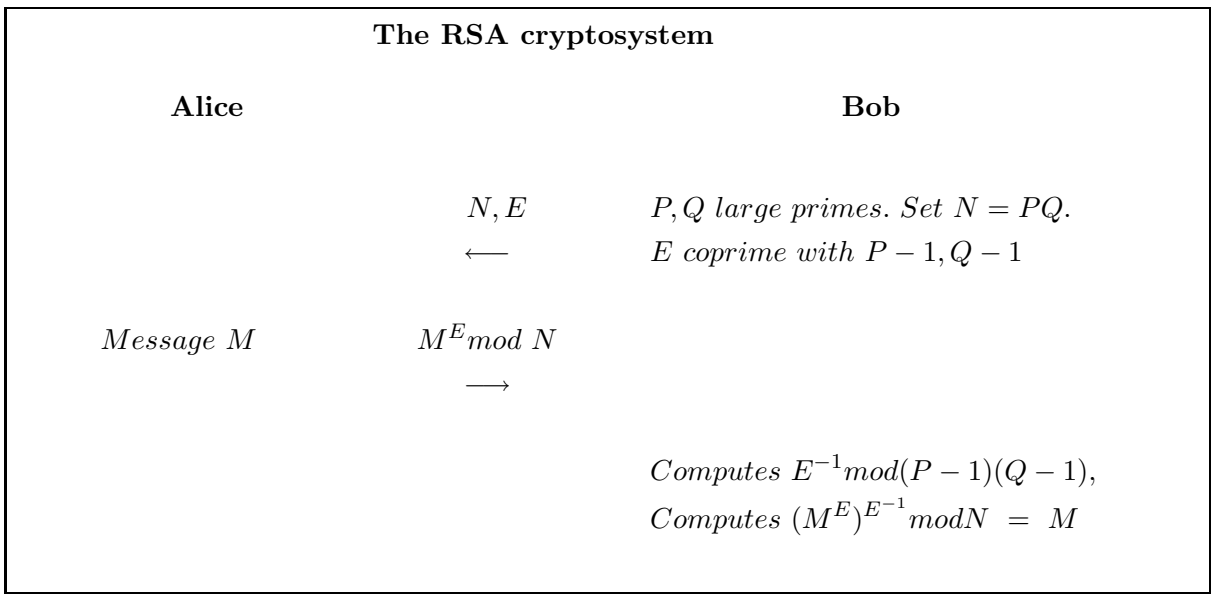
\frame{\begin{minipage}
[70mm]{160mm}
\raggedright~\\\raggedright
$~~~~~~~~~~~~~~~~~~~~~~~~~~~~~~~~~~~~~$
{\bf  The RSA cryptosystem}\\\raggedright~\\\raggedright
$~~~~~~$~~~~~~~~~~ {\bf Alice}$~~~~~~~~~~~~~~~~~~~~~~~~~~~~~~~~~~~~~~~~~~~~~~~~~~~~~~~$ ~~~~~
{\bf Bob}  \\\raggedright
\begin{eqnarray}
~~~~~~~~~~~~~~~~~~~~~~~~~~~~~~~~~ &~~~~~~~~~~~~N,E~~~~~~~~~~
&P,Q~ large~ primes.~
 Set ~ N=PQ.~~~~~~~~~~~~~~~~~\nonumber \\
 &\longleftarrow& E~ coprime~ with~ P-1, Q-1 \nonumber\\
~\nonumber\\
Message~ M~~~~~~~&M^E mod~N&\nonumber\\
&\longrightarrow&\nonumber\\
~\nonumber\\
&&Computes~ E^{-1} mod(P-1)(Q-1),\nonumber\\
&&Computes~ (M^{E})^{E^{-1}}modN~=~M\nonumber
\end{eqnarray}
\\~
\end{minipage}}

{~}

The key is chosen as follows:
Bob chooses two large primes $P$ and $Q$. 
He then computes $N=PQ$, and also picks an integer co-prime 
to $(P-1)(Q-1)=\phi(N),$ the number of co-primes to $N$ smaller than $N$. 
Bob sends $E$ and $N$ to the sender, Alice, 
using a public domain (newspaper, phone...)
The pair $(E,N)$ is called Bob's {\it public key.}
Bob keeps secret $D=E^{-1}mod (P-1)(Q-1)$, 
which he can compute easily knowing $P$ and $Q$, 
using the extended Euclid's algorithm\cite{fft}.
The pair $(N,D)$ is called Bob's {\it secret key.}
Alice computes her message, $M$, to the power of $E$, 
modulo $N$, and sends this number in a public channel to 
Bob. Note that Alice's computation is easy: taking a number $Y$ 
to the power of $X$ modulo $N$ is done by writing 
$X$ in binary representation: $X=x_1...x_n$. 
Then one can square
 $(Y^{x_i})$ $i$ times to get $(Y^{x_i})^{2^i}$,
 add the results for all $i$ and take the 
modulus over $N$. 
Bob decodes Alice's massage using his secret key 
by computing $(M^{E})^{D} mod N$.

Why does Bob get the correct message $M$?
This follows from Fermat's little theorem 
and the Chinese remainder theorem which together imply\cite{fft} that for 
          any $M$, $M^{k\phi(N)+1}=M~ mod ~N$. 
The security of this cryptosystem rests  on 
the difficulty of factoring large numbers.
If the eavesdropper has a factorization algorithm, he knows
 the factors $P,Q$, and he can simply play the role of Bob
in the last step of the cryptographic protocol. 
The converse statement, which asserts that in order 
to crack RSA one must have a factoring algorithm, is not proven.  
However, all known methods to crack  $RSA$ can be polynomially 
       converted to a factorization algorithm. 
Since factorization is assumed hard, classically, 
RSA is believed to be a secure cryptosystem to use. 
  In order to use RSA securely, one should work with integers that are a few hundreds digits in length, 
       since factoring smaller integers is still practical. 
       Integers of up to $130$ digits have been factorized by classical computers in no longer
       than  a few weeks. 
Due to the fact that the only classical factorization 
algorithm  is exponential, factorizing a number of twice the number of 
digits will take an eavesdropper not twice the time, but of the order
of million years.
If Alice and Bob  work with numbers of the order of hundreds of 
digits, they are presumably secure against classical eavesdroppers.

Shor's algorithm provides
 a quantum efficient way to break the RSA cryptosystem.
In fact,  Shor presented a quantum algorithm not for factoring, 
but for a different problem:
\begin{quote}{\bf Order modulo N:}\\
{\bf Input:} An integer $N$, and $Y$ coprime to $N$ \\
{\bf Output:} The order of $Y$, i.e. the minimal positive integer $r$ such that $Y^r=1~ mod~ N$.\nonumber
\end{quote}
   The problem of factorization can be polynomially reduced to the problem of finding the order 
        modulo $N$, using results from number theory.  
        I will not describe the reduction here; an 
explanation can be found in an excellent review on Shor's algorithm
\cite{ekert2}).
Instead, I will show a way\cite{cleve4} to crack RSA 
given an efficient algorithm to find the order modulo $N$:
Suppose the message sent is $M^E$. Find the order $r$ of  $M^E$ modulo $N$,
$r$ is also the order of $M$, since $E$ is coprime to $(P-1)(Q-1)=\phi(N)$.  
It is easy to find  efficiently
 the inverse 
of $E$, $D'=E^{-1}$ modulo $r$, using Euclid's algorithm.
Then simply, 
$(M^{E})^{D'}\equiv M~ mod ~N$, since $M^r\equiv~ 1~ mod ~N$.

Let me now present Shor's beautiful algorithm for finding the order of $Y$,
for any given $Y$, 
modulo $N$. 
The description follows\cite{ekert2}.  
In short, the idea of the algorithm is to create a state with 
periodicity $r$, and then apply Fourier transform 
over $Z_Q$, (the additive group of integers modulo $Q$), to reveal this periodicity. 
The Fourier transform over the group $Z_Q$ is defined as follows:
\begin{equation}
|a\ra \longmapsto \frac{1}{\sqrt{Q}}\sum_{b=0}^{Q-1} e^{2\pi iab/Q}  |b\ra=|\Psi_{Q,a}\ra
\end{equation}
The algorithm to compute this Fourier transform will be given 
in the next section, which is devoted entirely to Fourier transforms.
Again we will work with two registers. The first will hold a number between 
$1$ to $Q$. ($Q$ will be fixed later:  it is much larger than $N$, but
still polynomial in $N$.)
The second register will carry 
numbers between $1$ to $N$. 
Hence the two registers will consist of $O(\rm{log}(N))$ qubits

{~}

\frame{\begin{minipage}
[70mm]{165mm}
~\\\raggedright~\\\raggedright
$~~~~~~~~~~~~~~~~~~~~~~~~~~~$
{\bf   Shor's Algorithm}\\\raggedright~\\\raggedright
$~~~~~~|\stackrel{\rightarrow}{0}\ra\otimes |\stackrel{\rightarrow}{0}\ra$\\\raggedright~\\\raggedright
$~~~~~$ Apply Fourier Transform over $Z_Q$
   on the first register\\
 \center{ $\Downarrow$}\\~\\
\raggedright
\(~~~~~\frac{1}{\sqrt{Q}}\sum_{l=0}^{Q-1} |l\ra\otimes |\stackrel{\rightarrow}{0}\ra\)
\\~\\
\raggedright
$~~~~~$ Call subroutine
 which computes $|l\ra|d\ra \longmapsto|l\ra |d\oplus Y^l~ mod ~N\ra$ \\
 \center{ $\Downarrow$}\\~\\
\raggedright\(~~~~\frac{1}{\sqrt{Q}}\sum_{l=0}^{Q-1} |l\ra\otimes|Y^l mod N\ra\)
\\~\\\raggedright
$~~~~~$ Measure second register.\\
 \center{ $\Downarrow$}\\~\\
 \raggedright\(~~~~
\frac{1}{\sqrt{A}}\sum_{l=0|Y^l=Y^{l_0}}^{Q-1} |l\ra\otimes|Y^{l_0}\ra=
\frac{1}{\sqrt{A}}\sum_{j=0}^{A-1} |jr+l_0\ra\otimes|Y^{l_0}\ra
\)
\\~\\
\raggedright$~~~~~$
Apply Fourier Transform over $Z_Q$  on the first register\\
\center{$\Downarrow$}\\~\\
\raggedright\(~~~~
\frac{1}{\sqrt{Q}}\sum_{k=0}^{Q-1}\left(\frac{1}{\sqrt{A}} 
\sum_{j=0}^{A-1} e^{2\pi i (jr+l_0)k/Q}\right)
|k\ra\otimes|Y^{l_0}\ra\)\\
\center{ $\Downarrow$}\\~\\\raggedright
 $~~~~~$ Measure first register. Let $k_1$ be the outcome.\\ 
\raggedright
 $~~~~~$ Approximate the fraction $\frac{k_1}{Q}$ by a fraction with 
 denominator smaller than $N$, 
 \\ \raggedright $~~~~~$ using the 
(classical)  method of continued fractions.\\
\raggedright $~~~~~$
If the denominator $d$ doesn't satisfy $Y^d=1 mod ~N$, throw it away.\\
\raggedright$~~~~~$ Else call the denominator $r_1$.
 \\ \center{ $\Downarrow$}\\~\\
\raggedright
$~~~~~$ Repeat all previous steps $\rm{poly}(\rm{log}(N))$ times to get $r_1$, $r_2$,..\\
\raggedright
$~~~~~$  Output the minimal $r$. 
~\\~
\end{minipage}}

{~}

Let us now understand how this  algorithm works.
In the second step of the algorithm, all numbers between 
$0$ and $Q-1$ are present in the superposition, with 
equal weights.
In the third step of the algorithm, 
they are separated to sets, each has periodicity $r$. 
This is done as follows:
there are $r$ possible values written 
on the second register: $a\in\{Y^0,Y^1,....Y^{r-1}\}$.
The third state can thus be written 
as:
\[\frac{1}{\sqrt{Q}}\left( (\sum_{l=0|Y^l=Y}^{Q-1}
 |l\ra\otimes|Y\ra)+
(\sum_{l=0|Y^l=Y^2}^{Q-1} |l\ra\otimes|Y^2\ra)+....+
(\sum_{l=0|Y^l=Y^{r}}^{Q-1} |l\ra\otimes|Y^r=1\ra)\right)\]
Note that the values $l$ that give $Y^l=a$ 
have periodicity $r$:
If the smallest such $l$  is $l_0$, then $l=l_0+r,l_0+2r,..$  will 
also give $Y^l=a$. Hence each term in the brackets
 has periodicity $r$. 
Each set of $l'$s, 
with periodicity $r$, is attached to a different state 
of the second register. Before the computation of $Y^l$, 
all $l$'s appeared equally in the superposition. 
 Writing down the $Y^l$
 on the second register can be thought of as giving a different ``color''
to each periodic set in $[0,Q-1]$.
Visually, this can be viewed as follows:

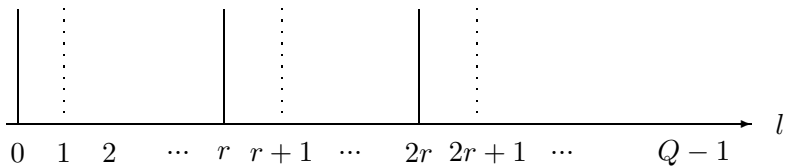
\setlength{\unitlength}{0.030in}

\begin{picture}(40,30)(-10,0)

\put(0,0){\vector(1,0){130}}

\put(2,-5){\makebox(0,0){$0$}}\put(2,0){\line(0,1){20}}
\put(10,-5){\makebox(0,0){$1$}}\qbezier[10](10,0)(10,10)(10,20)
\put(18,-5){\makebox(0,0){$2$}}
\put(30,-5){\makebox(0,0){$...$}}
\put(38,-5){\makebox(0,0){$r$}}\put(38,0){\line(0,1){20}}
\put(48,-5){\makebox(0,0){$r+1$}}\qbezier[10](48,0)(48,10)(48,20)
\put(60,-5){\makebox(0,0){$...$}}
\put(72,-5){\makebox(0,0){$2r$}}\put(72,0){\line(0,1){20}}
\put(84,-5){\makebox(0,0){$2r+1$}}\qbezier[10](82,0)(82,10)(82,20)
\put(97,-5){\makebox(0,0){$...$}}
\put(120,-5){\makebox(0,0){$Q-1$}}
\put(135,0){\makebox(0,0){$l$}}
\end{picture}

{~}

{~}

The measurement of the second register picks randomly one of these 
sets, and the state collapses to a superposition of $l'$s
with periodicity $r$,  with an arbitrary shift $l_0$.
Now, how to obtain the periodicity?  The first idea that 
comes to mind is to  measure 
the first register twice, in order to get two 
samples from the same periodic set, and somehow
deduce $r$ from these samples. 
However, the probability that the measurement of the second register yields the same shift 
in two runs of the algorithm, i.e. that the same periodic set is chosen twice, 
   is exponentially small. 
How to gain information about the periodicity
in the state without simply sampling  it?
This is done by the Fourier transform.
To understand the operation of the Fourier 
transform, we use a diagram again:

{~}

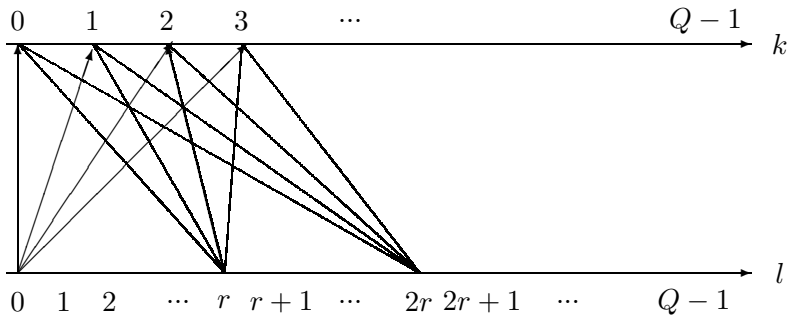
\setlength{\unitlength}{0.030in}

\begin{picture}(40,50)(-10,0)
\put(0,0){\vector(1,0){130}}
\put(0,40){\vector(1,0){130}}
\put(135,40){\makebox(0,0){$k$}}
\put(2,44){\makebox(0,0){$0$}}
\put(15,44){\makebox(0,0){$1$}}
\put(28,44){\makebox(0,0){$2$}}
\put(41,44){\makebox(0,0){$3$}}
\put(2,0){\vector(2,3){27}}
\put(2,0){\vector(0,1){40}}
\put(2,0){\vector(1,3){13}}
\put(2,0){\vector(1,1){40}}
\put(60,44){\makebox(0,0){$...$}}
\put(122,44){\makebox(0,0){$Q-1$}}
\put(2,-5){\makebox(0,0){$0$}}
\put(10,-5){\makebox(0,0){$1$}}
\put(18,-5){\makebox(0,0){$2$}}
\put(30,-5){\makebox(0,0){$...$}}
\put(38,-5){\makebox(0,0){$r$}}
\qbezier(38,0)(38,0)(2,40)
\qbezier(38,0)(38,0)(15,40)
\qbezier(38,0)(38,0)(28,40)
\qbezier(38,0)(38,0)(41,40)
\put(38,0){\vector(-1,4){10}}
\qbezier(72,0)(72,0)(2,40)
\qbezier(72,0)(72,0)(15,40)
\qbezier(72,0)(72,0)(28,40)
\qbezier(72,0)(72,0)(41,40)

\put(48,-5){\makebox(0,0){$r+1$}}
\put(60,-5){\makebox(0,0){$...$}}
\put(72,-5){\makebox(0,0){$2r$}}
\put(83,-5){\makebox(0,0){$2r+1$}}
\put(98,-5){\makebox(0,0){$...$}}
\put(120,-5){\makebox(0,0){$Q-1$}}
\put(135,0){\makebox(0,0){$l$}}
\end{picture}

{~}

{~}

Each edge in the diagram indicates that 
there is some probability amplitude to transform from
the bottom basis state to the upper one.
We now measure the first register, to obtain $k$. 
To find the probability to measure each $k$, we need to 
sum up the weights coming from all the $j'$s in the 
periodic set.

	\begin{equation}
	\rm{Prob}(k)=|\frac{1}{\sqrt{QA}} \sum_{j=0}^{A-1} e^{2\pi i k(jr+l_0)/Q}|^2=
	|\frac{1}{\sqrt{QA}} \sum_{j=0}^{A-1} (e^{2\pi i kr/Q})^j|^2
	\end{equation}

\noindent Hence, in order  to compute the probability to measure each 
$k$, we need to evaluate a geometrical series. Alternatively 
the geometric series is a sum over unit vectors in the complex plane.

{~}

\noindent {\bf Exact periodicity:} 
Let us assume for a second {\it exact periodicity}, 
i.e. that $r$ divides  $Q$ exactly. Then 
$A=Q/r$. 
In this case, the above geometrical series is equal to zero, unless 
$e^{2\pi i kr/Q}=1$. 
Thus we measure with probability $1$ only $k's$ such that
 $kr=0 ~mod~Q$. 
This is where destructive interference comes to play:
only ``good'' $k'$s, which satisfy  $kr=0 ~mod~Q$, 
remain, and all the others cancel out.
Why are such $k'$s ``good''?
We can write 
$kr=mQ$, for some integer $m$, or  $ k/Q=m/r$. 
We know $Q$, and we know $k$ since we have measured it. 
Therefore we can reduce the fraction $k/Q$. 
If $m$ and $r$ are coprime.
  the denominator 
will be exactly $r$ which we are looking for! 
By the prime number theorem,
there are approximately $n/log(n)$ numbers smaller than $n$ and  coprime  with $n$, so since 
 $m$ is chosen randomly,  repeating the experiment  a large enough 
number of  
times we will with very high probability 
eventually get $m$ coprime to $r$.

{~}

\noindent {\bf Imperfect periodicity:} 
In the general case, $r$ does not divide $Q$, 
and this means that the picture is less clear.
``Bad'' k's do not completely cancel out. 
We distinguish between two types of $k'$s, 
for which the geometrical series 
of  vectors in the complex plain looks as follows:

{~}

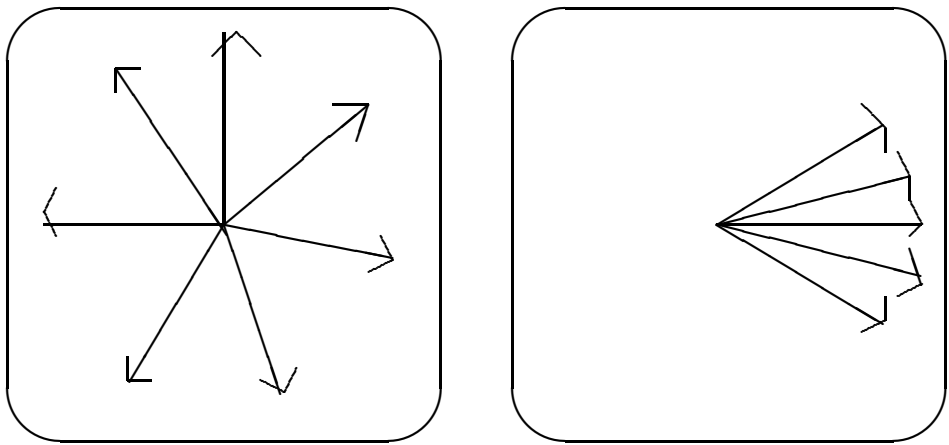
\setlength{\unitlength}{0.00083300in}%
\begingroup\makeatletter\ifx\SetFigFont\undefined
\def\x#1#2#3#4#5#6#7\relax{\def\x{#1#2#3#4#5#6}}%
\expandafter\x\fmtname xxxxxx\relax \def\y{splain}%
\ifx\x\y   
\gdef\SetFigFont#1#2#3{%
  \ifnum #1<17\tiny\else \ifnum #1<20\small\else
  \ifnum #1<24\normalsize\else \ifnum #1<29\large\else
  \ifnum #1<34\Large\else \ifnum #1<41\LARGE\else
     \huge\fi\fi\fi\fi\fi\fi
  \csname #3\endcsname}%
\else
\gdef\SetFigFont#1#2#3{\begingroup
  \count@#1\relax \ifnum 25<\count@\count@25\fi
  \def\x{\endgroup\@setsize\SetFigFont{#2pt}}%
  \expandafter\x
    \csname \romannumeral\the\count@ pt\expandafter\endcsname
    \csname @\romannumeral\the\count@ pt\endcsname
  \csname #3\endcsname}%
\fi
\fi\endgroup
\begin{picture}(5870,2718)(1866,-5095)
\thicklines
\put(3226,-3736){\oval(2704,2704)}
\put(6376,-3736){\oval(2704,2704)}
\put(3226,-3736){\line( 0, 1){1200}}
\put(3226,-3736){\line( 6, 5){900}}
\put(4126,-2986){\line(-1, 0){ 75}}
\put(3226,-3736){\line( 5,-1){1052.885}}
\put(3226,-3736){\line( 1,-3){352.500}}
\put(3226,-3736){\line(-3,-5){588.971}}
\put(3226,-3736){\line(-1, 0){1125}}
\put(2551,-2761){\line( 2,-3){692.308}}
\multiput(3301,-2536)(-6.00000,-6.00000){26}{\makebox(6.6667,10.0000){\SetFigFont{7}{8.4}{rm}.}}
\multiput(3301,-2536)(6.00000,-6.00000){26}{\makebox(6.6667,10.0000){\SetFigFont{7}{8.4}{rm}.}}
\put(3901,-2986){\line( 1, 0){225}}
\put(4126,-2986){\line(-1,-3){ 75}}
\multiput(4201,-3811)(3.75000,-7.50000){21}{\makebox(6.6667,10.0000){\SetFigFont{7}{8.4}{rm}.}}
\multiput(4276,-3961)(-7.50000,-3.75000){21}{\makebox(6.6667,10.0000){\SetFigFont{7}{8.4}{rm}.}}
\multiput(3676,-4636)(-3.75000,-7.50000){21}{\makebox(6.6667,10.0000){\SetFigFont{7}{8.4}{rm}.}}
\multiput(3601,-4786)(-7.50000,3.75000){21}{\makebox(6.6667,10.0000){\SetFigFont{7}{8.4}{rm}.}}
\put(2776,-4711){\line(-1, 0){150}}
\put(2626,-4711){\line( 0, 1){150}}
\multiput(2176,-3511)(-3.75000,-7.50000){21}{\makebox(6.6667,10.0000){\SetFigFont{7}{8.4}{rm}.}}
\multiput(2101,-3661)(3.75000,-7.50000){21}{\makebox(6.6667,10.0000){\SetFigFont{7}{8.4}{rm}.}}
\put(2551,-2911){\line( 0, 1){150}}
\put(2551,-2761){\line( 1, 0){150}}
\put(6301,-3736){\line( 1, 0){1275}}
\put(6301,-3736){\line( 4, 1){1200}}
\put(6301,-3736){\line( 4,-1){1270.588}}
\put(6301,-3736){\line( 5, 3){1036.765}}
\put(6301,-3736){\line( 5,-3){1036.765}}
\multiput(7201,-2986)(6.00000,-6.00000){26}{\makebox(6.6667,10.0000){\SetFigFont{7}{8.4}{rm}.}}
\put(7351,-3136){\line( 0,-1){150}}
\multiput(7426,-3286)(3.75000,-7.50000){21}{\makebox(6.6667,10.0000){\SetFigFont{7}{8.4}{rm}.}}
\put(7501,-3436){\line( 0,-1){150}}
\multiput(7501,-3586)(3.75000,-7.50000){21}{\makebox(6.6667,10.0000){\SetFigFont{7}{8.4}{rm}.}}
\multiput(7501,-3811)(6.25000,6.25000){13}{\makebox(6.6667,10.0000){\SetFigFont{7}{8.4}{rm}.}}
\put(7501,-3886){\line( 1,-3){ 75}}
\multiput(7576,-4111)(-7.50000,-3.75000){21}{\makebox(6.6667,10.0000){\SetFigFont{7}{8.4}{rm}.}}
\put(7351,-4186){\line( 0,-1){150}}
\multiput(7351,-4336)(-7.50000,-3.75000){21}{\makebox(6.6667,10.0000){\SetFigFont{7}{8.4}{rm}.}}
\end{picture}

{~}

In the left case, all vectors point in different 
directions, and they tend to cancel each other. 
This will cause destructive interference, which will 
cause the amplitude of such $k'$s to be small. 
In the right case, all vectors point almost to the same 
direction. In this case there will be constructive interference of all 
the vectors.
This happens when $e^{2\pi i kr/Q}$ is close to one, 
or when $kr ~mod ~Q$ is close to zero.
This means that with high probability, we will measure 
only $k's$ which satisfy an {\it approximate} 
criterion $kr\approx 0 ~mod ~Q$.
In particular, consider 
$k$'s which satisfy:
 \begin{equation}\label{criterion}
           -r/2 \le kr ~{\rm mod} ~Q \le r/2
            \end{equation}

There are exactly $r$ values of $k$
satisfying this requirement,   because $k$ runs from $0$ to $Q-1$, therefore $kr$ runs from $0$ to $(Q-1)r$,
           and this set of integers contains exactly 
$r$ multiples of $Q$.  Note, that for such $k'$s all the 
           complex vectors lie in the upper half of the complex plane,   so they are 
           instructively interfering. 
Now the probability to measure such a  $k$
is bounded below, by choosing the largest exponent possible:
\begin{eqnarray*}
               \rm{Prob}(k)=
		|\frac{1}{\sqrt{QA}} \sum_{j=0}^{A-1} (e^{2\pi i kr/Q})^j|^2\ge
		|\frac{1}{\sqrt{QA}} \sum_{j=0}^{A-1} (e^{i \pi r/Q})^j|^2 \\
		=\frac{1}{QA}|\frac{1-e^{\pi i rA/Q}}{1-e^{i\pi  r/Q}}|^2
                = \frac{1}{QA}|\frac{sin(\frac{\pi r A}{2 Q})}{sin(\frac{\pi r}{2 Q})}|^2
		\approx \frac{4}{\pi^2 r}
        	\end{eqnarray*}
Where the approximation is due to the fact that $Q$ is chosen 
to be much larger than $N> r$,   therefore the sinus in the  enumerator  is close to $1$ 
                                with negligible correction of the order of $r/Q$.
   In the denominator we use the approximation
                         $\rm{sin}(x)\approx x$  for small $x$, and the correction is again 
                        of the order of $r/Q$. The probability to measure any $k$
                        which satisfies \ref{criterion} is approximately $4/\pi^2$,
 since there are $r$                          such $k'$s. 

  Why are such $k'$s "good"?  Given an integer $k$ which satisfies the criterion 
               \ref{criterion}, we can find $r$ with reasonably high probability. 
Note that for
``good'' $k$'s, there exists an integer $m$ such that:
\[ |\frac{k}{Q}-\frac{m}{r}|\le \frac{1}{2Q}.\]

Remember that  $Q$ is chosen to be much larger than $N$, 
say $Q\ge N^2$. 
This means that  $\frac{k}{Q}$, 
a fraction with denominator $\ge N^2$,
can be approximated by $\frac{m}{r}$,  a fraction 
with denominator smaller than $N$, to within $\frac{1}{N^2}$
 There is only one fraction with such a small denominator that 
approximates a fraction so well with such large denominator.
Given $k/Q$, the approximating fraction, $\frac{m}{r}$,
 can be found efficiently,  using the method of continued 
fractions:  
\[a=a_0+\frac{1}{a_1+\frac{1}{a_2+...}},\]  where $a_i$ are all integers. 
Finding this fraction, the denominator will be $r$!
Well, not precisely.
Again, it might be the case that $m$ and $r$ are not coprime, 
and the number we find will be the denominator of the reduced 
fraction of $\frac{m}{r}$. In this case the number will 
fail the  test $Y^r=1$ which is included in Shor's algorithm,  and it will be thrown away. 
Fortunately, the probability for  $m$ to be  coprime to $r$ is 
large enough: it is greater than $1/log(r)$. 
 We repeat the experiment until this happens.

This concludes Shor's algorithm.
In the next chapter we will see an alternative  algorithm 
by Kitaev for finding the order modulo $N$.

\section{Fourier Transforms}  
The ability to  efficiently apply Fourier transforms over groups 
with exponentially many elements
 is unique to the quantum world. In fact, Fourier transforms are the 
{\it only} known tool in quantum computation which gives exponential
advantage. For this reason it is worthwhile to 
devote a whole chapter for Fourier transforms.  
The Fourier transform is defined as follows. 
Denote the additive group of integers modulo  $Q$ by $Z_Q$. 
         	Let $f$ be a function from the group $Z_Q$ to the complex numbers: 
\begin{equation}
f:a\longmapsto f(a)\in C
\end{equation}
The Fourier transform of this function is 
 another function from $Z_Q$ to the complex numbers:
\begin{equation}
	\hat{f}:a \longmapsto \hat{f}(a)=\frac{1}{\sqrt{Q}}
	\sum_{b\in Z_Q} e^{2\pi i ab/Q} f(b)\in C
	\end{equation}

The straight forward way to compute
the $Q$ Fourier coefficients of the function, $\hat{f}(a)$ $\forall a$,
 will take $O(Q^2)$ time.
When $Q$ is a factor of $2$,
 there is a way to shorten the trivial Fourier transform   algorithm using 
recursion. This is called
fast Fourier transform, or in short $FFT$, and it enables to compute  
the Fourier transform  within $O(Q\rm{log}(Q))$ time steps\cite{fft}. 
When $Q$ is very large, this still is a very slow operation.

In the quantum world, a function from the Abelian 
group $G=Z_Q$ to the complex numbers
$f:  a\longmapsto f(a)$ can be represented 
by a superposition $|f\ra=\sum_{a=0}^{Q-1} f(a)|a\ra$ (perhaps normalized.)
The Fourier transform of the function will be 
$|\hat{f}\ra=\sum_{a=0}^{Q-1} \hat{f}(a)|a\ra$.
Note that in the quantum setting, 
the function on $Q$ elements is represented compactly
as a superposition on $log(Q)$ qubits.
This compact representation allows in some cases to apply 
the transformation $|f\ra \longmapsto |\hat{f}\ra$ very efficiently, 
 in only $O(log(Q))$ time steps.
Indeed, measuring all the Fourier 
coefficients will still  take    time which is exponential in log($Q$) simply because the number 
          of coefficients is exponential. However, the actual transformation  from 
a superposition to its Fourier transform will be very fast. 

  In order to apply the Fourier transformation on general states, 
            it suffices to apply the following transformation on the basis states:
\begin{equation}
|a\ra \longmapsto 
|\Psi_{Q,a}\ra=\frac{1}{\sqrt{Q}}\sum_{b=0}^{Q-1}
 e^{2\pi i ab/Q}
|b\ra.
\end{equation}
  We will first consider the special case of $Q=2^m$, which is simpler than the 
            general case, since classical techniques
for fast Fourier transforms can be adopted\cite{shor1,cleve3,coppersmith,
deutsch3,griffiths} 
I will give here a  nice description by Cleve {\it et. al.}\cite{cleve4}. 
Later I'll describe Kitaev's\cite{kitaev1} more general quantum Fourier transform, for any Abelian group, which implies a beautiful
alternative factorization algorithm.
 

%

{~}

\noindent{\bf Quantum fast  Fourier transform.}
Let $Q=2^m$. 
An integer  $a\in\{0,1,...,2^m-1\}$ is represented in binary representation by 
$|a_1...a_m\ra$, 
so $a=a_{1}2^{m-1}+a_{2}2^{m-2}+....+a_{m-1} 2^1+a_m$.
Interestingly, the Fourier state in this case 
is not entangled, and can be written 
as a tensor product: 
\begin{equation}\label{ft}|\Psi_{Q,a}\ra=\frac{1}{\sqrt{Q}}\sum_{b=0}^{Q-1}e^{2\pi iab/Q}
|b\ra=\frac{1}{\sqrt{2^m}}(|0\ra+e^{2\pi i 0.a_m}|1\ra)
(|0\ra+e^{2\pi i 0.a_{m-1}a_m}|1\ra)...(|0\ra+e^{2\pi i 0.a_{1}...a_{m-1}a_m}
|1\ra)\end{equation}
We can see this by computing the coefficient of $b$ 
in this formula. 
In fact, what matters is that the phases in the coefficient of $b$
from both sides of the equality  
are equal (modulo $1$). 
To see this, observe that the phase of $|b\rangle$ in the left term is
             $2^{-m}ab=2^{-m}\sum_{i,j=1}^{m} a_i 2^{m-i} b_j 2^{m-j}$, which can be seen to  
be equal   modulo $1$ to $0.a_m\cdot b_1 + 0.a_{m-1}a_m\cdot b_2 +... +0.a_1...a_{m-1}a_m\cdot b_m$ which is the phase of $|b\rangle$ in the  right term.

To apply the QFFT, we will need only two gates. 
The first is  the Hadamard gate on one qubit.
The second gate is a  gate  on two qubits, which  applies
a conditioned phase shift on one qubit, given that the other qubit
is in state $|1\ra$. 
 $R_k$ denotes 
 the phase shift on one qubit by  $e^{2\pi i/2^k}$.
\begin{equation}\label{fast}
R_{k}=\left(\begin{array}{cc}
 1& 0\\
 0 & e^{2\pi i/2^k}
\end{array}\right) ~~~, ~~~ 
H=\left(\begin{array}{ll}
\frac{1}{\sqrt{2}}&\frac{1}{\sqrt{2}}\\
\frac{1}{\sqrt{2}}&-\frac{1}{\sqrt{2}}
\end{array}
\right)
\end{equation}
We will operate the following gate array:

{~}

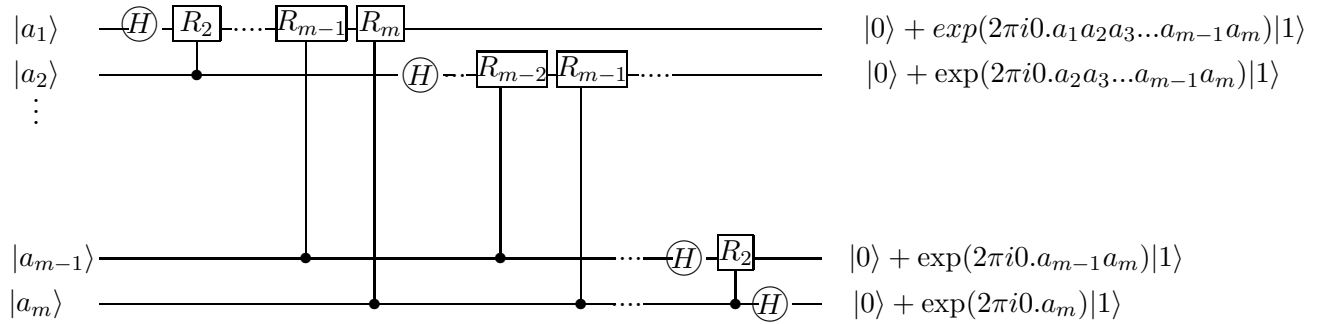
\setlength{\unitlength}{0.030in}

\begin{picture}(40,60)(-10,0)

\put(-11,0){\makebox(0,0){$|a_m\ra$}}
\put(0,0){\line(1,0){90}}
\put(92,0){\makebox(0,0){$....$}}
\put(95,0){\line(1,0){20}}\put(117,0){\circle{6}}
\put(117,0){\makebox(0,0){$H$}}
\put(121,0){\line(1,0){5}}
\put(155,0){\makebox(0,0){$|0\ra+\exp(2\pi i 0.a_m)|1\ra$}}

\put(-8,8){\makebox(0,0){$|a_{m-1}\ra$}}
\put(0,8){\line(1,0){90}}
\put(92,8){\makebox(0,0){$....$}}
\put(95,8){\line(1,0){5}}\put(102,8){\circle{6}}
\put(102,8){\makebox(0,0){$H$}}
\put(106,8){\line(1,0){2}}
\put(108,6){\framebox(6,6){$R_2$}}
\put(111,0){\line(0,1){6}}
\put(111,0){\circle*{2}}
\put(114,8){\line(1,0){12}}
\put(160,8){\makebox(0,0){$|0\ra+\exp(2\pi i 0.a_{m-1}a_m)|1\ra$}}

\put(-11,35){\makebox(0,0){$\vdots$}}

\put(-11,40){\makebox(0,0){$|a_{2}\ra$}}
\put(0,40){\line(1,0){52}}\put(56,40){\circle{6}}
\put(56,40){\makebox(0,0){$H$}}
\put(60,40){\line(1,0){2}}
\put(62,40){\makebox(0,0){$...$}}
\put(64,40){\line(1,0){2}}
\put(66,38){\framebox(12,6){$R_{m-2}$}}
\put(70,8){\line(0,1){30}}
\put(70,8){\circle*{2}}
\put(78,40){\line(1,0){2}}
\put(80,38){\framebox(12,6){$R_{m-1}$}}
\put(84,0){\line(0,1){38}}
\put(84,0){\circle*{2}}
\put(92,40){\line(1,0){2}}
\put(97,40){\makebox(0,0){$....$}}
\put(100,40){\line(1,0){26}}
\put(170,40){\makebox(0,0){$|0\ra+\exp(2\pi i 0.a_2a_3...a_{m-1}a_m)|1\ra$}}

\put(-11,48){\makebox(0,0){$|a_{1}\ra$}}
\put(0,48){\line(1,0){5}}\put(7,49){\circle{6}}
\put(7,49){\makebox(0,0){$H$}}
\put(11,48){\line(1,0){2}}
\put(13,46){\framebox(8,6){$R_{2}$}}
\put(17,40){\line(0,1){6}}
\put(17,40){\circle*{2}}
\put(21,48){\line(1,0){2}}
\put(26,48){\makebox(0,0){$....$}}
\put(29,48){\line(1,0){2}}
\put(31,46){\framebox(12,6){$R_{m-1}$}}
\put(36,8){\line(0,1){38}}
\put(36,8){\circle*{2}}
\put(43,48){\line(1,0){2}}
\put(45,46){\framebox(8,6){$R_{m}$}}
\put(48,0){\line(0,1){46}}
\put(48,0){\circle*{2}}
\put(53,48){\line(1,0){73}}
\put(172,48){\makebox(0,0){$|0\ra+exp(2\pi i 0.a_1a_2a_3...a_{m-1}a_m)|1\ra$}}

\end{picture}

{~}

{~}

 We claim that this gate array implements the FT, except that 
         the output is in reverse order of bits.  To prove this, we show  that each bit gains the phase it is supposed to
gain, according to  equation \ref{ft}.
The first $H$ on the first bit $a_1$ produces the state on $m$ qubits:
\[(|0\ra +e^{2\pi i (0.a_1)}|1\ra)|a_2...a_m\ra\]
and the next $R_2$ makes it
\[(|0\ra +e^{2\pi i (0.a_1a_2)}|1\ra)|a_2...a_m\ra,\]
and so on until the first qubit is in the correct state (of the last bit 
in equation \ref{ft}):
\[(|0\ra +e^{2\pi i (0.a_1a_2...a_m)}|1\ra)|a_2...a_m\ra.\]
In the same way the phases of the rest of the qubits are fixed,
 one by one. 
We now simply reverse the order of the bits to obtain the correct FT.

%

Note that the number of gates is $m(m-1)/2$ which is $O(log^2(Q))$.
In fact, many of these gates can be omitted, because 
 $R_{k}$ can be exponentially close to one. 
 omitting such  gates we still obtain 
a very good approximation of the Fourier transform\cite{coppersmith}.

{~}

\noindent{\bf Kitaev's algorithm}:
Kitaev's algorithm\cite{kitaev1} shows how to approximate efficiently
 the FT over the cyclic group $Z_Q$ for any $Q$ (a cyclic group is a group that is generated by one element).
The generalization to any Abelian group   is simple\cite{kitaev1}, but will not be described  here.
The sequence of operation is the following:

{~}

\frame{\begin{minipage}
[70mm]{160mm}
\raggedright~\\\raggedright
$~~~~~~~~~~~~~~~~~~~~~~~~~~~$
{\bf   Fourier Transform a la Kitaev}\\\raggedright~\\\raggedright
$~~~|a\ra\otimes|0\ra ~~\Longrightarrow ~~
|a\ra\otimes|\Psi_{Q,0}\ra ~~\Longrightarrow~~
|a\ra\otimes|\Psi_{Q,a}\ra~~\Longrightarrow ~~
|0\ra\otimes|\Psi_{Q,a}\ra~~ \Longrightarrow~~
|\Psi_{Q,a}\ra\otimes|0\ra ~~~$

{~}

\end{minipage}}

\noindent The most important and difficult step in this algorithm is 
the third step.  Let us understand how to perform each of the other steps
first:
\begin{enumerate}
\item   $|0\ra \longmapsto |\Psi_{Q,0}\ra$ is actually a classical operation.  
We pick an integer between $1$ and $Q$ uniformly at random using a recursive procedure.
Let $2^{n-1}<Q<2^n$.  Denote $Q_0=2^{n-1}$ and
$Q_1=Q-Q_0$. Apply the one qubit gate
\(
|0\ra \longmapsto \sqrt{\frac{Q_0}{Q}}|0\ra +\sqrt{\frac{Q_1}{Q}}|1\ra\).
Now,  conditioned on the first bit $x$,  create on the last $n-1$ bits,
the state $|\Psi_{Q_x,0}\ra$ recursively.

\item
 $|a\ra\otimes|\Psi_0\ra ~~\Longrightarrow~~
|a\ra\otimes|\Psi_a\ra$ is achieved by applying
 \(|a,b\ra \longmapsto e^{2\pi i ab/Q} |a,b\ra.\)
 
\item
The third operation is, perhaps surprisingly, 
the 
 most difficult part in the FT, and I will sketch the idea next. 
\item  The last operation is merely swapping the bits.
\end{enumerate}

To apply the third step, we note that  
the vectors $|\Psi_{Q,a}\ra$
are eigenvectors of the unitary operation $U: |g^m\ra\longmapsto |g^{m+1}\ra$,
where $g$ is the generator of the cyclic group, 
with eigenvalues $e^{- 2\pi i a/Q}$
The operation  \(|a\ra\otimes|\Psi_{Q,a}\ra~~\Longrightarrow ~~
|0\ra\otimes|\Psi_{Q,a}\ra\) is actually the reverse of computing 
the eigenvalue of an eigenvector.
We need to be able to write down the eigenvalues of a given 
unitary matrix. Kitaev has proved the following lemma:
\begin{lemm} (Kitaev)
Let $U$ be a unitary matrix on $n$ qubits such that   
$U,U^2,U^4...U^{2^n}$ can be applied efficiently. 
Let $ |\Psi_{\theta}\ra$ be $U$'s eigenvectors with corresponding 
eigenvalues $e^{i\theta}$.
Then the transformation  \(|\Psi_{\theta}\ra\otimes|0\rangle~~\Longrightarrow ~~|\Psi_{\theta}\ra\otimes|\theta\ra\) can be approximated to
exponential accuracy, efficiently. 
\end{lemm}

%
%
%

{\bf Proof:}
The idea that lies behind this theorem is {\it interference.}
The eigenvalues are phases, and in order
 to gain information about a phase we need to compare it with some
 reference phase, just like what happens in  
an interferometer. The implementation of this idea
in the setting of qubits  is done by adding a control qubit. We proceed as follows.
 We apply the Hadamard transform $H$ on the control qubit, which
 separates the state to two paths, 
one in which the control qubit is in state $|1\ra$
and the other in which it is $|0\ra$.
Now $U$ is applied  on $|\Psi_\theta\ra$, {\it conditioned} that 
the control qubit is $1$.
This adds a phase $e^{i\theta}$ on one of the paths, which can be compared
to the reference path. Finally, the controlled qubit is rotated again
by a Hadamard transform. The following diagram captures the idea schematically:

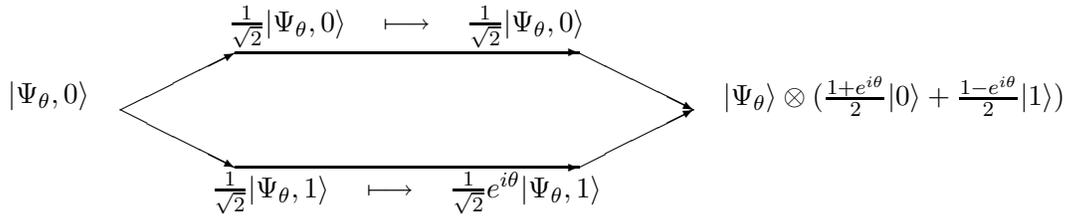
\setlength{\unitlength}{0.030in}
\begin{picture}(100,50)(-20,-20)
\put(-10,0){\makebox(5,5){$|\Psi_\theta,0\ra$}}
\put(5,0){\vector(2,1){20}}
\put(30,12){\makebox(50,5){$\frac{1}{\sqrt{2}}|\Psi_\theta,0\ra~~~
 \longmapsto~~~\frac{1}{\sqrt{2}}|\Psi_\theta,0\ra $}}
\put(30,-17){\makebox(50,5){$\frac{1}{\sqrt{2}}|\Psi_\theta,1\ra ~~~
\longmapsto~~~\frac{1}{\sqrt{2}}e^{i\theta}|\Psi_\theta,1\ra$}}
\put(5,0){\vector(2,-1){20}}
\put(25,10){\vector(1,0){60}}
\put(25,-10){\vector(1,0){60}}
\put(85,10){\vector(2,-1){20}}
\put(85,-10){\vector(2,1){20}}
\put(115,0){\makebox(50,5){$|\Psi_\theta\ra
\otimes(\frac{1+e^{i\theta}}{2}|0\ra+\frac{1-e^{i\theta}}{2}|1\ra)$}}
\end{picture}

{~}

The control qubit is now in a state 
$|\beta\ra=(\frac{1+e^{i\theta}}{2}|0\ra+\frac{1-e^{i\theta}}{2}|1\ra)$,
 which is a qubit biased 
according to the eigenvalue. If we measure this qubit, it behaves
 like a coin flip
 with bias $p=|1-e^{i\theta}|^2/4=\frac{1-\rm{cos}\theta}{2}$.

	The idea is to create many control qubits, and measure all of them. 
        This is like performing many independent coin tosses.  
        We can deduce $\theta$ from the ratio between the number of times we got $1$ 
        and the number of times we got $0$. For this, we will apply a classical algorithm 
        on the outcomes of the measurements.  
        However, there are two problems with this idea. One is that the outcome of the algorithm will 
       be classical, while we want to create 
        a unitary transformation which writes down the eigenvalues
       and can be applied on superpositions.  We will deal with 
       this problem later. A more severe problem is that the algorithm should find $\theta$ with 
        exponential accuracy (polynomially many bits), since there are exponentially many eigenvalues. 
       To achieve  exponential accuracy in $\theta$  we need exponentially many coin tosses; 
       By Chernoff's inequality\cite{fft},  exponentially many coin tosses are required in order to
        achieve exponential accuracy in $\theta$. 
       Since we are limited to  polynomial algorithms, we can only deduce $\theta$ with
       polynomial accuracy.  The solution to this problem takes advantage 
        of the fact that the powers of $U$ can be applied efficiently. 
         To deduce $\theta$ to higher accuracy, 
	we slightly modify the interference scheme: instead of $U$,
        we apply  $U^2$. This will generate another set of  biased qubits, 
        from which we can deduce $2\theta$ with polynomial accuracy. 
         The same thing can be done using \(U^4,...,U^{2^n},\)
        and this will generate $n$ sets of $m=poly(n)$ biased qubits. 
        From the outcomes of the measurements of the $j'$th set, we compute
         $2^j\theta$ with polynomial accuracy. It is easy to construct  a 
        polynomial classical algorithm that computes $\theta$ with exponential  precision 
         (which is what we need) from the polynomial approximations of  
         $\theta$,$2\theta$,$4\theta$,... $2^n \theta$.
        
        It is left to show how the above computation can be made unitary. 
        The idea is that it is not necessary to measure each set of qubits, 
        in order to count the number of $1'$s. 
Instead of measuring these bits, we will apply a unitary transformation
that counts the portion  of $1$'s  out of $m$
 and writes this portion down on a {\it counting} register.  
If we denote by $w(i)$ the number of $1'$s in a string $i$, or the 
{\it weight} of the string, then this transformation will be:
\begin{equation}\label{statekitaev}
|i\ra|0\ra \longmapsto |i\ra|w(i)/m\ra.
\end{equation}
The resulting state will look something like:
\begin{equation}\label{stateprob}
|\Psi\ra\otimes \sum_i \sqrt{p^w(i)(1-p)^{m-w(i)}} |i\ra|w(i)\ra
\end{equation}
with perhaps extra phases.
Most of the weight in this state is concentrated on strings with 
approximately $pm$ $1'$s, like in a Bernoulli experiment. 
For each set of control qubits, we obtain some portion, written on 
           the counting register of that set. We denote the $n$ portions by 
          \(w_\theta,w_{2\theta}...w_{2^n\theta}. \)
           We can now apply  the unitary version of the classical algorithm which 
          computes an exponentially 
          close approximation of $\theta$ given the portions $w$.
          If we call this procedure $T$, we have: 

	\begin{equation}
	|w_\theta\ra|w_{2\theta}\ra\cdots|w_{2^n\theta}\ra|0\ra 
        \stackrel{T}{\longrightarrow}
	|w_\theta\ra|w_{2\theta}\ra\cdots|w_{2^n\theta}\ra|\theta\ra 
	\end{equation}
        
           We now have $\theta$ written down on the last register. 
          Let us denote by $Q'$ the unitary operation which the algorithm applies so far.  
         It is tempting to think that $Q'$ is  the desired    transformation,  
              \(|\Psi_{\theta}\ra\otimes|0\rangle~~\Longrightarrow ~~|\Psi_{\theta}\ra\otimes|\theta\ra.\) 
         This is not true. Actually, $Q'$ is exponentially close
        to 
\(|\Psi_{\theta}\ra\otimes|0\rangle\otimes|0\rangle~~\stackrel{Q'}{\Longrightarrow}
 ~~|\Psi_{\theta}\ra\otimes|\theta\ra\otimes|garbage_\theta\rangle,\)
              
        where the last register consists of all the control qubits and ancilla qubits which we have used
        during the computation. 
        The reason for the fact that $Q'$ is not exactly $Q$, is that 
        in the classical coin tossing, there is an exponentially 
      small probability to get result which is very far from the expected number of $1'$s, $mp$. 
       This translates  in
       equation \ref{stateprob} to the appearance, with exponentially small 
       weight, of strings $i$ which are very far from the 
      expected number of $1'$s $mp$. We now want to ask, why do the garbage qubits matter.  
        These  qubits carry information which is  no longer needed, 
           but never the less are entangled with the rest of the computer. 
        The point is that their existence might  prevent interference in future computation.
         We will develop tools to think about interference in section $9$, but roughly, garbage has the 
        same effect as interaction with the environment, which is known to cause decoherence.
       How to get rid of the garbage? 
             The problem is that we cannot simply erase  the garbage by setting all the garbage qubits
            to $|0\rangle$, because             the transformation that takes a general 
       state to  $|0\rangle$ is not   unitary.
           Fortunately, in our case there is a unitary transformation that erases the garbage.
            We do the following: 
            We copy $\theta$, which is written on the last register,
             on an extra register which is initialized
          in the state $|0\rangle$. The copying is done bit by bit,  using polynomially many
           $CNOT$ gates.  We now  apply in reverse order the reverse of all transformations done so far
            in the algorithm, except for the $CNOT$ gates. 
           The overall transformation is exponentially close to the following sequence
            of operations:
           apply $Q$, then copy $\theta$ and then apply $Q^{-1}$. 
            This sequence of operation indeed achieves the desired transformation 
           without garbage:
    
        \begin{eqnarray*}    
        |\Psi_{\theta}\ra\otimes|0\rangle\otimes|0\rangle\otimes|0\rangle~~\stackrel{Q}{\Longrightarrow}
         ~~|\Psi_{\theta}\ra\otimes|\theta\ra\otimes|garbage_\theta\rangle\otimes|0\rangle
       ~~\stackrel{CNOT~~gates} {\Longrightarrow}\\
  ~~|\Psi_{\theta}\ra\otimes|\theta\ra\otimes|garbage_\theta\rangle\otimes|\theta\rangle
   ~~\stackrel{Q^{-1}} {\Longrightarrow}
   ~~|\Psi_{\theta}\ra\otimes|0\ra\otimes|0\rangle\otimes|\theta\rangle.
      \end{eqnarray*}
      
      One can save many qubits by erasing garbage in the middle of the computation, when it 
     is no longer needed, and using these erased qubits as register in the rest of the computation. 
A different proof of this lemma can be found in \cite{cleve4}, where 
QFFT over $Z_2^n$ is used.
$\Box$

{~}

This concludes the Fourier transform algorithm. 
Kitaev's procedure of writing the eigenvalue down
 implies a very simple alternative factorization algorithm.
The way an integer $N$ is factorized is done again by finding the order of 
a number $Y$ which is coprime to $N$. 
(Recall that the order of $Y$ is the least $r$ such that $Y^r=1 ~ mod ~N$.) 
Consider the unitary transformation 
\(U:|g\ra \longmapsto |gY ~mod ~N\ra\).
The eigenvectors of $U$, $\{|\Psi\ra\}$,
 are exactly the linear superpositions of 
all configurations in the subgroup $\{Y, Y^2, Y^3,...Y^r\}$,
or any coset of this subgroup,  $\{gY, gY^2,gY^3,...gY^r\}$,
with appropriate phases:
\[U|\Psi_a\ra=U(\sum_j e^{2\pi i ja/r} |gY^j\ra)=e^{-2\pi i a/r}
(\sum_j e^{2\pi i ja/r} |gY^j\ra) .\] 
The eigenvalues of $U$ hold information about 
$r$! The idea would be to apply Kitaev's lemma, 
write down $\theta=2\pi a/r$ and deduce $r$ 
from it.

We start with the basis state $|0\rangle$, which can  be written as an equal superposition of all 
eigenvectors: $|0\ra=\sum_a |\Psi_a\ra$, 
as you can  easily check. Applying Kitaev's lemma on the state $|0\ra$
 we get on the second register all eigenvalues written
with uniform probability. We now measure this 
       register, which carries an exponentially close approximation of $2\pi a/r$.
       We divide by $2\pi$ to get $c$,  an exponentially good approximation of $a/r$.
       Now, using the method of continued fraction, like in Shor's algorithm, we find 
       the closest fraction to $c$ with denominator less than $N$. With high enough probability 
$a$ and $r$ are coprime, so we  get $r$ in the denominator. If not, the denominator does not 
satisfy $Y^r=1~mod N$, and we repeat the experiment again. Here is a summary of the algorithm: 

{~}

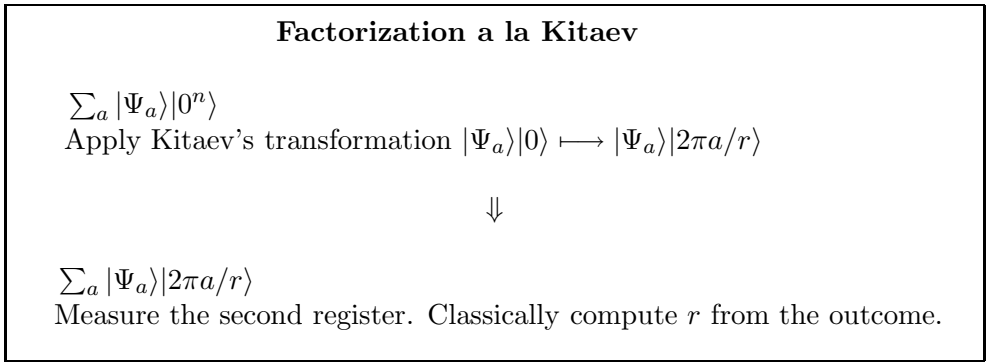
\frame{\begin{minipage}
[70mm]{130mm}
\raggedright~\\\raggedright
$~~~~~~~~~~~~~~~~~~~~~~~~~~~$
{\bf   Factorization a la Kitaev}\\~\\
\raggedright
 $~~~~~~\sum_{a} |\Psi_a\ra|0^n\ra $\\
\raggedright
 $~~~~~$ Apply Kitaev's transformation $|\Psi_a\ra|0\ra \longmapsto 
 |\Psi_a\ra|2\pi a/r\ra$
  \\ \center{ $\Downarrow$}\\~\\
\raggedright
$~~~~~\sum_{a} |\Psi_a\ra|2\pi a/r\ra$\\
\raggedright
 $~~~~~$Measure the second register. Classically compute $r$ from the outcome. \\ 
{~}

\end{minipage}}

{~}

Factorization can be viewed as finding the order of elements 
in Abelian groups.
Many people tried to generalize Shor's and Kitaev's algorithms 
to non-Abelian groups.  It is conjectured that 
Fourier transforms over non-Abelian groups would be helpful 
tools, however they are much more complicated operations
since the Fourier coefficients are complex 
{\it matrices}, and not complex numbers!
  Beals\cite{beals1} made the first (and only) step in this direction
by   discovering
  an efficient quantum Fourier transform 
algorithm for the non-Abelian permutations group, $S_n$,
building on the classical FFT over $S_n$\cite{diaconis,clausen}. 
Beals was motivated by an old hard problem in computer science:
Given two graphs, can we say whether they are isomorphic (i.e 
one is simply a permutation of the other) or not. 
This problem is not known to be $NP-$ complete, but the best 
known algorithm is exponential. 
It is still not known whether Beals' Fourier transform 
 can be used for solving graph isomorphism.
A very interesting open question is whether efficient quantum Fourier 
transforms can be done over any group, and can they be used 
to solve other problems.

\section{Grover's Algorithm for Finding a Needle in  a Haystack}
Grover's  algorithm is surprising and counter 
intuitive at first sight, though it achieves only a polynomial (quadratic)
improvement over classical algorithms. It deals with the {\it database 
search problem.}
Suppose you have access to an unsorted database of size $N$. 
You are looking for an item $i$ which satisfies some property.
It is easy to check whether the property is satisfied or not.
 How long will it take you to find such an item, if it exists?
If you are using classical computation, obviously it can take you $N$ 
steps. 
If you are using probabilistic classical computation, you can reduce it 
to $N/2$ expected steps.
But if you are using a quantum computer, you can find the item 
in $O(\sqrt{N})$ steps!
I will present here the algorithm which was found by Grover\cite{grover1} in 
$1995$. However, I will use here a different representation of 
the algorithm, which is mainly based on the geometrical 
interpretation by Boyer {\it et.al.}
\cite{boyer1,brassard1}.

The algorithm works as follows. 
Set  $\rm{log}(N)=n$, and let us define a function
\(
f:\{0,1\}^n\longmapsto \{0,1\}\)
where   $f(i)=0$ if the $i'th$ item does not satisfy
the desired property, and $f(i)=1$ in the case it does. 
Let $t$ be the number of items such that $f(i)=1$.
For the moment, we assume  that $t=1$.
The algorithm operates in the Hilbert space of $n$ qubits. 
Its main part actually works in a subspace of dimension $2$ of this space.
This subspace is the one which is spanned by the two vectors:
\begin{equation}
|a\ra=\frac{1}{\sqrt{N}}\sum_{i=0}^{2^n-1}|i\ra~~~,~~~ 
|b\ra= \frac{1}{\sqrt{N-1}}\sum_{i=0|f(i)=0}^{2^n-1}|i\ra.
\end{equation}

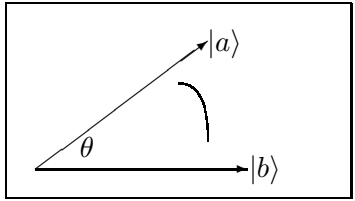
\setlength{\unitlength}{0.030in}
\begin{picture}(100,40)(-60,-5)
\put(0,0){\framebox(60,34)}
\put(5,5){\vector(4,3){30}}
\put(5,5){\vector(1,0){37}}
\put(45,5){\makebox(0,0){$|b\ra$}}
\put(14,9){\makebox(0,0){$\theta$}}
\put(38,27){\makebox(0,0){$|a\ra$}}
\qbezier(35,10)(35,20)(30,20)
\end{picture}

We begin by applying
a FT on $|0\ra$ which generates the uniform vector $ |a\ra$, 
using $n$ Hadamard gates.
We now want to rotate the vector in the two dimensional subspace
spanned by $ |a\ra$ and $|b\ra$, so that eventually we have 
large projection on the direction orthogonal to
 $|b\ra$, which is exactly the item we want.
The idea is that a rotation by the angle $2\theta$, 
is equivalent to two {\it reflections},
 first with respect to $ |a\ra$, 
and then with respect
 to $|b\ra$.
We define a Boolean function $g(i)$ to be $0$ only for $i=0$, and $1$ 
for the rest. A reflection around $|0\ra$ is obtained by 
$R_0: |i\ra \longmapsto (-1)^{g(i)} |i\ra$.
A reflection around $|a\ra$ is achieved by: $R_a=FT\circ R_0\circ FT$.
To reflect around $|b\ra$, apply the transformation:
$R_b:|i\ra \longmapsto (-1)^{f(i)} |i\ra$.
A rotation by an angle $2\theta$ is achieved by applying $R_aR_b$.

{~}

\frame{\begin{minipage}
[30mm]{100mm}
$~~~~~~~~$\center{\bf{Grover's algorithm}}\\
\raggedright $~~~~$ Apply Fourier transform on $|0\ra$ to get $|a\ra$.\\
\raggedright $~~~~$ Apply $R_aR_b ~~~\sqrt{N}\pi/4$ times.\\
 \raggedright $~~~~$ Measure all bits.\\
{~}\\
\end{minipage}}

{~}

\noindent The crucial point is that $\theta$  satisfies
$\rm{cos}(\theta)=\sqrt{\frac{N-1}{N}}$  
so for large $N$, we have 
\begin{equation}
\theta\approx \rm{sin}(\theta)=\frac{1}{\sqrt{N}}
\end{equation}
Therefore after $O(\sqrt{N})$ rotations,
 with high probability the measurement yields
 an item satisfying $f(i)=1$.
Note that this algorithm  relies heavily on the assumption
 that the number of ``good'' items is one.
If for example the number of ``good'' items is $t=2$ , we will have 
almost $0$ probability to measure a ``good'' item, exactly when we expect this probability to be 
almost one! 
There are several ways to generalize this algorithm to the general case where 
the number of ``good'' items is not known. 
One is a known classical reduction\cite{valiant3}. 
%
Another generalization was suggested in \cite{boyer1}.
This suggestion not only finds a ``good''  item regardless 
of what the number, $t$, of ``good'' items is, but also 
gives a good estimation of $t$.
The idea is that the probability to measure a ``good'' item is a 
periodic function in the number of Grover's iteration, where this period 
depends on $t$ in a well defined way. 
The period can be found using ideas similar to 
what is used in Shor's algorithm, by Fourier transforms. 
Grover's algorithm can be used to solve $NP$ complete problems 
in time $\sqrt{2^n}$, instead of the classical $2^n$, 
which simply goes over all the $2^n$ items in the database.

{~}

\noindent
Grover's algorithm provides a quadratic advantage over any possible classical 
algorithm, which is optimal, due to Bennett {\it et.al.}\cite{bbbv,boyer1,zalka2}, a result 
which I will  discuss when dealing with quantum lower bounds in
section \ref{bounds}. 
Let me now describe several variants on Grover's algorithm,
 all using Grover's iteration as the basic step.
(These variants and others can be found in Refs. 
 \cite{brassard3, grover2, durr, grover3,brassard4,boyer1} and \cite{grover4}.)


{~}

\noindent {\bf Estimating the median to a precision $\epsilon$.}\cite{grover4,grover2}
\begin{quote}
$f$ is a function from $\{1,..N\}$ to $\{1,..N\}$ where 
$N$ is extremely large. We are given $\epsilon>0$,
We want to find the median $M$, where we allow a deviation by $\epsilon$, i.e. the
 number of items smaller than $M$ should be between 
$\frac{(1\pm\epsilon)N}{2}$.
   We also allow an exponentially small (in $1/\epsilon$)  probability for an error.
\end{quote}
We assume that $N$ is very large, and so only 
polylog($N$) operations are considered feasible. 
Classically, this means that the Median cannot be computed exactly but 
only  estimated probabilistically. A classical probabilistic algorithm 
         cannot do better than sample random elements $f(i)$, and compute their median. 
         An error would occur if more than half the elements are chosen from the last 
         $\frac{1+\epsilon}{2}$ items, or from the first $\frac{1-\epsilon}{2}$ items.
         For these events to have exponentially small probability, we need 
         $O(\frac{1}{\epsilon^2})$ samples, by Chernoff's law\cite{fft}. 
The following quantum algorithm performs the task in 
$O(\frac{1}{\epsilon})$ steps.

The idea is to find $M$ by binary search, 
starting with some value, $M_0$, as a guess. We
will estimate up to precision $\epsilon$, the number 
 $|\eta|$ such that $(1+\eta)N/2$ items satisfy $f(i)> M_0$,
This will take us $O(\frac{1}{\epsilon})$ steps. 
We can now continue the binary search of $M$, according 
to the   $\eta$ which we have found. Note that since we do not have information about the sign of $\eta$, a simple binary search will not do, but a slight modification 
will. 
Each step  reduces the possible range of $M$ by a factor of half, 
and thus the search will take polylog($N$)$O(\frac{1}{\epsilon})$ steps. 
It is therefore enough to estimate $|\eta|$ 
in  $O(\frac{1}{\epsilon})$ steps, given 
a guess for the median, $M_0$. Here is how it is done.  

We  define
 $f_0(i)=1 $ if $f(i)>M_0$, 
and   $f_0(i)=0 $ if $f(i)\le M_0$. 
Our basic iteration will be a 
 rotation in the subspace spanned by two vectors:

 \begin{equation}
|\alpha\ra=\frac{1}{\sqrt{N}}\sum_{i=0}^{2^n-1}|i\ra,~~~~ 
|\beta\ra= \frac{1}{\sqrt{N}}\sum_{i=0}^{2^n} (-1)^{f_0(i)}|i\ra
\end{equation}
    Let $|\gamma\rangle$ be a vector orthogonal to $|\beta\rangle$ in 
	the two dimensional subspace.
The angle between $|\alpha\ra$ and  $|\gamma\ra$, is $\theta\approx \rm{sin}(\theta)=\eta$. 
Rotation by $2\theta$ can be done like in Grover's algorithm.
We start with $|\alpha\ra$ 
and rotate by $2\theta$ $\frac{1}{2\epsilon}$ times. 
The angle between our vector and $|\alpha\ra$
is  $\eta/\epsilon$.
We can now project on $|\alpha\ra$ (by rotating $|\alpha\ra$ to $|0\ra$ and projecting on $|0\ra$). The result is distributed like a coin flip 
with bias  $\cos^2(\eta/\epsilon)$.
We can repeat this experiment poly($\frac{1}{\epsilon}$)   number of times.
This will allow us to  estimate the bias $\cos^2(\eta/\epsilon)$ and
 from it
$|\eta|/\epsilon$, up to a $1/4$, with exponentially small error probability. 
	Thus we can estimate $|\eta|$ up to $\epsilon/4$ in $O(\frac{1}{\epsilon})$ time.

{~}

\noindent {\bf Estimating the mean to a precision $\epsilon$.}
\begin{quote}
$f$ is a function from  $\{1,..N\}$ to $[-0.5,0.5]$, where $N$ is assumed to be
very
large. We are given $\epsilon>0$,
We want to estimate the mean  $M$ up to a precision $\epsilon$.
\end{quote}
Again, classically, this will take $O(\frac{1}{\epsilon^2})$, 
assuming that $N$ is extremely large.
Grover  suggested a quantum algorithm to solve this problem 
in $O(\frac{1}{\epsilon})$ steps\cite{grover2}.
Instead of showing Grover's version, I will
 show a simple classical reduction\cite{wigderson}
which allows solving the mean estimation problem given the median algorithm.  
 The idea is that for Boolean functions the 
mean and median problems coincide.
We write the real number $f(i)$, which is between 
$-0.5$ to $0.5$ in its binary representation: 
$f(i)=0.f_1(i)f_2(i)f_3(i).....$ up to $\rm{log}(\frac{2}{\epsilon})$ digits, 
where $f_j(i)$ is the $j'$th bit of $f(i)$.
 Hence, $f_j(i)$ are Boolean functions. 
We can denote by  $M_j$ the mean of $f_j$, which can be estimated
by the median algorithm.
The mean of $f$ can be computed from
  $\frac{1}{N}\sum_i f(i)=\sum_j 2^{-j} (\frac{1}{N}\sum_i f_j(i))=
\sum_j 2^{-j} M_j$. Cutting the number of digits causes at most 
$\frac{\epsilon}{2}$ error in $M$.
Each $M_j$ will be estimated to precision $\epsilon/2$, and this will 
cause $\frac{\epsilon}{2}$
 additional error all together. 

{~}

\noindent {\bf Finding the minimum}
\begin{quote}
$f$ is a function from $\{1,..N\}$ to $\{1,..N\}$ . We want to find 
$i$ such that $f(i)$ is minimal.
\end{quote}
Classically, this will take $O(N)$, if the database is not sorted. 
Durr and Hoyer\cite{durr} show a quantum algorithm which finds the minimum in 
$O(\sqrt{N})$.
This is done by a binary search of the minimum:
At each step $j$ , we have a  threshold $\theta_j$. This defines a function: 
$f_j(i)=1$ if $f(i)<\theta_j$, and   $f_j(i)=0$ otherwise.
$\theta_0$ is fixed
 to be $N/2$, i.e. in the middle of the interval $[1,...N]$. 
Then we apply Grover's search, to find an $i$ such that 
 $f_0(i)=1$.
If we find such an $i$, we fix the new threshold, $\theta_1$ to be $f(i)$. 
Else, we fix  $\theta_1=3N/4$, i.e. in the middle of the interval 
$[N/2,...N]$.
We continue this binary search until the current interval
has shrunk to the size of one number.
This is the minimum. 

{~}

 Grover's iteration can be used to achieve a quadratic gap also 
between quantum and classical communication complexity\cite{buhrman},
 an issue which is beyond of the scope of this review.

\section{What Gives Quantum Computers their (Possible) Extra Power}\label{path}
Let us ask ourselves why quantum computers 
can perform tasks which seem hard or impossible to do efficiently 
by classical machines.
This is a delicate question which is still an issue of debate.
One way to look at this question is using Feynman's path integrals. 
We will associate a diagram with a computation, in 
which 
 the vertical
 axis will run over all $2^n$ possible classical configurations, and 
the horizontal axis will be time. Here is an example of 
such a diagram:

{~}

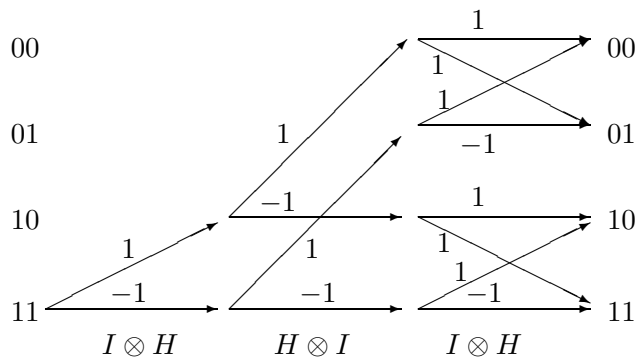
\setlength{\unitlength}{0.030in}

\begin{picture}(100,60)(-50,0)
\put(0,0){\vector(2,1){30}}
\put(12,0){\makebox(5,5){$-1$}}
\put(0,0){\vector(1,0){30}}
\put(12,8){\makebox(5,5){$1$}}

\put(-6,-3){\makebox(5,5){$11$}}
\put(-6,13){\makebox(5,5){$10$}}
\put(-6,28){\makebox(5,5){$01$}}
\put(-6,43){\makebox(5,5){$00$}}

\put(98,-3){\makebox(5,5){$11$}}
\put(98,13){\makebox(5,5){$10$}}
\put(98,28){\makebox(5,5){$01$}}
\put(98,43){\makebox(5,5){$00$}}

\put(14,-9){\makebox(5,5){$I\otimes H$}}

\put(32,16){\vector(1,1){31}}\put(39,28){\makebox(5,5){$1$}}
\put(32,16){\vector(1,0){30}}\put(38,16){\makebox(5,5){$-1$}}
\put(32,0){\vector(1,1){30}}\put(44,8){\makebox(5,5){$1$}}
\put(32,0){\vector(1,0){30}}\put(45,0){\makebox(5,5){$-1$}}

\put(44,-9){\makebox(5,5){$H\otimes I$}}

\put(65,16){\vector(1,0){30}}\put(73,17){\makebox(5,5){$1$}}
\put(65,16){\vector(2,-1){30}}\put(67,9){\makebox(5,5){$1$}}

\put(65,47){\vector(1,0){30}}\put(73,48){\makebox(5,5){$1$}}
\put(65,47){\vector(2,-1){30}}\put(66,40){\makebox(5,5){$1$}}

\put(65,0){\vector(1,0){30}}\put(74,0){\makebox(5,5){$-1$}}
\put(65,0){\vector(2,1){30}}\put(70,4){\makebox(5,5){$1$}}

\put(65,32){\vector(2,1){30}}\put(73,26){\makebox(5,5){$-1$}}
\put(65,32){\vector(1,0){30}}\put(67,34){\makebox(5,5){$1$}}

\put(74,-9){\makebox(5,5){$I\otimes H$}}

\end{picture}

{~}

{~}

In this diagram, the state is  initially $|11\ra$. 
The operation  $H$ is applied thrice: First
on the first bit, then on the second bit 
and then again on the first bit.
The numbers near the edges indicate the probability amplitude 
to transform between configurations weights:
$-1$ corresponds to $-\frac{1}{\sqrt{2}}$ and 
$1$ corresponds to  $\frac{1}{\sqrt{2}}$.
Let us now compute the weight of each basis state in the final superposition.
This weight is   the sum  of the weights of all paths 
leading from the initial configuration to the final one, 
where the weight of each path is the product of the weights on 
the edges of the path. 

\begin{equation}
\rm{Quantum}: ~~~~\rm{Prob}(j)=|\sum_{d:i\mapsto j} w(d)|^2
\end{equation}  
One can see that in the above diagram the weights of $10$ and $00$
in the final superposition 
are zero, because the two paths leading to each one of 
these states cancel one another.

What can we learn from this diagram?
In order to analyze this diagram, I would like to define a classical 
computation model, called {\it stochastic circuits}
 which can be associated with very similar diagrams. 
The comparison between the two models is quite instructive.
The nodes in a stochastic circuit
 have an equal number of inputs and outputs, 
like nodes in a quantum circuit. Instead of unitary matrices,
 the nodes  will be associated with
stochastic matrices, 
which means that the entries of the matrices are positive reals, and 
 the columns are probability distributions.
Such matrices correspond to applying stochastic transformations on the 
bits, i.e. a string $i$ transforms to string $j$ with the probability 
which is equal to the matrix entry $R_{i,j}$.
For example, let $R$ be the stochastic matrix on one bit:
\begin{equation}
R=\left(\begin{array}{cc}
\frac{1}{2} &\frac{1}{2}\\
\frac{1}{2} &\frac{1}{2}
\end{array}\right)
\end{equation}
This matrix takes any input to a uniformly random bit. 
Consider the probabilistic computation on two bits, 
where we apply $R$ on the first bit, then on the second bit, 
and then again on the first bit.
The diagram we get is:

{~}

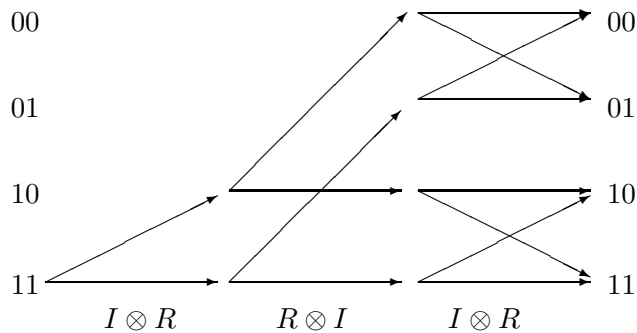
\setlength{\unitlength}{0.030in}

\begin{picture}(100,60)(-50,-10)
\put(0,0){\vector(2,1){30}}
\put(0,0){\vector(1,0){30}}

\put(-6,-3){\makebox(5,5){$11$}}
\put(-6,13){\makebox(5,5){$10$}}
\put(-6,28){\makebox(5,5){$01$}}
\put(-6,43){\makebox(5,5){$00$}}

\put(98,-3){\makebox(5,5){$11$}}
\put(98,13){\makebox(5,5){$10$}}
\put(98,28){\makebox(5,5){$01$}}
\put(98,43){\makebox(5,5){$00$}}

\put(14,-9){\makebox(5,5){$I\otimes R$}}

\put(32,16){\vector(1,1){31}}
\put(32,16){\vector(1,0){30}}
\put(32,0){\vector(1,1){30}}
\put(32,0){\vector(1,0){30}}

\put(44,-9){\makebox(5,5){$R\otimes I$}}

\put(65,16){\vector(1,0){30}}
\put(65,16){\vector(2,-1){30}}

\put(65,47){\vector(1,0){30}}
\put(65,47){\vector(2,-1){30}}

\put(65,0){\vector(1,0){30}}
\put(65,0){\vector(2,1){30}}

\put(65,32){\vector(2,1){30}}
\put(65,32){\vector(1,0){30}}

\put(74,-9){\makebox(5,5){$I\otimes R$}}

\end{picture}

{~}

{~}
\noindent  where the weights of all  edges are  $\frac{1}{2}$.
Just like in quantum computation, the probability for a configuration in the 
final state is computed by summing over the weights of all paths leading 
to that configuration, where the weight of each path is the product of 
the weights of the edges participating in the path.

\begin{equation}
\rm{Stochastic}: ~~~~\rm{Prob}(j)=\sum_{d:i\mapsto j} \rm{Prob}(d)
\end{equation}

In this diagram all the configurations in the final state 
 have probability $\frac{1}{4}$.

We now have two models which are very similar. It can be easily 
seen that stochastic circuits are equivalent to 
probabilistic TM. 
 This means that we can find the advantage of  quantum computation
over classical computation 
 in the difference
 between quantum circuits and stochastic circuits.
It is sometimes tempting to say that quantum computation is powerful 
 because it has exponential parallelism.
For $n$ particles, the vertical axis will run over 
 $2^n$ possible classical states.
But this will also be true in the diagram of 
stochastic computation on $n$ bits!
The difference between quantum and classical computations
is therefore more subtle.

To  reduce the difference between the two models even further, 
it can be shown\cite{bv} that the complex numbers in quantum computation
can be replaced with real numbers, without damaging the
computational power.
This is done by adding an extra qubit to the entire 
circuit, which will carry the information
of whether we are working in the real or imaginary part of the numbers.
The correspondence between the superpositions of the complex circuit 
to the real circuit  will be:
\begin{equation}
\sum_i c_i|i\ra \longmapsto \sum_i Re(c_i)|i,0\ra + Im(c_i)|i,1\ra
\end{equation}

Hence quantum computers maintain their computational 
power even if they use only real valued unitary gates. 
 There are two differences between these gates and  stochastic gates. 
           One is that stochastic gates have positive entries while real unitary gates have 
           positive and negative entries. The other difference is that unitary gates preserve 
           the $L_2$ norm of vectors, while stochastic gates preserve $L_1$ norm.
%
The difference between the quantum and classical models
can therefore be summarized in the following table:

\begin{eqnarray*}
~~~~~~~ \underline{Quantum} ~~~~~~~ & ~~~\underline{Stochastic}\\
Negative ~+~ Positive & ~~~Positive\\
~~~~~~L_2~ Norm~~~~~~ & ~~~L_1~ Norm
\end{eqnarray*}

Why are negative numbers so important?
The fact that weights can be negative allows different paths 
to cancel each other. 
We can have many non-zero paths leading to the same 
final configuration, all cancelling each other, causing 
destructive interference. 
This is exactly what happens in Deutsch and Jozsa's algorithm, 
Simon's algorithm and Shor's algorithm, where 
the paths that lead to ``bad'' strings in the last step of the algorithm
are destructively interfering, and 
at the same time paths that lead to ``good'' strings
are constructively interfering. 
In the probabilistic case, interference cannot occur.
Paths do not {\it talk} to each other, 
there is no influence of one path on the other.
Probabilistic computation 
has  the power of exponentiality, but lacks the power of 
  interference offered by computation that uses negative numbers.
An exponential advantage in
 computational power of negative numbers is already familiar
 from classical complexity theory, when comparing Boolean circuits with 
monotone Boolean circuits\cite{valiant2}.

There are other computational models which exhibit interference, 
such as optical computers.  However, these models do not exhibit exponentiality. 
It is only the quantum model which combines the two 
features of  exponential space
which can be explored in polynomial time,
 together with the 
ability of interference.
(See also \cite{aharonov5}.)

Another point of view of the origin of the power of quantum computation
is quantum correlations, or {\it entanglement}. 
Two qubits are said to be entangled if their state is not in tensor 
product, for example the EPR pair $\frac{1}{\sqrt{2}}(|00\ra+|11\ra)$.
In a system of $n$ qubits, the entanglement can be spread over 
macroscopic range, like in the state $\frac{1}{\sqrt{2}}(|0^n\ra+|1^n\ra)$, 
or it can be concentrated between pairs of particles like
in the state 
$\bigotimes_{n/2} \frac{1}{\sqrt{2}}(|00\ra+|11\ra)$.
 It can be shown that quantum computational
power exists only when the entanglement is spread over macroscopically 
many particles. If the entanglement is not macroscopically spread,
 the system can be easily 
simulated by a classical computer\cite{aharonov2}.
For the importance of entanglement see for example
 Jozsa's review\cite{jozsa2}.
This macroscopic spread of entanglement lies in the essence of 
another important topic, quantum error correcting codes, 
which we will encounter later.


\section{What We Cannot Do with Quantum Computers}\label {bounds}
Now that we have all this repertoire of algorithms in our hands, 
it is tempting to try and solve everything on a quantum computer!
Before doing that, it is worthwhile to understand the limitations
of this model. 
The first thing to know is that this model cannot solve any question 
which is undecidable by a classical machine.
This is simply due to the fact that anything that can be done in this model 
can be simulated on a classical machine by computing the coefficients of the superposition 
and writing them down. This will take an  exponential amount of time, 
but finally  will solve anything which can be done quantumly.
Therefore the only difference between classical and quantum computation lies
 in the computational cost.

The trivial simulation of quantum computers by classical machines
is exponential both in time and space.
Bernstein and Vazirani\cite{bv} showed that classical Turing machines  can  simulate 
quantum computers in polynomial space, although still in exponential time: 
 \begin{theo}(Bernstein, Vazirani)
\(BQP\subseteq Pspace\)
\end{theo}

The theorem means that anything that 
can be done on a quantum machine 
can be done by a classical machine which uses only polynomial 
space. To prove this result, have another look  
 on the Feynman path graph 
presented in  Sec. $9$. 
To compute the weight  of one path, we need only polynomial space. 
We can run over all paths leading to the same configuration, 
computing the weight one by one, and adding them up. 
This will give the probability of one configuration.  
To compute the probability 
to measure  $0$, we add the probabilities of 
all the configurations with the result bit 
being $0$. This again will take exponential 
time, but only polynomial space.  $\bbox$

Valiant improved this result\cite{bv} to show that $BQP$ 
is contained in  a complexity class  which is weaker than $Pspace$, 
                  namely  $P^{\#P}$, which I will not define here.
It might still be that quantum computation is much less powerful, but 
we still do not have a proof for that. 
 In particular, the relation between $BQP$ and $NP$ is not known yet. 

We do understand a lot about the following question:
\begin{quote}
	Can quantum computation be much more efficient than classical 
	computation in terms of number of accesses to the input?
	\end{quote}
Consider accessing the $n$ input bits $X_1,...X_n$,
 for a problem or a function via an oracle, i.e. by applying  the unitary transformation:
\begin{equation}
|i\ra|0\ra \longmapsto  |i\ra|X_i\ra
\end{equation}

\noindent This unitary transformation corresponds
to the classical operation of asking:
``what is the $i$'th bit? `` and getting the answer $X_i$.
One might hope to make use of quantum parallelism, 
and query the oracle by 
the superposition $1/\sqrt{N}\sum_i|i\ra|0\ra \longmapsto 1/\sqrt{N}
 \sum_i|i\ra|X_i\ra$.
In one query to the oracle, the algorithm can read all the $N$ bits, 
so intuitively no quantum  algorithm needs more than one query to the oracle.
It turns out that this intuition  is completely wrong.
It can be shown, using the notion of von Neumann entropy (see \cite{peres})
 that there are no more than $log(N)$ bits of information in the 
 state $1/\sqrt{N}
 \sum_i|i\ra|X_i\ra$.
Bennett {\it et.al.}\cite{bbbv} show that if the quantum algorithm is supposed to 
compute the $OR$ of the oracle bits $X_1,...X_n$, then at least 
$O(\sqrt{N})$ queries are needed.
Note that $OR$ is exactly the function computed by Grover's database
search. Hence this gives a lower bound of $O(\sqrt{N})$ for 
database search, and shows that Grover's algorithm is optimal.
\begin{theo}
Any quantum algorithm that computes 
$OR(X_1...X_N)$ requires at least $O(\sqrt{N})$ steps.
\end{theo}

\noindent The idea of the proof 
 is that if the number of the queries to the oracle is small, 
 there exists at least one index 
$i$, such the algorithm will   be almost indifferent to $X_i$, and 
so will not distinguish between the case of all bits $0$ and the case 
that all bits are zero except $X_i=1$.
Since the function which the algorithm computes is OR, this is a 
contradiction.

\noindent Beals {\it et. al.}\cite{beals2} recently generalized the above result 
             building on classical results by Nisan and Szegedi\cite{nisan}.
Beals {\it et.al.}  compare the minimal number of queries to  the oracle
which are needed in a quantum algorithm, with the minimal number of queries which are needed 
in a classical algorithm.
Let us denote by $D(f)$ and $Q(f)$
the minimal number of queries in  a classical and  quantum algorithm
respectively. 
Beals {\it et.al.}\cite{beals2} show that $D(f)$ is at most polynomial in $Q(f)$.

\begin{theo}
\(D(f)=O(Q(f)^6)\)
\end{theo}
Beals {\it et. al.} use similar methods to give lower bounds on the time 
required to quantumly compute 
the functions  MAJORITY, PARITY\cite{farhi}, OR and AND:

{~}

$~~~~~~~~~~~~~~~~~~~$\begin{tabular}{|l|l|}\hline
OR & \(\Theta(\sqrt{N})\)\\ \hline
AND & $\Theta(\sqrt{N})$\\ \hline
PARITY & $N/2$\\ \hline
MAJORITY & $\Theta(N)$\\ \hline
\end{tabular}

{~}

(Here $f=\Theta (g)$ means that $f$ and $g$ behave the same asymptotically.)
 The lower bounds are achieved by showing that the number of times the
             algorithm is required to access the input is large. 
This is intuitive,
since these functions are  very sensitive to their input bits. 
For example, the string $0^N$ satisfies $OR(0^N)=0$,  but flipping any bit 
will give $OR(0^{N-1}1)=1$.

	The meaning of these results, is that in terms of the number of accesses to the input,  
	quantum algorithms have no more than polynomial advantage
	over classical algorithms\cite{ozhi}.
	This polynomial relation can give us a hint when looking for 
	computational problems in which quantum algorithms may have  
	an exponential advantage over classical algorithms.
 	These  problems will have the property that in a  classical algorithm that solves them, 
 	the bottle neck is the information processing, while the  number of accesses to the 
	input can be very small.  Factorization is exactly such a problem. $D(f)$ is $log(N)$, because 
	the algorithm simply needs to read the number $N$ in binary representation, 
	but the classical information processing takes exponential in $log(N)$ steps. 
 	Shor's quantum algorithm enables an exponential speed up in the information processing.
	An opposite example is the database search. Here,  the bottle neck in classical computation
	 is not the  information processing but simply the fact that the size of the input 
          is very large.  Indeed, in this case,  quantum computers have only quadratic advantage 
          over classical computers.

Now that we  understand some of the limitations and advantages of the 
quantum model, let us go on to the subject of quantum noise.

\section{Worries about Decoherence, Precision and Inaccuracies}
Learning about the possibilities which lie in 
quantum computation gave rise to a lot of enthusiasm, 
but many physicist\cite{landauer1,unroh1,decoherence,barenco7}
were at the same time very sceptic about the entire field. 
The reason was that all quantum algorithms achieve their advantage over 
classical algorithms when assuming that the gates and wires operate 
 without any inaccuracies or errors.   Unfortunately, in reality we cannot expect any system  to be ideal.  
            Quantum systems in particular tend to lose their quantum nature easily. 
 	Inaccuracies and errors may cause the damage  to accumulate exponentially fast during 
          the time of the computation\cite{decoherence,decoherence2,barenco6,barenco7, miquel1}. 
         In order to perform computations, one must be able to reduce the effects of 
       inaccuracies and errors, and to correct the quantum state.
 
         Let us try to understand the types of errors and inaccuracies that 
might occur in a quantum 
         computer. The simplest problem is that the gates  
          perform unitary operations which slightly deviate from the correct ones. 
Indeed, it was shown by Bernstein and Vazirani\cite{bv} that 
it suffices that  the entries of the gates are  precise only up to 
 $1/n$, where $n$ is the size of the computation. 
However, 
 it is not reasonable to assume 
that inaccuracies decrease as $1/n$.
What seems to be reasonable to assume is that the devices 
we will use in the laboratory have some finite precision, independent of 
the size of the computation. Errors, that might occur, 
will behave,  presumably, according to the same law 
of constant probability for error per element  per time step.
 Perhaps the most severe problem was that of
         {\it decoherence}\cite{mott,stern1,zurek1,palma,gardiner}. 
           Decoherence is the physical process, 
           in which quantum system lose some of their quantum characteristics  due to 
           interactions with environment. 
           Such interactions are inevitable because no system can be kept entirely isolated
             from the environment.
            The effect of entanglement with the environment can be viewed as if the environment applied 
            a partial measurement on the system, which caused the wave function to collapse, 
            with certain probability. 
This collapse of the wave function seems to be an irreversible process
by definition.
How can we correct a wave function which has collapsed?

In order to solve the problem of correcting the effects of noise, 
we have to give a formal description 
of the noise process.
Observe that the most general quantum 
operation on a system is a unitary operation on the system and its 
environment.
Noise, inaccuracies, and decoherence can all 
be described in this form.
Formally, the model of noise is 
that in between the time steps, we will allow a ``noise''
operator to operate on the system and an environment. 
We will assume that the environment is renewed each time step, 
              so there are no correlations between the noise processes at different times. 
              Another crucial assumption is that the noise is {\it local}. This means that 
each qubit interacts with 
its own environment during the noise process, 
and that there are no interactions or correlations between these 
environments.
In other words,  the noise operator on $n$ qubits, at each time step, 
 can be written as a tensor product of 
$n$ local noise operators, each operating on one 
qubit:
\[{\cal E}= {\cal E}_1\otimes{\cal E}_2\otimes\cdots\otimes{\cal E}_n.\]

If the 
qubits were correlated in the last time step by a quantum gate,
 the local noise operator  operates 
on all the qubits participating in one gate together.
This noise model assumes that correlations between errors on different qubits
can only appear  due to the qubits interacting through a gate. 
 Otherwise, each qubit interacts with its own environment.

     The most general noise operator on one qubit is a general unitary transformation on the qubit
           and its environment: 
\begin{eqnarray}
|e\rangle |0\rangle \rightarrow 
|e_{0}\rangle|0\rangle  + \ |e^b_{0}\rangle|1\rangle  \\ \nonumber  
|e\rangle|1\rangle \rightarrow
|e_{1}\rangle |1\rangle  + \ |e^b_{1}\rangle|0\ra
\label{env}
\end{eqnarray}
   When qubits interact via a gate, the most general noise operation would            
             be a general unitary transformation on the qubit participating in the gate 
           and their  environments.

When dealing with noise, it is more convenient to use the language 
of density matrices, instead of vectors in the Hilbert space. I
will define them here, 
so that I can explain the notion of  ``amount of noise'' in the system,
however they will rarely  be used again later in this review. The 
density matrix describing a system in the state $|\alpha\ra$ is 
$\rho=|\alpha\ra\la \alpha|$. 
The density matrix of part $A$ of the system can be derived from 
$\rho$ by tracing out, or integrating, the degrees of freedom 
which are not in $A$.
The unitary operation 
on the environment and the system, which corresponds 
to quantum noise, can be viewed as a linear operator on
the density matrix describing only the system. 
As a   metric on density matrices 
we can use the fidelity\cite{wootters1}, or the trace metric\cite{aharonov4}, 
where the exact definition does not matter now. 
Two quantum operations are said to 
be close if when operating on the same density matrix, 
they generate two close density matrices. 
We  will say that the {\it noise rate} in the system is 
$\eta$ if each of the local noise operators is within $\eta$ distance
from the identity map on density matrices.

 We now want to find a way to compute
           fault tolerantly in the presence of noise rate $\eta$, where we do not 
           want to assume any knowledge about the noise operators, except the noise rate.
  We will first  concentrate on a simple special case, in which the computation consists 
              of one time step which computes the identity operator on all qubits. 
               This problem is actually
              equivalent to the problem of communicating with noisy channels. 
          In order to understand the subtle points when trying to communicate 
with noisy channels, 
          let us consider the classical analogous case. 
Classical information is presented by a string of 
bits instead of qubits, and the error
model is simply that each bit flips with probability $\eta$.

Suppose Alice wants to send Bob a string of bits, and the channel 
which they use is noisy, with noise rate $\eta$, 
i.e. each bit flips with probability $\eta$.
 In order to protect information against noise, 
Alice can use redundancy. Instead of sending $k$ bits, 
Alice will encode her bits on more bits, say $n$, such that Bob 
can apply some recovery operation to get the original $k$ bits.
The idea is that to first approximation, most of the bits will 
not be damaged, and  
the encoded bits,  sometimes called the 
 the {\it logical bits}, can be recovered.
The simplest example of a classical code is  
 the majority code, which encodes one logical bit on three 
bits. 
\[0 \longmapsto 0_L=000 ~~~~,~~~~~ 1\longmapsto 1_L=111\]
This classical code corrects one error, because if 
 one bit has flipped, taking the majority vote of the three bits still 
recovers the logical bit.
However, if more then one bit has flipped, the logical bit can no longer 
be recovered.
 If the probability for a bit flip is $\eta$, 
then the probability that the three bits cannot be recovered, 
i.e. the effective noise rate $\eta_e$, equals:
\[\eta_e=3\eta^2(1-\eta)+\eta^3.\]
If we require that we gain some advantage in reliability by the code, 
then $\eta_e< \eta$ implies a {\it threshold}
on $\eta$, which is $\eta<0.5$. 
If $\eta$ is above the threshold, using the code will only decrease
the reliability.

 The majority code becomes extremely non efficient when Alice wants to send long  messages.  
         If we require that Bob receives all the logical bits with high probability of being correct,
          Alice will have to use exponential redundancy for each bit. 
        However, there are error correcting codes which map $k$ bits to $m=O(k)$ bits,
          such that the probability for Bob to get the original message 
          of $k$ bits correct is high, even when $k$ tends to infinity. 
          A very useful class of error correcting codes are the {\it linear} codes, 
          for which the mapping from $k$ bits to $n$ bits
 is linear, and the set of {\it code words},
           i.e. the image of the mapping, is a linear subspace of $F_2^m$. 
A code is said to correct $d$ errors if a recovery
 operation exists 
 even if $d$ bits have flipped.
The {\it Hamming distance} between two strings is defined to be the number of 
coordinates by which the two strings differ.  
Being able to recover the string after $d$ bit flips have occurred
 implies that the distance 
between two possible code words is at least $2d+1$, so that 
each word is corrected uniquely. 
 For an introduction to the subject of classical error correcting codes, see van Lint\cite{lint}.

   We define a quantum code in a similar way. The state of $k$ qubits is mapped into the state 
       of $m$ qubits. The term {\it logical  state} 
is used for the original state of the $k$ qubits. 
        We say that such a code corrects $d$ errors, if there exists a recovery operation 
          such that 
 if not more than $d$ qubits were damaged, the logical state can still be
 recovered.  It is important here that Bob has no control on the environment
         with which the qubits interacted during the noise process. Therefore  
      we require that the recovery operation does  not operate on the environment but merely  
        on the $m$ qubits carrying the message and perhaps some ancilla qubits. 
The image of the map in the Hilbert space of $m$ qubits 
will be called a {\it quantum code}.

Let us now try to construct a quantum code. Suppose that Alice  wants to send Bob a qubit in the state 
$c_0|0\ra+c_1|1\ra.$ 
How can she encode the information?
One way to do this is simply to send the classical information describing 
$c_0$ and $c_1$ up to the desired accuracy.
We will not be interested in this way, because when Alice wants to send 
Bob a state of $n$ qubits,  the amount of classical bits  that needs to be sent
grows exponentially with $n$.
We will want to encode qubits on qubits, to prevent this exponential 
overhead.
The simplest idea that comes to mind is that 
  Alice generates a few copies of the same state,
 and sends the following state to Bob:
\[c_0|0\ra+c_1|1\ra\longmapsto \left(c_0|0\ra+c_1|1\ra\right)\otimes
\left(c_0|0\ra+c_1|1\ra\right)\otimes
\left(c_0|0\ra+c_1|1\ra\right).\]
Then Bob is supposed to apply some majority vote among the qubits.
Unfortunately,  a quantum majority vote among general quantum states is not a linear 
      operation. Therefore, simple redundancy will not do. 
Let us try another
 quantum analog of the classical majority code:
\[ c_0|0\ra+c_1|1\ra
\longmapsto c_0|000\ra+c_1|111\ra\]
This code turns out to be another bad quantum code.
It 
does not protect the quantum information even against 
one error.
Consider for example, the local noise operator  which operates on the first qubit in the encoded state  $c_0|000\ra+c_1|111\ra$. It does nothing to that qubit, but it changes the 
state of the environment according to whether this bit is $0$ or $1$:
\begin{eqnarray}
 |0\ra\otimes |e\ra & \longmapsto & |0\ra \otimes |e_0\ra \\
 |1\ra\otimes |e\ra & \longmapsto & |1\ra \otimes |e_1\ra \nonumber
\end{eqnarray}
Here $\la e_0|e_1\ra =0$.
Even though only an identity operation was applied on the first bit, the fact 
that the environment changed according to the state of this bit 
is equivalent to the environment {\it measuring }
the state of the first qubit. This measurement is an irreversible process.
After the noise operation, the environment is no 
longer in a tensor product with the state.
 Bob can only apply local operations on his system, and 
cannot control the environment. This means that the entanglement
 between the state of the first
qubit, and the environment cannot be broken during the recovery 
operation; the coherence of the state is lost.
A theorem due to Schumacher and Nielsen\cite{schumacher2}
 formalizes this intuition. 
The claim is that if the reduced density matrix of the environment 
is  different for different code words, 
then there is no  unitary operation that operates on the system 
and recovers the logical state. 


\begin{theo}
It is impossible to recover the logical state, if information about it has leaked to the environment 
via the noise process.
\end{theo}

This theorem underlines the main distinction between quantum error correcting 
codes and classical error correcting codes.  
 Quantum codes
try to {\it hide} information from the environment,  In contrast,
the protection of classical information from noise,
is completely orthogonal to the question of hiding secrets.
The theorem gives us insight as to  the basic idea 
in quantum computation:
The idea is to spread the quantum information 
over more than $d$ qubits, in a non-local way, 
such that 
the environment which can access only a small number of qubits can gain  
no information about the quantum logical state,
 and this information will
be protected. Now, that we have some intuition about the requirements 
             from quantum codes, we can proceed to show how to construct such codes.

\section{Correcting Quantum Noise}
In order to succeed in correcting quantum noise, 
we need to consider more carefully the process 
of noise.  
The first and most crucial step is 
the discovery that quantum noise can be treated as discrete.
In the quantum setting, we assume all qubits undergo a noise of size $\eta$. 
We want to replace this with the case in which a few qubits are completely 
damaged, but the rest of the qubits are completely fine.
This can be done by rewriting 
 the effect of a general noise operator.
Let the state of  $m$ qubits be  $|\alpha\ra$.
If the noise rate is $\eta$,
we can develop the operation of a general
 noise operator operating on $|\alpha\ra$
 by orders of magnitude of $\eta$:
\begin{equation}\label{dis}\begin{array}{l}
{\cal E}_1 {\cal E}_2....{\cal E}_m|\alpha\ra
=\\
(I_1+\eta {\cal E'}_1)(I_2+\eta {\cal E'}_2)...
(I_m+\eta {\cal E'}_m)|\alpha\ra=\\
I_1I_2...I_m |\alpha\ra+
 \eta \left({\cal E'}_1I_2...I_m+...+I_1I_2...I_{m-1}{\cal E'}_m\right)|\alpha\ra+
....+
\eta^m\left({\cal E'}_1{\cal E'}_2...{\cal E'}_m\right)|\alpha\ra.
\end{array}\end{equation}

The lower orders in $\eta$ correspond to a small 
number of qubits being operated upon, and higher orders 
in $\eta$ correspond to more qubits being contaminated.
This way of writing the noise operator is the beginning of discretization
of the quantum noise, because in each term 
a qubit is either damaged or not.
For small $\eta$, we can neglect higher order terms and 
 concentrate  in the lower orders, where only
one or two qubits are damaged out of $m$.
A special case of this model is the probabilistic model, 
in which the local noise operator applies a certain operation with 
probability $\eta$ and the identity operation with probability
$(1-\eta)$. In this model, if the quantum system consists of $m$ 
qubits, we can assume that with high probability only a few of the qubits
went through some noise process. 
There are noise 
operators, such as amplitude damping, which do not obey this 
probabilistic behavior. However their description by equation 
(\ref{dis}) shows that we can treat them in the same discrete manner.

The second step is the discretization of the 
noise operation itself.
The most general quantum operation on the $k'$th qubit
and it's environment is described by:

\begin{eqnarray}
|e\rangle |0_k\rangle \rightarrow 
|e_{0}\rangle|0_k\rangle  + \ |e^b_{0}\rangle|1_k\rangle  \\ \nonumber  
|e\rangle|1_k\rangle \rightarrow
|e_{1}\rangle |1_k\rangle  + \ |e^b_{1}\rangle|0_k\ra
\end{eqnarray}

This operation, applied on any logical state $c_0|0_L\ra +c_1|1_L\ra$, acts as
the following operator:

\begin{equation}
(c_0|0_L\ra +c_1|1_L\ra)\rightarrow \Big (|e_+\rangle {\cal I} + |e_-\rangle \sigma_z^k + 
|e^b_+\rangle \sigma_x^k - |e^b_-\rangle i\sigma_y^k \Big )(c_0|0_L\ra +c_1|1_L\ra)
 \,,
\label{pauli}
\end{equation}
Where $ \sigma_i^k$ are the Pauli operators acting on the $k$'th qubit:
\begin{equation}
{\cal I} =\left(\begin{array}{cc}1& 0 \\0 & 1\end{array}\right),
 \sigma_x=\left(\begin{array}{cc}0 & 1 \\1 & 0\end{array}\right),
\sigma_y=\left(\begin{array}{cc}0 & -i \\i & 0\end{array}\right),
\sigma_z=\left(\begin{array}{cc}1 & 0 \\0 & -1\end{array}\right).
\end{equation}
The environment states are
 defined as  $|e_\pm\rangle=(|e_0\rangle \pm |e_1\rangle)/2$, 
$|e^b_{\pm}\rangle=(|e^b_0\rangle \pm |e^b_1\rangle)/2$.
The most crucial observations, which enables to correct quantum errors,
hide in equation \ref{pauli}.
The first observation is that everything that can happen to a qubit
is composed of four basic operations, so it is enough to correct 
        for these four errors\cite{bennett14,ekert3,knill3}. This resembles a discrete model 
more than a continuous one, and gives hope that such discrete 
errors can be corrected.
The second crucial point is
 that the  states of the environment which are entangled with the system
after the operation of noise,  
are {\it independent} of $(c_0|0_L\ra +c_1|1_L\ra)$
and depend only on which of the four operations 
 $ \sigma_i^k$ were applied.
In particular, for any superposition of the logical states $|0_L\ra, |1_L\ra$,
the operator will look the same.
 This suggests  the following scheme of breaking the entanglement of the system
 with the environment.
The idea is to measure which
 one  of the four possible operators was 
applied. This is called the {\it syndrome} of the error. 
Measuring the syndrome will collapse the system to a state which 
is one of the following tensor products of the system and the environment:

\begin{equation}
 \Big(|e_+\rangle {\cal I} + |e_-\rangle \sigma_z^k + 
|e^b_+\rangle \sigma_x^k - |e^b_-\rangle i\sigma_y^k \Big)
(c_0|0_L\ra+c_1|1_L)\ra\stackrel{measure}{\longrightarrow}
\left\{\begin{array}{l}
 |e_+\rangle {\cal I}\Big(c_0|0_L\ra+c_1|1_L\ra\Big)\\
|e_-\rangle \sigma_z^k \Big(c_0|0_L\ra+c_1|1_L\ra\Big)\\
|e^b_+\rangle \sigma_x^k\Big(c_0|0_L\ra+c_1|1_L\ra\Big)\\
|e^b_-\rangle i\sigma_y^k\Big(c_0|0_L\ra+c_1|1_L\ra\Big)
\end{array}\right.
\label{collapse}
\end{equation}

After we know which of the operators had occurred, we can simply apply its reverse,
and the state  $c_0|0_L\ra +c_1|1_L\ra$ will be recovered.  
This reduces the problem of error correction 
 to being able to detect which of the four operators
had occurred.
The operator $\sigma_x$ corresponds to a {\it bit flip}, 
which is a classical error. 
This suggests the following idea:
If  the superposition of the encoded 
state, is a sum of strings $|i\ra$ where the $i'$s are 
strings from a classical code, 
then bit flips can be detected by applying classical techniques.
 Correcting the noise  operator  $\sigma_z$, which is a {\it phase flip},
seems harder, but an important observation is  that 
$\sigma_z=H \sigma_x H$, 
where $H$ is the Hadamard transform.
Therefore, phase flips correspond to bit flips occurring in the Fourier 
transform of the state!
If the Fourier transform of the state is also a superposition of strings 
in a classical code, this enables a correction of phase flips by 
 correcting the bit flips in the Fourier transform basis.
This idea was discovered by Calderbank and Shor\cite{calshor} and 
Steane\cite{steane1}.

 A simple version of the recipe they discovered for cooking a quantum code goes as follows. 
            Let $C\subset F_2^m$ be a linear classical code, which corrects $d$ errors, such that 
             $C^\perp$, the set of all strings orthogonal over $F_2$ 
            to all vectors in $C$, is strictly contained in $C$. 
          We look at the cosets of  $C^\perp$ in $C$, i.e. we partition
             $C$ to non intersecting sets which are translations of $C^\perp$
             of the form $C^\perp+v$. The set of vectors in $C$, with the  identification of  
             $w$ with $w'$ when $w-w'\in C^{\perp}$   is called $C/ C^\perp$. 
            For each $w\in C/ C^\perp$ we associate  a code word:
\begin{equation}\label{code}
	|w\ra \longmapsto |w_L\ra  =  \sum_{i\in C^\perp} |i+w\ra  
	\end{equation}
   where we omit overall normalization factors. 
            Note that all the strings which appear in the superposition 
are  vectors in the code  $C$.
          It is easy to check that  
       the same is true for the  Fourier transform over $Z_2^m$ of the code words, which is achieved  
            by applying the Hadamard gate, $H,$
              on each qubit:
\begin{equation}
H\otimes H\otimes ....\otimes H|w_L\ra= \sum_{j\in C} (-1)^{w\cdot j} |j\ra.
\end{equation}

 The error correction goes as follows. To detect bit flips, we apply the classical error correction  according to the classical code $C$, on 
the states in equation (\ref{code}). This operation, 
 computes the syndrome (in parallel for all strings)
 and writes it on some 
 ancilla qubits. 
Measuring the ancilla will collapse the state to a state with 
a specific syndrome, and we can compute according 
to the result of the measurement which 
qubits were affected by a bit flip, and apply $NOT$ on those qubits.
To detect phase flips we apply Fourier transform on the entire state, 
and correct  bit flips classically according to the code  $C$.
Then we apply the reverse of the Fourier transform.
This operation will correct phase flips.
$\sigma_y$ is a combination of a bit flip and a phase flip, and is corrected 
by the above sequence of error corrections\cite{calshor}.  

            The number of qubits which can be encoded by this code is the logarithm with base $2$ 
             of the dimension of the space spanned by the code words. To calculate this
      dimension, observe 
          that the code words 
          for different $w$'s in  $C/ C^\perp$ are perpendicular. 
          The dimension of the quantum  code is equal to the 
            number of different words in $C/ C^\perp$,
          which is $2^{dim(C/ C^\perp)}$. Hence the number of qubits which can be encoded 
          by this quantum code is $dim(C/ C^\perp)$. 
          
          Here is an example, due to Steane\cite{steane1}. Steane's code
 encodes one qubit on seven qubits, 
and corrects one error. It is constructed from the classical code
known as the Hamming code,
which is the subspace of $F_2^7$ spanned by the four vectors:
\newline\(C=span\{1010101,0110011,0001111,1111111\}\).
 $C^\perp$ is spanned by the three vectors:
$1010101,0110011,0001111$.
 Since $C$ is of dimension $4$, and $C^\perp$ is of dimension $3$, the number of qubits 
           which we can encode is $1$.  The two code words are:
\newpage
\begin{eqnarray}
|0_L)=|0000000\ra+|1010101\ra+|0110011\ra+|1100110\ra \\ \nonumber
+|0001111\ra+|1011010\ra+|0111100\ra+|1101001\ra\\ \nonumber
|1_L)=|1111111\ra+|0101010\ra+|1001100\ra+|0011001\ra \\ \nonumber
+|1110000\ra+|0100101\ra+|1000011\ra+|0010110\ra 
\end{eqnarray}

 Observe that the minimal Hamming distance between  two words in $C$ is $3$, so 
           one bit flip and one phase flip can be corrected.

         One qubit cannot be encoded on less than $5$ qubits, if we require that 
          an error correction of one general error can be done. This was shown by 
         Knill and Laflamme\cite{knill3}. Such a code, called a perfect quantum code,
          was found by Bennett et 
	al\cite{bennett14} and by Laflamme {\it et.al.} \cite{laflamme2}. If we  restrict the error, 
       e.g. only bit flips or only phase flips occur than one qubit can 
         be encoded on less than $5$ qubits. 

The theory of quantum error correcting codes has further developed.
 A group theoretical structure was 
discovered \cite{calderbank3,gf4,gottesman1,gottesman2,knill3,shor4},
 which most of the known  quantum error correcting 
codes obey.
 Codes that obey this structure are called stabilizer codes\cite{gottesman1,gottesman2},   
and their group theoretical structure gives a recipe for constructing more quantum codes.
Quantum codes are used for purposes of quantum communication 
	with noisy channels, which is out of the scope of this review. 
           For an overview on the subject of quantum communication 
         consult Refs.  \cite{barnum3,optic} and \cite{lloyd5}.
We now have the tools to deal with the question of 
quantum  computation in the presence of noise, which I will 
discuss in the next section.


\section{Fault Tolerant Computation}

In order to protect quantum computation, the idea is that one should compute 
on encoded states. The entire operation will occur in the protected 
subspace, and every once in a while 
an error correction procedure will be applied, 
to ensure that errors do not accumulate.
The original quantum circuit will be replaced by a quantum 
circuit which operates on encoded state. 
Suppose we use a quantum code which encodes one 
qubit into a block of $5$ qubits. 
Then in the new circuit, each wire will be replaced by five wires, 
and the state of the new circuit will encode the state of the
 original circuit.  
In order to apply computation on encoded states, the original gates 
will be replaced by procedures which apply the corresponding 
operation. If $\Phi$ is the encoding,  
$U$ is a quantum gate, then $\Phi(U)$ should be the 
``encoded gate'' $U$, which preserves the encoding.
In other words, the following diagram should be commutative:

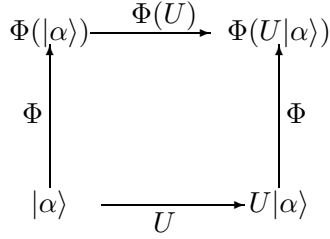
\setlength{\unitlength}{0.030in}

\begin{picture}(40,60)(-50,0)

\put(10,10){\makebox(0,0){$|\alpha\ra$}}
\put(10,40){\makebox(0,0){$\Phi(|\alpha\ra)$}}
\put(10,13){\vector(0,1){25}}
\put(7,26){\makebox(0,0){$\Phi$}}

\put(50,10){\makebox(0,0){$U|\alpha\ra$}}
\put(50,13){\vector(0,1){25}}
\put(50,40){\makebox(0,0){$\Phi(U|\alpha\ra)$}}
\put(53,26){\makebox(0,0){$\Phi$}}

\put(17,40){\vector(1,0){21}}
\put(19,10){\vector(1,0){25}}
\put(30,43){\makebox(0,0){$\Phi(U)$}}
\put(30,7){\makebox(0,0){$U$}}

\end{picture}

Hence, using a code $\Phi$, which takes one qubit to $m$ 
qubits, we replace a quantum circuit by another circuit
which operates on encoded states,
in this circuit 
\begin{itemize}
\item 1 qubit $\longmapsto $ $m$ qubits
\item A gate $U$ $\longmapsto$ $\Phi(U)$
\item Every few time steps, an error correction procedure 
  is applied.
\end{itemize}

However, this naive scheme encounters deep problems.
Since quantum gates create interactions between qubits, errors may propagate through the gates.
Even a small number of errors might  spread to more qubits 
than the error correction can recover. 
Moreover,
we can no longer assume that the recovery operation is 
error free.
The correction procedure might cause more damage
than it recovers.
Consider, for example, a code $\Phi$ that takes one qubit to 
$5$ qubits. A gate on two qubits, $U$, 
 is replaced in the encoded  circuit by the
 encoded gate $\Phi(U)$ which operates on $10$ qubits. 
Let us consider two scenarios:

{~}

{~}

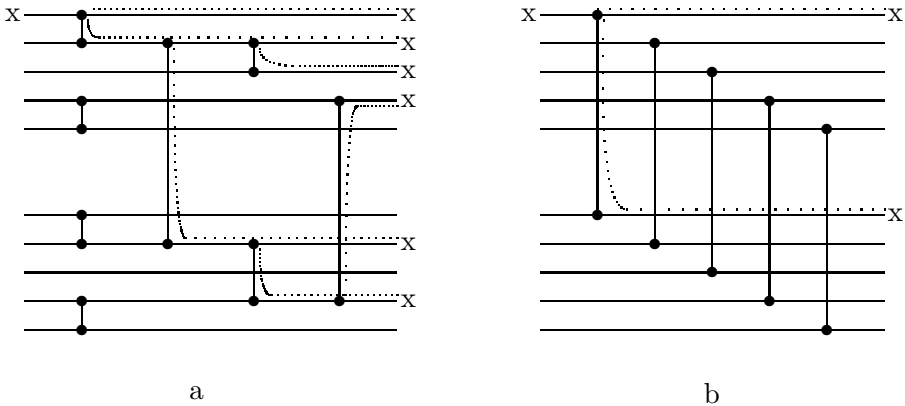
\setlength{\unitlength}{0.030in}

\begin{picture}(40,60)(-10,0)

\put(-2,60){\makebox(0,0){x}}

\put(0,5){\line(1,0){65}}
\put(0,10){\line(1,0){65}}
\put(0,15){\line(1,0){65}}
\put(0,20){\line(1,0){65}}
\put(0,25){\line(1,0){65}}

\qbezier[40](12,61)(32,61)(65,61)

\qbezier[10](41,54)(41,51)(48,51)
\qbezier[25](48,51)(59,51)(65,51)

\qbezier[10](11,59)(11,56)(13,56)
\qbezier[30](13,56)(25,56)(65,56)

\qbezier[30](26,54)(26,21)(28,21)
\qbezier[25](28,21)(50,21)(65,21)

\qbezier[20](41,19)(41,11)(43,11)
\qbezier[20](43,11)(55,11)(65,11)

\qbezier[25](56,11)(56,44)(58,44)
\qbezier[10](58,44)(61,44)(65,44)

\put(67,60){\makebox(0,0){x}}
\put(67,50){\makebox(0,0){x}}
\put(67,45){\makebox(0,0){x}}
\put(67,20){\makebox(0,0){x}}
\put(67,55){\makebox(0,0){x}}
\put(67,10){\makebox(0,0){x}}
\put(0,40){\line(1,0){65}}
\put(0,45){\line(1,0){65}}
\put(0,50){\line(1,0){65}}
\put(0,55){\line(1,0){65}}
\put(0,60){\line(1,0){65}}

\put(88,60){\makebox(0,0){x}}
\put(10,55){\circle*{2}}
\put(10,60){\circle*{2}}
\put(10,55){\line(0,1){5}}

\put(10,40){\circle*{2}}
\put(10,45){\circle*{2}}
\put(10,40){\line(0,1){5}}

\put(10,5){\circle*{2}}
\put(10,10){\circle*{2}}
\put(10,5){\line(0,1){5}}

\put(10,20){\circle*{2}}
\put(10,25){\circle*{2}}
\put(10,20){\line(0,1){5}}

\put(25,20){\circle*{2}}
\put(25,55){\circle*{2}}
\put(25,20){\line(0,1){35}}

\put(40,10){\circle*{2}}
\put(40,20){\circle*{2}}
\put(40,10){\line(0,1){10}}

\put(40,50){\circle*{2}}
\put(40,55){\circle*{2}}
\put(40,50){\line(0,1){5}}

\put(55,10){\circle*{2}}
\put(55,45){\circle*{2}}
\put(55,10){\line(0,1){35}}

\put(30,-6){\makebox(0,0){a}}

\put(90,5){\line(1,0){60}}
\put(90,10){\line(1,0){60}}
\put(90,15){\line(1,0){60}}
\put(90,20){\line(1,0){60}}
\put(90,25){\line(1,0){60}}

\put(90,40){\line(1,0){60}}
\put(90,45){\line(1,0){60}}
\put(90,50){\line(1,0){60}}
\put(90,55){\line(1,0){60}}
\put(90,60){\line(1,0){60}}

\put(100,25){\circle*{2}}
\put(100,60){\circle*{2}}
\put(100,25){\line(0,1){35}}

\put(110,20){\circle*{2}}
\put(110,55){\circle*{2}}
\put(110,20){\line(0,1){35}}

\put(120,15){\circle*{2}}
\put(120,50){\circle*{2}}
\put(120,15){\line(0,1){35}}

\put(130,10){\circle*{2}}
\put(130,45){\circle*{2}}
\put(130,10){\line(0,1){35}}

\put(140,5){\circle*{2}}
\put(140,40){\circle*{2}}
\put(140,5){\line(0,1){35}}

\qbezier[25](100,61)(125,61)(150,61)
\qbezier[20](101,59)(100,26)(105,26)
\qbezier[20](105,26)(130,26)(150,26)

\put(152,25){\makebox(0,0){x}}
\put(152,60){\makebox(0,0){x}}

\put(120,-6){\makebox(0,0){b}}

\end{picture}

{~}

{~}

In figure $(a)$, the encoded gate is a  gate array 
with large connectivity. An error which occurred in the first qubit, 
will propagate through the gates to five more qubits. At the end of 
the procedure, the number of damaged qubits is too large 
for any error correction to take care of. 
Such procedure will not tolerate even one error!
In figure $(b)$, we see an alternative way to implement $\Phi(U)$, 
in which the error cannot propagate to more than one qubit 
 in each block. 
If the  gate is encoded such  that 
one error effects only one qubit in each block, 
we say that the encoded gate is implemented  {\it distributively}.
Such damage will be corrected during the error corrections.
Of course, the error correction procedures should also be implemented in 
a distributed manner. Otherwise the errors generated during the 
correction procedure itself will contaminate the state.

Probably the simplest gate to implement distributively is the encoded 
 NOT gate on Steane's code.
The encoded NOT is simply achieved by applying a NOT gate bitwise on each qubit
in the code. 
The implementation of the XOR gate is applied bitwise as well, and the network 
is the same as that in figure $(b)$, only on $7$ qubits instead of five.
However, for other gates much more work needs to be done.
Shor\cite{shor3}, showed a way to implement a universal set of gates 
in this way, where the implementation of some of the gates, and Toffoli's 
gate in particular, require 
some hard work and the use of additional ``ancilla'' or ``working'' 
 qubits. 
Together with the set of universal encoded gates, one also needs 
an error correction procedure, an encoding procedure 
to be used in the beginning of the computation, and a decoding procedure
to be used  at the end. All these procedures should be 
 implemented distributively, 
to prevent propagation of errors.
A code
 which is accompanied by a set of universal gates, encoding, decoding
 and correction procedures, all implemented distributively, will be called a {\it quantum computation code}.
Since Shor's suggestion, other computation codes were found\cite{aharonov1,knill2}. Gottesman\cite{gottesman2} has generalized these results
and showed  how to construct a computation code from any 
 stabilizer code.

Is the encoded circuit more reliable?
The {\it effective noise rate}, $\eta_e$ of the encoded circuit,
is the probability for an encoded gate to
 suffer a number of errors which cannot be corrected. 
In the case of figure $(b)$, one error is still recoverable, but 
two are not. 
The effective noise rate is thus the probability for two or more
 errors to occur in $U(\Phi)$. Let $A$ denote the  
 number of places in the implementation of $U(\Phi)$ where errors can 
occur. $A$ stands for the {\it area} of $U(\Phi)$. 
The probability for more than $d$ errors to occur can be bounded
 from above, using simple counting arguments:  
\begin{equation}\label{noise}
\eta_e\le \left(\begin{array}{c}
A\\d+1\end{array}\right)\eta^{d+1}
\end{equation}
We will refer to this bound as the {\it  effective noise rate.}
To make a computation of size $n$ reliable, we need an effective noise rate of the 
order of $\frac{1}{n}$. 
Using a code with blocks of $\rm{log}(n)$ qubits, Shor\cite{shor3}
 managed to show that 
 the computation will be reliable, with
polynomial cost. However, Shor had to assume that
$\eta$ is as small as $O(\frac{1}{\log^4(n)})$.
This assumption is not  physically reasonable ,
 since $\eta$ is a parameter of the system, 
independent of the computation size.
The reader is urged to play with the parameters of equation \ref{noise}
in order to be convinced that  assuming $\eta$ to be constant cannot lead to a 
polynomially small effective noise rate, as required.

Another idea, which was found independently by several groups
\cite{ aharonov1,knill2, kitaev2,gottesman5}
 was needed to close the gap, and to show that 
computation in the presence of constant noise rate and finite precision 
is possible. 
The idea is simple. Apply Shor's scheme recursively, gaining small improvement in
the effective noise rate 
each level . 
Each circuit is replaced by a slightly more reliable circuit, 
which is replaced again by yet another circuit.
If each level gains only a slight improvement from $\eta$ to $\eta^{1+\epsilon}$, 
then the final circuit  
 which is the one implemented in the laboratory, 
will have an effective noise rate exponentially smaller:
\[\eta\longmapsto \eta^{1+\epsilon}\longmapsto  (\eta^{1+\epsilon})^{1+\epsilon}...
\longmapsto  \eta^{(1+\epsilon)^r}\]
The number of levels the recursion should be applied 
to get a polynomially small effective noise rate is only  $O(\log(\log(n)))$.
The cost in time and space is thus only polylogarithmic.
 A similar concatanation scheme was used in the  context of classical 
self correcting cellular automata\cite{tsirelson,gacs}.

The requirement that the noise rate is improved
 from one level to the next 
 imposes a threshold requirement on $\eta$: 

\[ \left(\begin{array}{c}
A\\d+1\end{array}\right)\eta^{d+1} < \eta\]

If $\eta$ satisfies the above requirement, fault tolerant computation can be achieved.  This is known as the threshold result\cite{ aharonov1,knill2, kitaev0,gottesman5}:

\begin{theo}\label{fault} {\bf Fault tolerance: }
Quantum computation of any length
 can be applied efficiently with arbitrary level of confidence, if the 
noise rate is smaller than the threshold $ \eta_c$.
\end{theo}

The {\it threshold} $\eta_c$,
 depends on the parameters of the computation code: 
$A$, the largest procedure's area, and $d$, the number
 of errors which the code 
can correct. 
Estimations\cite{aharonov1,knill2,gottesman2,gottesman5,
knill4, preskill2} of $\eta_c$ 
are in the range between $ 10^{-4}$ and $10^{-6}$. 
Presumably the
correct threshold is much higher.  
The highest noise rate in which  
 fault tolerance is possible is not known yet.

The rigorous proof of the threshold theorem is quite complicated.
 To gain some insight we can view the
 final $r'$th circuit  as a multi scaled system, 
where computation and error correction are
 applied 
in many scales simultaneously.
The largest procedures,  computing on the largest (highest level) blocks, 
 correspond to operations on the logical qubits, i.e. qubits in the original circuit. 
The smaller procedures,  operating on smaller blocks,
correspond to computation in lower levels.
Note, that each level simulates the error corrections in the 
previous level, and adds error corrections in the current level.
The final circuit, thus, includes error corrections of all the levels,
where during the computation of error corrections of larger blocks 
 smaller blocks of lower levels are being 
corrected.
The lower the level, the more often 
error corrections of this level are applied, which is in correspondence with 
the fact that smaller blocks
 are more likely to be quickly damaged.

The actual system consists of 
$m=n\log^c(n)$ qubits (where $n$ is the size of the 
original circuit), with a  Hilbert space
 ${\cal H}=C^{2^m}$.
In this Hilbert space we find a subspace, isomorphic to 
$C^{2^n}$, which is protected against noise. 
This subspace is a complicated  multi-scaled construction, 
which is  small in dimensions, compared to the 
Hilbert space of the system, 
but not negligible. The subspace is  
protected against noise for almost as long as we wish, and 
the quantum computation is done  exactly in this protected subspace.
The rate by which the state 
increases its distance from this subspace
corresponds to the noise rate. 
The efficiency of the 
 error correction determines the rate 
by which the distance from this subspace decreases. 
The threshold in the noise rate is the point 
where distance is decreases faster than it increases. 
In a sense, the situation can be viewed as the operation of a 
renormalization group,
 the change in the noise rate being the renormalization flow.

{~}

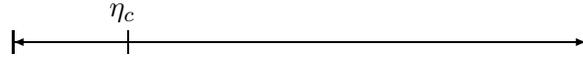
\setlength{\unitlength}{0.030in}
\begin{picture}(40,0)(-40,0)
\put(20,0){\vector(-1,0){20}}
\put(20,0){\vector(1,0){80}}
\put(0,-2){\line(0,1){4}}
\put(20,-2){\line(0,1){4}}
\put(100,-2){\line(0,1){4}}
\put(19,5){\makebox(0,0){$\eta_c$}}

\end{picture}

{~}

It should be noted that along the proof of fault tolerance, 
a few implicit assumptions were made \cite{steane7}.
The ancilla qubits that we need in the middle of the computation
for error correction are assumed to 
 be prepared in state $|0\ra$ {\it when needed}, and not at the beginning
of the computation. 
This requires the ability to cool part of the system constantly. 
It was shown by Aharonov {\it et. al.}\cite{aharonov4} that if all operations are unitary, 
the system keeps warming (in the sense of getting
more noise)  with no way to cool, and
the rate in which the system warms up is {\it exponential}. 
 Fault tolerant quantum computation requires using 
non-unitary gates which enables to cool a qubit.
This ability to cool qubits is used implicitly 
in all fault tolerant schemes.
Another point which should be mentioned is that 
 fault tolerant computation uses immense parallelism, i.e. 
there are many gates which are applied at the same time.
Again, this implicit assumption is essential. 
If operation were sequential, fault tolerant computation would 
have been impossible, as was shown  by Aharonov and Ben-Or\cite{aharonov1}.
However, with mass parallelism, constant supply of cold qubits 
and a noise rate which is smaller than $\eta_c$, 
it is possible to perform fault tolerant computation.

The fault tolerance result holds for the general 
local noise model, as defined before, 
 and this includes 
probabilistic collapses, 
inaccuracies, systematic errors, decoherence, etc.
One can compute fault tolerantly 
 also with quantum circuits which are
 allowed to operate only on nearest neighbor qubits\cite{aharonov1}
( In this case the threshold $\eta_c$ will be smaller, because the 
procedures are bigger when only 
nearest neighbor interactions are allowed. )
In a sense, the question of noisy quantum computation is theoretically closed.
But a question still ponders our minds:
Are the assumptions on the  noise correct?
Dealing with non-local noise is an open and challenging problem.

\section{Conclusions and Fundamental Questions}
We cannot foresee  which goals will be 
 achieved, if quantum computers be  the
 next step in the evolution of  
computation\cite{haroche}. This question involves two directions of research.
From the negative side, 
 we are still very far from understanding the 
limitations of quantum computers as 
computation devices. 
It is possible that quantum Fourier transforms are the only real
 powerful tool in quantum computation. Up to now, this is the only tool
which implies exponential advantage over classical algorithms. However, 
 such a strong statement of the uniqueness of the Fourier transform 
is not known.  
Taking a more positive view, the goal is to find other techniques
in  addition to the  Fourier transform. 
One of the main directions of research in quantum algorithms is 
finding an efficient solutions for a number of problems which 
are not known to be NP complete, but  
do not have a known efficient classical solution. 
Such is the problem of checking whether two graphs are isomorphic,
known as {\it Graph Isomorphism}.
Another important direction in quantum algorithms is 
finding algorithms that simulate quantum physical systems 
more efficiently. 
The field of quantum complexity is still in its infancy.

Hand in hand with the complexity questions, arise 
deep fundamental questions about quantum physics.
The computational power of all classical systems seem 
to be equivalent, 
whereas quantum complexity, in light of the above results,
 seems inherently different. 
     If it is true that quantum systems are exponentially better 
            computation devices than classical systems, this 
           can give rise to a new definition of quantum versus classical
             physics, and might lead to a change in the 
            way we understand the transition 
             from  quantum to classical physics.
The ``phase diagram'' of quantum versus classical 
behavior can be viewed as follows:


{~}

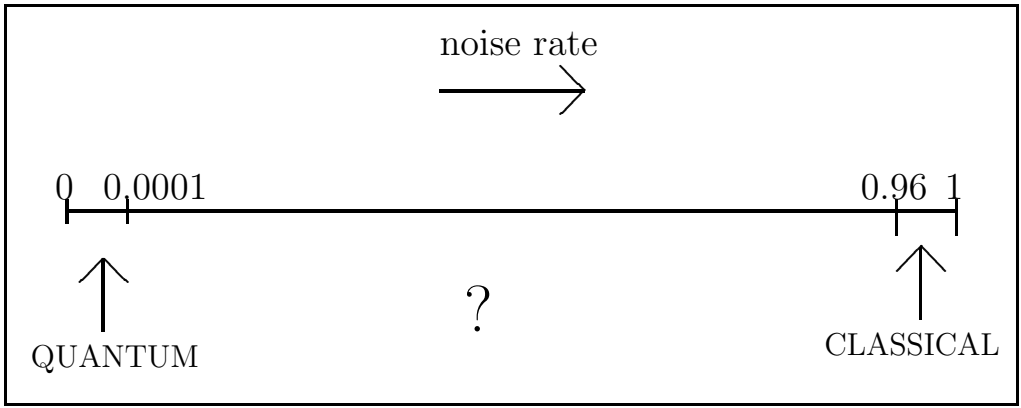
\setlength{\unitlength}{0.00083300in}%
\begingroup\makeatletter\ifx\SetFigFont\undefined
\def\x#1#2#3#4#5#6#7\relax{\def\x{#1#2#3#4#5#6}}%
\expandafter\x\fmtname xxxxxx\relax \def\y{splain}%
\ifx\x\y   
\gdef\SetFigFont#1#2#3{%
  \ifnum #1<17\tiny\else \ifnum #1<20\small\else
  \ifnum #1<24\normalsize\else \ifnum #1<29\large\else
  \ifnum #1<34\Large\else \ifnum #1<41\LARGE\else
     \huge\fi\fi\fi\fi\fi\fi
  \csname #3\endcsname}%
\else
\gdef\SetFigFont#1#2#3{\begingroup
  \count@#1\relax \ifnum 25<\count@\count@25\fi
  \def\x{\endgroup\@setsize\SetFigFont{#2pt}}%
  \expandafter\x
    \csname \romannumeral\the\count@ pt\expandafter\endcsname
    \csname @\romannumeral\the\count@ pt\endcsname
  \csname #3\endcsname}%
\fi
\fi\endgroup
\begin{picture}(6324,2499)(1639,-4648)
\thicklines
\put(2026,-3436){\line( 1, 0){5550}}
\put(2401,-3361){\line( 0,-1){150}}
\put(2251,-3736){\line( 0,-1){450}}
\multiput(2101,-3886)(6.00000,6.00000){26}{\makebox(6.6667,10.0000){\SetFigFont{7}{8.4}{rm}.}}
\multiput(2251,-3736)(6.00000,-6.00000){26}{\makebox(6.6667,10.0000){\SetFigFont{7}{8.4}{rm}.}}
\put(2026,-3361){\line( 0,-1){150}}
\put(7201,-3361){\line( 0,-1){225}}
\put(7576,-3361){\line( 0,-1){225}}
\put(7351,-3661){\line( 0,-1){450}}
\multiput(7201,-3811)(6.00000,6.00000){26}{\makebox(6.6667,10.0000){\SetFigFont{7}{8.4}{rm}.}}
\multiput(7351,-3661)(6.00000,-6.00000){26}{\makebox(6.6667,10.0000){\SetFigFont{7}{8.4}{rm}.}}
\multiput(5101,-2536)(6.00000,-6.00000){26}{\makebox(6.6667,10.0000){\SetFigFont{7}{8.4}{rm}.}}
\multiput(5101,-2836)(6.00000,6.00000){26}{\makebox(6.6667,10.0000){\SetFigFont{7}{8.4}{rm}.}}
\put(4351,-2686){\line( 1, 0){900}}
\put(1651,-4636){\framebox(6300,2475){}}
\put(1801,-4411){\makebox(0,0)[lb]{\smash{\SetFigFont{12}{14.4}{rm}QUANTUM}}}
\put(6751,-4336){\makebox(0,0)[lb]{\smash{\SetFigFont{12}{14.4}{rm}CLASSICAL}}}
\put(4501,-4186){\makebox(0,0)[lb]{\smash{\SetFigFont{34}{40.8}{rm}?}}}
\put(4351,-2461){\makebox(0,0)[lb]{\smash{\SetFigFont{14}{16.8}{rm}noise rate}}}
\put(1951,-3361){\makebox(0,0)[lb]{\smash{\SetFigFont{14}{16.8}{rm}0}}}
\put(2251,-3361){\makebox(0,0)[lb]{\smash{\SetFigFont{14}{16.8}{rm}0.0001}}}
\put(6976,-3361){\makebox(0,0)[lb]{\smash{\SetFigFont{14}{16.8}{rm}0.96}}}
\put(7501,-3361){\makebox(0,0)[lb]{\smash{\SetFigFont{14}{16.8}{rm}1}}}
\end{picture}

{~}

Changing the noise rate, the system transforms from quantum behavior
to classical behavior. 
As was shown by Aharonov and Ben-Or\cite{aharonov2}, there is a
 constant $\eta$ bounded away from $1$
where the system cannot perform quantum computation at all. 
Fault tolerance shows that there is a constant $\eta$ 
bounded away from $0$ for which quantum systems exhibit their full 
quantum computation power.
The regimes are characterized by the range of quantum entanglement, 
where in the quantum regime this range is macroscopic,
and quantum computation is possible. 
On the right, ``classical'', range, entanglement is confined to microscopic
clusters.
A very interesting question is how does the transition between 
the two regimes occur. 
In  \cite{aharonov2} we gave 
indications to the fact
 that the transition is sharp and  has many characteristics of a  
 phase transition (and see also \cite{tsirelson1}.) 
The order parameter corresponds to the range of entanglement, 
or to the size of entangled clusters of qubits. 
Unfortunately,  we 
were unable yet to prove the existence of such a phase transition, 
presumably because of lack of the correct definition 
of an order parameter that
 quantifies ``quantumness over large scales''. 
Never the less I conjecture that the transition from 
macroscopic quantum behavior to macroscopic  classical behavior
occurs as  a phase transition.
The idea that the transition from quantum to classical physics
is abrupt stands  in contrast to the standard 
view of a gradual transition due to decoherence\cite{zurek1}. 
I believe that the flippant frontier between quantum 
and classical physics will be better understood
if we  gain better understanding of the transition from 
quantum to classical computational behavior.

An interesting conclusion of the threshold result is that 
one dimensional quantum systems can exhibit a non trivial 
phase transition at a critical noise rate $\eta_c$, below which 
 the mixing time of the system is exponential,
 but above which the system mixes rapidly. 
This phase transition might be different from the transition
from classical to quantum behavior, or it might be the same.
This existence of a one dimensional 
phase transition is interesting because 
 one dimensional phase transitions 
are rare, also in classical systems, though there exist
several complicated examples\cite{mukamel,
gacs1}. 

Perhaps a vague, but
 deeper, and more thought provoking question is that of the postulates
of quantum mechanics.
The possibility that the model will be realized
will enable a thorough test of
 some of the more philosophical aspects of quantum theory, such 
as understanding the collapse of the wave function, 
the process of measurement, and other elements which are used as 
everyday tools in quantum algorithms.
It might be that the realization of quantum computation
will reveal the fact that what we understand in quantum physics
 is merely an approximation holding only for small number of
 particles, which we extrapolated to 
 many particles. 
Such questions are appealing motivations for 
this extremely challenging task of
 realizing the quantum computation model physically.
It seems that successes, and also failures, in achieving this 
ambitious task,  will open new exciting paths and possibilities 
 in both computer science and fundamental physics.

\section{Acknowledgments}
I am most grateful to Michael Ben Or who introduced me to this 
beautiful subject. We had a lot of fascinating discussions together
on many of the things I presented. 
 Noam Nisan taught me a lot 
simply by asking the right questions, with his clear point of view. 
It was a pleasure not to know the answers. 
It was   great  fun to  argue with Avi Wigderson on quantum 
computation and other things. 
It is a special pleasure to thank my colleagues Peter Hoyer, 
Lidror Troyanski, Ran Raz and particularly Michael Nielsen. 
They all  
 read the manuscript, corrected
 many errors, and had extremely helpful suggestions.
 Finally, I thank Ehud Friedgut for direct and 
                     indirect contributions to this review.

\footnotesize
\bibliographystyle{plain}

\end{document}